\documentclass[PAPER, atlasdraft=false, cernpreprint, USenglish, texmf, orcidlogo]{atlasdoc}
\usepackage[subcaption,subfig, multirow, backend=biber,default]{atlaspackage}
\usepackage{caption}
\usepackage{lscape}
\usepackage{bigstrut}
\usepackage{atlasbiblatex}

\usepackage{atlasphysics}
\ifthenelse{\boolean{AtlasDraft}}{\usepackage[output=true, shift=true]{}}{}

\graphicspath{{logos/}{figures/}}

\AtlasTitle{Search for events with one displaced vertex from long-lived neutral particles decaying into hadronic jets in the ATLAS muon spectrometer in $pp$ collisions at $\sqrt{s}=13$~TeV}

\AtlasAbstract{%
A search for events with one displaced vertex from long-lived particles using data collected by the ATLAS detector at the Large Hadron Collider is presented, using 140~fb$^{-1}$ of proton--proton collision data at $\sqrt{s} = 13$~TeV~recorded in \mbox{2015--2018}. The search employs techniques for reconstructing vertices of long-lived particles decaying into hadronic jets in the muon spectrometer displaced between 3~m and 14~m from the primary interaction vertex. The observed number of events is consistent with the expected background and limits for several benchmark signals are determined. A scalar-portal model and a Higgs-boson-portal baryogenesis model are considered. A dedicated analysis channel is employed to target  Z-boson associated long-lived particle production, including an axion-like particle and a dark photon model. For the Higgs boson model, branching fractions above 1\% are excluded at 95\% confidence level for long-lived particle proper decay lengths ranging from 5~cm to~40 m. For the photo-phobic axion-like particle model considered, this search produces the strongest limits to date for proper decay lengths greater than $\mathcal{O}(10)$~cm.
}

\AtlasRefCode{EXOT-2022-17}

\PreprintIdNumber{CERN-EP-2025-062}

\AtlasJournalRef{Phys.\ Rev.\ D.\ 112,\ 092001}
\AtlasDOI{https://doi.org/10.1103/cmql-s9sq}

\AtlasCoverEgroupAnalysisTeam{atlas-exot-2022-17-analysis-team@cern.ch}

\hypersetup{pdftitle={ATLAS document},pdfauthor={The ATLAS Collaboration}}

\begin{document}

\maketitle

\tableofcontents

\section{Introduction}

The discovery of the Higgs boson at the Large Hadron Collider (LHC)~\cite{Evans:2008zzb} marked a significant milestone in completing the Standard Model (SM) of elementary particles. However, many fundamental aspects of our universe remain unexplained, including dark matter, neutrino mass, matter-antimatter asymmetry, and the hierarchy problem.

Addressing these unresolved issues has led to the development of numerous theoretical frameworks beyond the SM (BSM). Many of these frameworks predict the existence of long-lived particles (LLPs) with a proper lifetime times the speed of light ($c\tau$) between a few centimeters and hundreds of meters. These LLPs can then decay into SM particles far from the interaction point (IP), resulting in displaced vertices (DVs). Examples include various supersymmetric (SUSY) models addressing the hierarchy problem like mini-split SUSY~\cite{Arvanitaki:2012ps, ArkaniHamed:2012gw}, gauge-mediated SUSY breaking~\cite{Giudice:1998bp}, $R$-parity-violating SUSY~\cite{Barbier:2004ez, Csaki:2013jza}, Stealth SUSY~\cite{Josh1,Josh2}, and Neutral Naturalness~\cite{Chacko:2015fbc,Burdman:2006tz, Cai:2008au, Chacko:2005pe}, as well as Hidden Valley~\cite{Strassler:2006im,Strassler:2006ri} models and others addressing dark matter~\cite{Baumgart:2009tn, Kaplan:2009ag, Chan:2011aa, Dienes:2011ja, Dienes:2012yz} and the matter--antimatter asymmetry of the universe~\cite{CuiSundrum:2012,Cui:2013,Cui:2014twa}. Models that aim to explain the origin of neutrino masses~\cite{Helo:2013esa,Batell:2016zod} frequently include LLPs as well.

While searches for LLPs decaying into final states containing jets have been conducted at various colliders, including LEP~\cite{LLPDelphi}, the Tevatron~\cite{CDF, D0}, and the LHC~\cite{HVAtlas,LHCb,HVAtlas8TeV,SUSY-2014-02,CMS,CalRatio, LHCb-PAPER-2016-065,LHCb-1612.00945,CMS-EXO-16-003,CMS-EXO-17-018, CMS-EXO-18-007, EXOT-2017-05, EXOT-2019-24, EXOT-2017-24, SUSY-2016-08, EXOT-2017-25,LHCb-2011.05263,LHCb-1612.00945,LHCb-1705.07332}, no evidence of BSM neutral LLPs has been observed to date.

This search focuses on identifying DVs occurring in the ATLAS muon spectrometer (MS) using 140 fb$^{-1}$ of $\sqrt{s} = 13$~\TeV\ $pp$ collision data. This study builds on the previous searches for one- and two-DVs~\cite{EXOT-2017-05, EXOT-2019-24}, using an improved background estimation methodology and a refined signal efficiency extrapolation based on the LLP lifetime. The event selection criteria and vertex reconstruction algorithms are designed to target candidate events with a single MS DV. The analysis primarily interprets results in terms of scalar portal, baryogenesis, axion-like particle (ALP), and dark photon models, though it remains sensitive to additional models.

The previous searches for two DVs~\cite{EXOT-2017-05, EXOT-2019-24} in the MS benefit from very low background and have excellent sensitivity for $c\tau$ values of tens of meters, but the requirement that both particles decay within the MS reduces the sensitivity for longer $c\tau$ values, with exclusion limits on cross-sections scaling as $(c\tau)^{-2}$. By requiring only one DV in the MS, this analysis significantly extends the sensitivity to longer-lived LLPs, with exclusion limits scaling as $(c\tau)^{-1}$, and adds sensitivity to models producing single LLPs.

The paper proceeds with a detailed description of the ATLAS detector in Section~\ref{sec:detector}, an examination of the theoretical models in Section~\ref{sec:models}, an overview of the analysis strategy in Section~\ref{sec:analysis}, a description of the $pp$ collision data and Monte Carlo simulation in Section~\ref{sec:datamc}, and an explanation of the specialized trigger and displaced vertex reconstruction algorithms in Section~\ref{sec:trigvtxreco}. Section~\ref{sec:eventselection} is dedicated to describe the signal event selection, while Section~\ref{sec:Bkg1MSVx} describes the background estimation. The systematic uncertainties are discussed in Section~\ref{sec:syst}, and the results are presented in Section~\ref{sec:result}, including the combination with the search for two DVs~\cite{EXOT-2019-24}.

\section{ATLAS detector}
\label{sec:detector}

The ATLAS detector~\cite{PERF-2007-01}, which has nearly 4$\pi$ steradian coverage,\footnote{ATLAS uses a right-handed coordinate system with its origin at the nominal IP in the center of the detector and the $z$-axis along the beam pipe. The $x$-axis points from the IP to the center of the LHC ring, and the $y$-axis points upwards. Polar coordinates $(r,\phi)$ are used in the transverse plane, $\phi$ being the azimuth angle around the $z$-axis. The pseudorapidity is defined in terms of the polar angle $\theta$ as $\eta = - \ln \tan(\theta/2)$ and is equal to the rapidity $y = \frac{1}{2} \ln \left ( \frac{E+p_z}{E-p_z}\right )$ in the relativistic limit. Angular distance is measured in units of $\Delta R \equiv \sqrt{(\Delta y)^{2} + (\Delta\phi)^{2}}$.} is a multipurpose detector consisting of an inner tracking detector (ID) surrounded by a superconducting solenoid providing a \SI{2}{\tesla} axial magnetic field, electromagnetic and hadronic calorimeters, and a muon spectrometer based on three large air-core toroidal superconducting magnets, each with eight coils. The field integral of the toroids ranges between \num{2.0} and \SI{6.0}{\tesla\metre}
across most of the detector.

The ID covers the pseudorapidity range \(|\eta| < 2.5\). It consists of a silicon pixel detector, a silicon microstrip detector, and a straw-tube transition-radiation tracker.

The calorimeter system covers the pseudorapidity range $|\eta| < 4.9$. It consists of a high-granularity electromagnetic calorimeter (ECal) surrounded by a hadronic calorimeter (HCal). Within the region $|\eta|< 3.2$, the ECal comprises liquid-argon (LAr) barrel and endcap electromagnetic calorimeters with lead absorbers. An additional thin LAr presampler covering $|\eta| < 1.8$ is used to correct for energy loss in material upstream of the calorimeters. The ECal extends from 1.5~m to 2.0~m in $r$ in the barrel and from 3.6~m to 4.25~m in $|z|$ in the endcaps.
The HCal is a steel/scintillator-tile calorimeter that is segmented into three barrel structures within $|\eta| < 1.7$, and two copper/LAr hadronic calorimeters in the endcaps ($1.5 < |\eta| < 3.2$). The HCal covers the region from 2.25~m to 4.25~m in $r$ in the barrel (although the HCal active material extends only up to 3.9~m) and from 4.3~m to 6.05~m in $|z|$ in the endcaps. The solid angle coverage is completed with forward copper/LAr and tungsten/LAr calorimeter modules optimized for electromagnetic and hadronic measurements, respectively. Together the ECal and HCal have a thickness of 9.7~interaction lengths at $\eta=0$.

The MS comprises three stations of separate trigger and tracking chambers that measure the deflection of muons in a magnetic field generated by the air-core toroid magnets. The barrel and endcap chamber systems are subdivided into 16 sectors: eight large sectors and eight small sectors. For the barrel, the small sectors are inside each of the eight magnet coils and the large sectors are between the coils. Three stations of resistive-plate chambers (RPC) and thin-gap chambers (TGC) are used for triggering in the MS barrel and endcaps, respectively. The first two RPC stations, which are radially separated by 0.5~m, begin at a radius of either 7~m (large sectors) or 8~m (small sectors). The third station is located at a radius of either 9~m (large sectors) or 10~m (small sectors). The TGC chambers are arranged into 24 or 48 sectors depending on their $\eta$ and radial position. The first TGC station is located at $|z|=13$~m. The other two stations start at $|z|=14$~m and $|z|=14.5$~m, respectively.

The muon tracking chamber system covers the region $|\eta| < 2.7$ with three stations of monitored drift tubes (MDT), complemented by cathode-strip chambers (CSC) in the forward region. The barrel chambers are arranged in three concentric cylindrical shells around the beam axis at radii of roughly 5~m, 7.5~m, and 10~m. In the endcap regions, muon chambers are configured as large, wheel-shaped structures perpendicular to the $z$-axis. These are located at distances of approximately $|z| = 7.4$~m, 10.8~m, 14~m, and 21.5~m from the interaction point, ensuring comprehensive muon detection coverage across the detector. The MDT chambers consist of two multilayers separated by a distance ranging from 6.5~mm to 317~mm. Each multilayer consists of three or four layers of drift tubes. The individual drift tubes are 30~mm in diameter and have lengths of 2--5~m (barrel) and 2--6.5~m (endcaps) depending on the location of the chamber in the spectrometer.
In each multilayer, charged-particle track-segment reconstruction entails finding the line that is tangent to the drift circles. These single-multilayer segments provide local measurements of the position and direction of the charged particle. Because of its design, the MDT measurement provides only a very coarse $\phi$ position of the track hit. To reconstruct the $\phi$ position and direction, the MDT measurements are combined with the $\phi$-coordinate measurements from the trigger chambers.

The ATLAS trigger and data acquisition system~\cite{TRIG-2016-01} consists of a hardware-based first-level (L1) trigger followed by a software-based high-level trigger (HLT) that reduces the rate of events selected for offline storage to 1~kHz.

The implementation of the L1 muon trigger logic is similar for the RPC and TGC systems. Each of the three stations of the RPC system and the two outermost stations of the TGC system consists of a doublet of independent detector layers.  The first TGC station contains a triplet of detector layers. The transverse momentum (\pT) of the muon candidate is measured by the L1 muon trigger, using different algorithms for low-\pT and high-\pT triggers. In the barrel, a low-\pT (${<}10$~\GeV) muon region-of-interest (RoI) is generated by requiring a coincidence of hits in at least three of the four layers of the two inner RPC stations. In the endcaps, the trigger requires hits in the two outer TGC stations. A high-$\pT$ muon RoI in the barrel requires additional hits in at least one of the two layers of the outer RPC station, while for the endcaps, additional hits in two of the three layers of the innermost TGC station are required.
The muon RoIs have an angular extent of $0.2\times0.2$ in $\Delta\eta \times \Delta\phi$ in the MS barrel and $0.1\times0.1$ in $\Delta\eta \times \Delta\phi$ in the MS endcaps. The muon trigger system covers the range $|\eta| < 2.4$.

An extensive software suite~\cite{SOFT-2022-02} is used in the reconstruction and analysis of real and simulated data, in detector operations, and in the trigger and data acquisition systems of the experiment.

\section{Signal model}
\label{sec:models}

While sensitive to a broad spectrum of models, the results of the previous analysis~\cite{EXOT-2017-05, EXOT-2019-24} were interpreted in the context of a \emph{scalar portal}~\cite{Strassler:2006ri,NonSMCoupling2} model. The present analysis expands the scope by additionally considering models motivated by \emph{baryogenesis}~\cite{Cui:2014twa}, \emph{axion-like particles}~\cite{Brivio_2017}, and \emph{dark photons}~\cite{darkZ}.

\begin{figure}[tb]
\begin{subfigure}[b]{0.49\textwidth}
\includegraphics[width=0.9\textwidth]{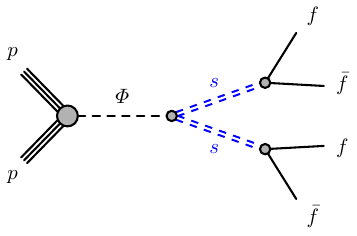}
\caption{}\label{fig:HiggsHadronicJets_diagram}
\end{subfigure}
\begin{subfigure}[b]{0.49\textwidth}
\includegraphics[width=0.9\textwidth]{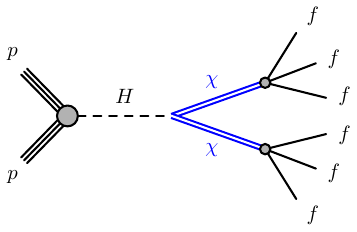}
\caption{}\label{fig:BaryogenesisH_diagram}
\end{subfigure}
\caption{Diagrams of the (a) scalar portal and (b) baryogenesis models. The LLPs are represented by double lines and labeled $s/\chi$, and the final-state SM fermions are labeled as $f$~\cite{Strassler:2006ri}.}
\end{figure}

The scalar portal model considered here is described in Refs.~\cite{EXOT-2017-05, EXOT-2019-24}. Its key features include the production of a Higgs boson, $H$, with a mass of 125~\GeV, or an alternative scalar boson, denoted by $\varPhi$, which decays into two long-lived scalars, as illustrated in Figure~\ref{fig:HiggsHadronicJets_diagram}.  In these scenarios, the 125 GeV Higgs boson or the alternative scalar boson weakly mixes with a scalar particle from the hidden sector.  This mixing can produce pair production of hidden sector scalars or pseudoscalars that carry no SM quantum numbers.  A weak coupling of the decay particles to the SM gives rise to the macroscopic decay lengths of $\gtrsim \mathcal{O}(1)$~cm.

Indirect upper limits on the branching fractions of BSM Higgs boson decays, of the order of 12\%, can be inferred from combined studies of Higgs boson production and decay~\cite{HIGG-2021-23}, under various assumptions, permitting substantial branching fractions for decays into non-SM particles. Long-lived scalars with a mass from 5 to 30~\GeV\ are also well motivated by naturalness models that are generic extensions of hidden-valley portal models~\cite{Uncolored}. Illustrated in Figure~\ref{fig:HiggsHadronicJets_diagram}, the mechanism for LLP production entails the decay of a scalar boson $H/\varPhi$ with some effective coupling into a pair of long-lived scalars, $s$. These scalars then decay into SM particles. In this model, the couplings of the scalar $s$ to SM fermions are dictated by the Higgs boson Yukawa couplings. Similar to the previous analysis assumptions~\cite{EXOT-2019-24}, each long-lived scalar primarily decays into the heaviest kinematically accessible fermion pair. For $m_s \gtrsim 25$~\GeV, the branching fractions into $b\overline{b}$, $c\overline{c}$, and $\tau^{+}\tau^{-}$ are nearly independent of $m_s$ and equal to 85\%, 5\%, and 8\%, respectively. While the previous ATLAS MS DV publication~\cite{EXOT-2017-05} only considered these decay modes, in this work the decay into $t\overline{t}$ is added, which becomes dominant once kinematically feasible. In this model, the branching fraction for $H/\varPhi$ decaying into a pair of long-lived scalars remains unconstrained, so it is set to 100\%. Given its dominance, this analysis solely considers the gluon--gluon fusion production mode.

The origin of the cosmic asymmetric abundance of baryons remains one of the most prominent questions in physics and requires new physics beyond the SM\@. Electroweak baryogenesis is one possible mechanism that could account for the baryon asymmetry and can be explored at the LHC\@. In the baryogenesis model~\cite{Cui:2014twa}, the lowest-dimension operator coupling a singlet $\chi$ to the SM is the Higgs boson portal. The simplest realization of this interaction is with a scalar, $\varPhi$, which mixes with the SM Higgs boson~\cite{NonSMCoupling2}. If $\varPhi$ has a Yukawa coupling to a pair of $\chi$, this leads to the Higgs boson portal production of $\chi$ via exchange of a single Higgs boson after mixing, $pp \rightarrow H \rightarrow \chi\chi$, as shown in Figure~\ref{fig:BaryogenesisH_diagram}. The decay modes of the $\chi$ must violate baryon or lepton number conservation, which generates the baryonic asymmetry. The lowest-dimensional interactions of this type allow $\chi$ to decay into three SM fermions. Similarly to the scalar portal, the small coupling between the $\chi$ and the SM is responsible for the possible macroscopic decay length of the $\chi$. Since LHC experiments have established the existence of a Higgs boson with a mass of 125~\GeV~\cite{HIGG-2019-16,CMS-HIG-19-004}, a minimal scenario is considered. The model assumes that the additional scalar $\varPhi$ is heavy and decouples, and focuses on the production channel via the SM Higgs boson portal.

\begin{figure}[tb]
\centering
\includegraphics[width=0.441\textwidth]{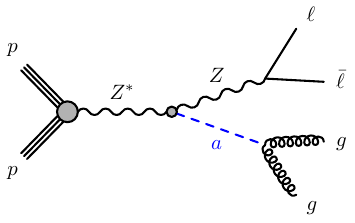}
\caption{Diagram of the ALP production through an intermediate, off-shell $\Zboson$ boson. The long-lived ALP is represented by a dashed line and labeled $a$, while the final-state SM charged leptons are labeled as $\ell$, and the SM gluons as $g$.}
\label{fig:ZALPFeyn}
\end{figure}

\begin{figure}[tb]
\centering
\begin{subfigure}[b]{0.49\textwidth}
\includegraphics[width=0.9\textwidth]{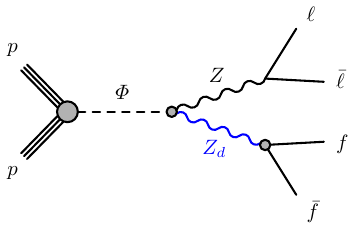}
\caption{}
\end{subfigure}
\begin{subfigure}[b]{0.49\textwidth}
\includegraphics[width=0.9\textwidth]{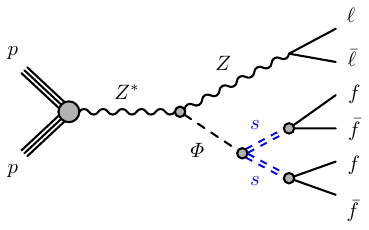}
\caption{}
\end{subfigure}
\caption{Diagrams of the (a) dark photon and (b) \Zboson-associated scalar models. The long-lived dark photon is represented by $Z_d$, the long-lived scalars are labeled $s$, the final-state SM charged leptons are labeled as $\ell$, and the final-state SM fermions are labeled as $f$.}
\label{fig:ZH_HZZD_Feyn}
\end{figure}

A second set of models is considered that involve the production of a neutral LLP in association with a prompt \Zboson. The first of these is the ALP model, shown in Figure~\ref{fig:ZALPFeyn}. Other examples include the dark photon and \Zboson-associated scalar models, which are illustrated in Figure~\ref{fig:ZH_HZZD_Feyn}. ALPs are new pseudo-scalars that are associated with the breaking of a global symmetry~\cite{Brivio_2017}. ALPs can couple to many SM particles, resulting in a variety of final states: photons, gluons, quarks, electroweak gauge bosons and the Higgs boson. While the ALPs can couple to either \Zboson or \Wboson bosons, the analysis targets only \Zboson-associated production due to the additional background rejection possible via selecting both outgoing leptons.

The production and decay of the ALP particle is controlled by its mass $m_a$, the couplings to the primary fields of the SM mediators ($C_{\tilde{G}}, C_{\tilde{B}}$, and $C_{\tilde{W}}$) and the ALP energy scale $f_a$. Several searches have addressed ALP signals decaying into photons, resulting in no excess above the SM background~\cite{fips}. For this search, due to the strong existing limits on photon-coupled models, a photo-phobic scenario is assumed in which the coupling of ALPs to photons $g_{a \gamma}$, derived from the couplings to the primary fields, is set to zero by imposing the condition $C_{\tilde{B}} = - \tan^2 (\theta_W) C_{\tilde{W}}$. For photo-phobic ALPs, the production cross-section is proportional to $C_{\tilde{W}}^2$, thus $C_{\tilde{W}}=1$ is set to maximize the simulated cross-section. Further, $f_a$ is fixed to $\SI{1}{\TeV}$.
Therefore, two distinct parameters ($m_a$ and $C_{\tilde{G}}$) regulate the cross-section of the process and the lifetime of the ALP candidate. For the models considered, the branching fraction of the ALP decay into gluons is assumed to be 100\% and only $\Zboson \rightarrow e \bar{e}, \mu \bar{\mu}$ decays are considered.

Another prompt-\Zboson containing model is the long-lived dark photon ($Z_d$) model~\cite{darkZ}, where the $Z_d$ is produced along with a SM \Zboson during the decay of an initial-state scalar~\cite{higgsDarkSector}, similarly to the scalar portal model. The associated production of a \Zboson and a dark photon is predicted by the inclusion of a new $U(1)_d$ dark symmetry to the SM\@. In this model, the lifetime of the dark photon is a free parameter and only $\Zboson \rightarrow e \bar{e}, \mu \bar{\mu}$ are simulated. As with the scalar portal model, the $Z_d$ is assumed to decay primarily into quark pairs.

Lastly, simulations of the scalar portal model are extended to include the production of the initial scalar and an associated \Zboson, shown alongside the dark photon process in Figure~\ref{fig:ZH_HZZD_Feyn}. For this model, all decays of the \Zboson to charged leptons are considered. While the presence of the additional \Zboson boson greatly reduces the background contribution, the reduced cross-section results in generally lower sensitivity to SM Higgs boson branching to long-lived scalars. The $\Zboson+$ALP, $Z_d$, and \Zboson-associated scalar portal models were considered in previous ATLAS searches for displaced jets in the ATLAS calorimeter in association with prompt leptons~\cite{EXOT-2022-04} and for displaced vertices in the ATLAS inner detector~\cite{EXOT-2021-32}. Extending these searches to vertices in the muon spectrometer improves sensitivity at longer lifetimes, due to the positioning of the calorimeters and inner detector closer to the collision point.

Within the two sets of models, the efficiency is primarily impacted by the kinematic properties of the LLPs, set by their specific production and decay modes, and by the relative masses of the particles involved. All details about the relative masses and lifetimes simulated for the signal models, along with comprehensive information about Monte Carlo (MC) event generation are reported in Section~\ref{sec:datamc}.

\section{Analysis strategy}
\label{sec:analysis}

The analysis presented here searches for events with one DV in the MS\@. %
The analysis is divided into two channels targeting events with or without an additional prompt \Zboson boson.

Candidate events with no associated boson production are selected by the Muon RoI Cluster HLT trigger~\cite{MuonRoITrigger}, described in Section~\ref{sec:trigvtxreco}, while events with prompt \Zboson bosons are selected using single- or di- lepton triggers~\cite{TRIG-2018-01,TRIG-2018-05}. In both channels a specialized algorithm to reconstruct DVs in the MS~\cite{PERF-2013-01} is used. Additional selection criteria are used to maximize the analysis sensitivity and are detailed in Section~\ref{sec:trigvtxreco}.

The primary background for LLPs decaying into hadronic jets in the MS stems from hadronic or electromagnetic showers that are not fully confined within the calorimeter volume, leading to tracks reconstructed in the MS — commonly referred to as ``punch-through jet''. Multijet events giving rise to vertices in the MS often exhibit ID tracks and calorimeter jets directed towards the MS DV\@. To minimize the acceptance of such spurious vertices originating from multijet events, isolation criteria are applied requiring separation in both $\eta$ and $\phi$ from ID tracks and calorimeter jets. Multijet events that include jets with mismeasured energy or direction, often due to incomplete detector coverage or misalignment, represent an additional background in LLP analyses. These events can satisfy the isolation criteria and be misidentified as signal.
Even though the \Zboson selection applied in the lepton-triggered channel greatly reduces the rate of punch-through jet background, \Zboson + jets processes where the jets punch through to the MS still form the dominant background.

Further background, termed \emph{non-collision background}, may arise from electronic noise in the MDT and RPC/TGC chambers, cosmic-ray muons, and beam-induced background~\cite{BIB} stemming from hadronic and electromagnetic showers caused by beam protons interacting with collimators or residual gas molecules within the vacuum pipe.

\section{Data and Monte Carlo simulation}
\label{sec:datamc}

The analysis presented uses \mbox{proton--proton} ($pp$) collision data at a center-of-mass energy of $\sqrt{s} = 13$~\TeV\ recorded by the ATLAS detector during the LHC 2015--2018 data-taking periods~\cite{DAPR-2018-01}. To ensure data integrity, events are exclusively selected from periods characterized by stable LHC beams and operational detector subsystems. Following stringent data quality requirements~\cite{DAPR-2018-01}, the total integrated luminosity amounts to 140~fb$^{-1}$.

Signal MC simulation samples for the scalar, and dark photon models were generated by employing the Hidden Abelian Higgs Model~\cite{NonSMCoupling2} at leading-order using \MGNLO~\cite{MadGraph5}, considering only the gluon--gluon fusion production mechanism. The baryogenesis and ALP model samples were generated at leading-order using \MGNLO~\cite{MadGraph5} following the \texttt{WIMP\textunderscore{}BG\textunderscore{}higgsportal\textunderscore{}full\textunderscore{}loop}~\cite{Cui:2014twa} model and \texttt{ALP\textunderscore{}linear\textunderscore{}UFO\textunderscore{}WIDTH}~\cite{Brivio_2017} model respectively. For all samples, the generation was interfaced to the \PYTHIA[8.2]~\cite{Sjostrand:2014zea} parton shower model with the A14 parameter tune~\cite{ATL-PHYS-PUB-2014-021}. While the \NNPDF[3.1lo] parton distribution function (PDF) set~\cite{Ball:2012cx} was employed for the scalar and baryogenesis models, the \NNPDF[2.3lo] PDF set was used in all models featuring \Zboson-boson associated LLP production. The \EVTGEN[1.2.0] program~\cite{EvtGent} was used to model $b$- and $c$-hadron decays for all samples. The transverse momentum $p_T$ distribution of the scalar boson $H/\varPhi$ was re-weighted to match that obtained for the corresponding next-to-leading-order (NLO) Higgs boson samples simulated using the \MGNLO FxFx approach~\cite{Frederix:2012ps} for matching and merging.
To ensure comprehensive coverage of the accessible parameter space a range of masses was selected, as outlined in Tables~\ref{tab:MCSamples} and~\ref{tab:MCSamples2}. The mean proper decay length in each sample was tuned to optimize the occurrence of decays across the entirety of the ATLAS detector volume, encompassing a range of 0.13 to 6 meters, depending on the specific sample.

The generated events were processed through a full simulation of the ATLAS detector geometry and response~\cite{GEANT4b} using the \GEANT~\cite{GEANT4a} toolkit. The simulation includes multiple $pp$ interactions per bunch crossing (pileup) and the effect on the detector response due to interactions from bunch crossings before or after the one containing the hard interaction. Pileup was simulated with the soft strong-interaction processes of \PYTHIA[8.1]~\cite{Sjostrand:2007gs} using the A3 tune~\cite{ATL-PHYS-PUB-2016-017} and the \NNPDF[2.3lo]~\cite{Ball:2012cx} PDF set. Per-event weights were applied to the simulated events to match the distribution of the average number of interactions per bunch crossing measured in data.

A multijet background sample was simulated using \PYTHIA[8.230]~\cite{Sjostrand:2014zea} with leading-order matrix elements for dijet production which were matched to the parton shower. The \NNPDF[2.3lo] PDF set was used in the matrix element generation, the parton shower, and the simulation of the multi-parton interactions. The A14 tune was used. Perturbative uncertainties were estimated through event weights~\cite{Mrenna:2016sih} that encompass variations of the scales at which the strong coupling constant is evaluated in the initial- and final-state showers and the PDF uncertainty in the shower and the non-singular part of the splitting functions.

\begin{table}[tb]
\centering
\caption{Parameters used for the simulated scalar portal based on the Hidden Abelian Higgs Model~\cite{NonSMCoupling2}, the baryogenesis model~\cite{Cui:2014twa}, and the photo-phobic ALP model~\cite{Brivio_2017}. }
\begin{tabular}{cccl}
\toprule
\toprule
Model & $m_{H/\varPhi}$ [\GeV]           & $m_s$ [\GeV]     & Proper decay length [m] \\
\hline
\multirow{14}{*}{Scalar portal} & \multirow{4}{*}{125}  & ~~5  &  0.127, 0.411  \\
&   & 16  &  0.580  \\
&   & 35  &  1.310, 2.630  \\
&   & 55  &  1.050, 5.320  \\ \cline{2-4}
& \multirow{2}{*}{~~60}    & ~~5       &  0.217 \\
&                          & 16        &  0.661 \\ \cline{2-4}
& \multirow{1}{*}{200}     & 50        &  1.255 \\ \cline{2-4}
& \multirow{1}{*}{400}     & 100~~     &  1.608 \\ \cline{2-4}
& \multirow{3}{*}{600}     & 50        &  0.590 \\
&                          & 150~~     &  1.840, 3.309  \\
&                          & 275~~     &  4.288 \\ \cline{2-4}
& \multirow{3}{*}{1000~~}  & 50        &  0.406 \\
&                          & 275~~     &  2.399, 4.328  \\
&                          & 475~~     &  6.039 \\
\hline \hline
& $m_\chi$ [\GeV]  & $\chi$ decay channel & Proper decay length [m] \\
\hline
\multirow{3}{*}{Higgs boson portal baryogenesis} & 10     & \multirow{3}{*}{$\tau^+ \, \tau^- \, \nu$ , $c \, b \, s$ ,  $\nu \, b \, \bar{b}$} &  0.920 \\
& 55     &  &  5.550 \\
& 100    &  &  3.500 \\

\hline\hline
& $m_a$ [\GeV]          & $C_{\tilde{G}}$       & Proper decay length [m] \\
\hline
\multirow{7}{*}{$\Zboson+$ALP}  & \multirow{2}{*}{0.1}  & $10^{-2}$  & 0.003 \\
&                       & $5 \times 10^{-3}$   & 0.012 \\ \cline{2-4}
& 1                     & $10^{-4}$             & 0.031 \\
\cline{2-4}
& \multirow{3}{*}{10}   & $3.2 \times 10^{-6}$  & 0.030 \\
&                       & $10^{-6}$             & 0.310 \\
&                       & $7.5 \times 10^{-7}$  & 0.551 \\
\cline{2-4}
& 40                    &  $10^{-7}$            & 0.481 \\
\bottomrule
\bottomrule
\end{tabular}\label{tab:MCSamples}
\end{table}

\begin{table}[tb]
\centering
\caption{Parameters used for the simulated \Zboson-associated scalar portal and dark photon models.}
\begin{tabular}{cccl}
\toprule
\toprule
Model & $m_{H/\varPhi}$ [\GeV]           & $m_s$ [\GeV]     & Proper decay length [m] \\
\hline
\multirow{16}{*}{\shortstack{\Zboson-associated \\ scalar portal}} & \multirow{4}{*}{125}  & ~~5  &  0.100, 0.300  \\
&   & 16  &  0.300  \\
&   & 35  &  0.750, 2.500  \\
&   & 55  &  3.500, 1.000  \\ \cline{2-4}
& \multirow{2}{*}{~~60}    & 5       &  0.12 \\
&                          & 15        &  0.25 \\ \cline{2-4}
& \multirow{2}{*}{200}     & 50        &  1.25 \\
&                          & 80        &  2.00 \\ \cline{2-4}
& \multirow{2}{*}{400}     & 100~~     &  1.25 \\
&                          & 175       &  2.50 \\ \cline{2-4}
& \multirow{3}{*}{600}     & 50        &  0.40 \\
&                          & 150~~     &  1.50, 3.50  \\
&                          & 275~~     &  2.50 \\ \cline{2-4}
& \multirow{3}{*}{1000~~}  & 50        &  0.30 \\
&                          & 275~~     &  1.50, 3.50  \\
&                          & 475~~     &  4.50 \\
\hline \hline
& $m_{H/\varPhi}$ [\GeV]           & $m_{Z_d}$ [\GeV]     & Proper decay length [m] \\
\hline
\multirow{8}{*}{Dark photon} & \multirow{2}{*}{125}  & ~~5  &  0.60  \\
&   & 15  &  1.60, 3.00  \\ \cline{2-4}
& \multirow{2}{*}{250}     & 50        &  1.60 \\
&                          & 100       &  3.40 \\ \cline{2-4}
& \multirow{2}{*}{400}     & 100       &  1.60 \\
&                          & 200       &  4.00 \\ \cline{2-4}
& \multirow{2}{*}{600}     & 150~~     &  1.60, 4.00 \\
&                          & 400       &  4.00 \\
\bottomrule
\bottomrule
\end{tabular}\label{tab:MCSamples2}
\end{table}

Given the wide range of mean proper decay lengths pertinent to this analysis, producing numerous samples to encompass this breadth would demand excessive processing time. Thus, a MC method employing pseudo-data was used to extrapolate the expected event count across mean proper decay lengths ranging from 0.001 to 1000 meters.
For each LLP in the MC sample, a decay time from an exponential distribution corresponding to a specified proper decay length was generated. Next, the physical decay distance in the detector for each simulated LLP based on its four-momentum was calculated. The overall probability of an event to satisfy the signal selection criteria, parameterized by the LLP decay position and boost, was then evaluated from the LLP trigger and vertex efficiencies detailed in Sections \ref{sec:RoICluTrig} and \ref{sec:1MSVxEff}, respectively. To validate the lifetime extrapolation procedure, secondary sets of samples with different lifetimes for select LLP masses were generated.

\section{Trigger selection and event reconstruction}
\label{sec:trigvtxreco}

Hadronic decays of LLPs in the MS typically generate narrow, high-multiplicity hadronic showers. The track multiplicity and shower width depend on the mass and boost of the decaying LLP and the final states to which the LLP decays. To address this, specialized  vertex~\cite{PERF-2013-01} algorithms were devised for the online selection and reconstruction of displaced decays in the MS, and specialized trigger~\cite{MuonRoITrigger} algorithms were devised for events without additional prompt objects. Due to the substantial material present in the calorimeter, only decays occurring within or after the last sampling layer of the hadronic calorimeter yield enough hits in the MS for DV reconstruction.

\subsection{Muon RoI Cluster trigger}\label{sec:RoICluTrig}

The Muon RoI Cluster trigger is a signature-driven trigger aimed at identifying candidate events associated with LLP decays in the MS~\cite{MuonRoITrigger} for signatures without additional prompt objects to use for triggering. To qualify, events must contain a minimum of two L1 muon RoIs with \pT higher than 10~\GeV. At the HLT, a cluster of three (four) muon RoIs, positioned within a $\Delta R=0.4$ cone centered on the L1 object, is required in the barrel (endcaps) region. The isolation criteria for jets and tracks outlined in Ref.~\cite{MuonRoITrigger}, which mitigate contributions from punch-through jets, are not used in the analysis presented herein. Consequently, the trigger encompasses both isolated, signal-like events and non-isolated, \mbox{background-like} events utilized for data-driven background estimates.

\begin{figure}[tb]
\begin{subfigure}[b]{0.49\textwidth}
\includegraphics[width=\textwidth]{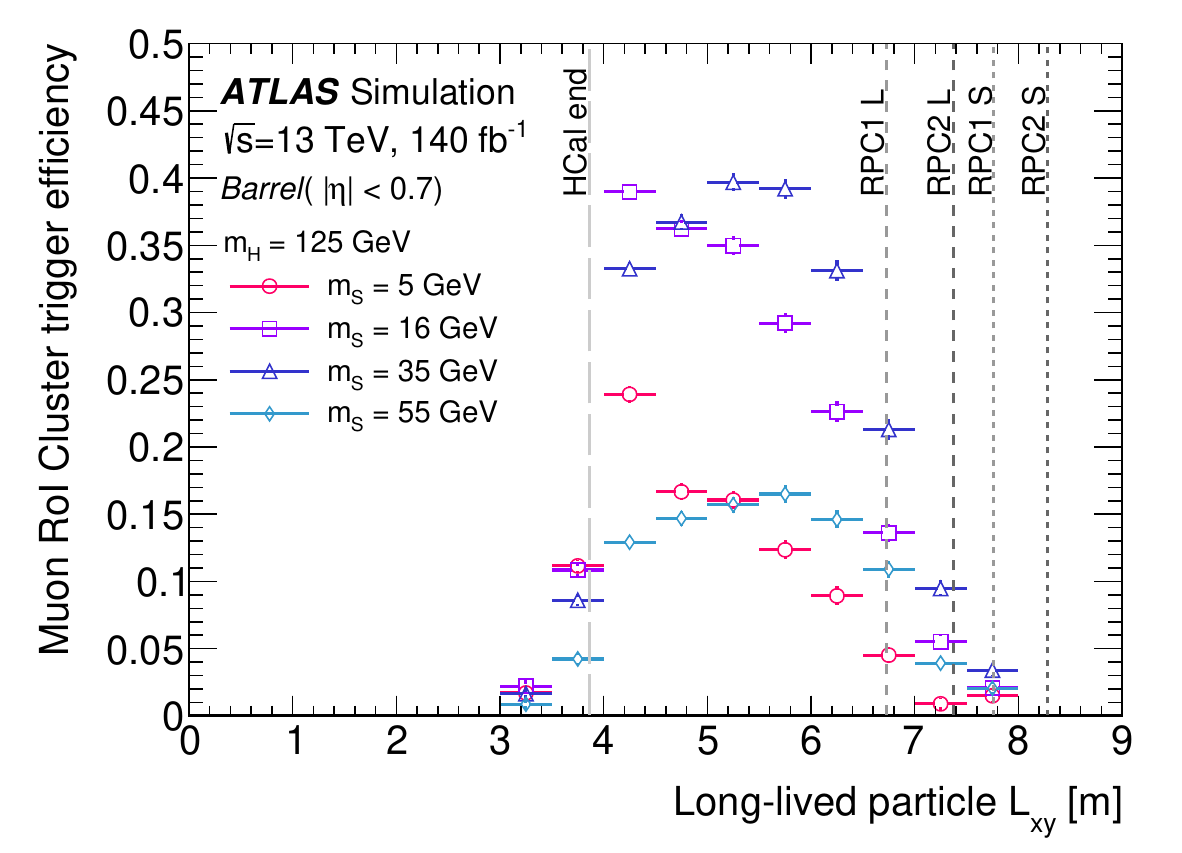}
\caption{}\label{fig:MuonClusterTriggerEffBarrel}
\end{subfigure}
\begin{subfigure}[b]{0.49\textwidth}
\includegraphics[width=\textwidth]{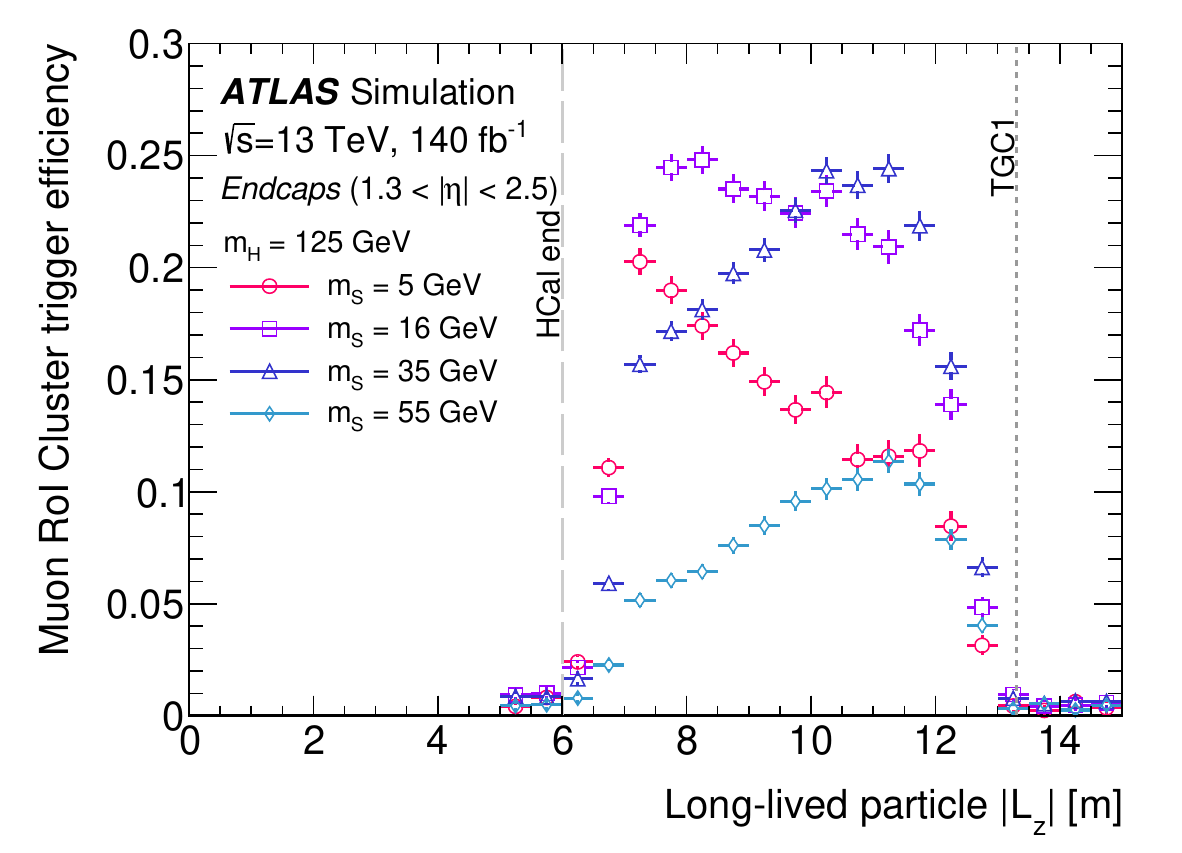}
\caption{}\label{fig:MuonClusterTriggerEffEndcaps}
\end{subfigure}
\caption{Efficiency for the Muon RoI Cluster trigger against the LLP decay position for some scalar portal samples in the (a) MS barrel and (b) MS endcaps. The distributions include events in which only one LLP decays in the MS, satisfying the data quality requirements and featuring a reconstructed primary vertex. These efficiency profiles are derived exclusively from MC simulation and do not undergo any corrections for mis-modeling, as outlined in Section~\ref{sec:RoICluTrig}. The vertical lines in the plots denote relevant detector boundaries. ``HCal end'' signifies the outer boundary of the hadronic calorimeter, while ``RPC 1/2'' indicates the first/second stations of RPC chambers. ``TGC 1'' represents the first stations of TGC chambers, and ``L/S'' denotes the Large or Small sectors.}
\end{figure}

Figures~\ref{fig:MuonClusterTriggerEffBarrel} and~\ref{fig:MuonClusterTriggerEffEndcaps} depict the trigger efficiency, in the MS barrel and endcap regions respectively, for four MC-simulated benchmark samples with LLP decays. To ensure orthogonality with respect to the results reported in Ref.~\cite{EXOT-2019-24}, only events in which a single LLP decays in the MS are considered. The efficiency is quantified as the fraction of LLP decays within the fiducial volume of the MS that are selected by the trigger, as a function of the LLP decay position and is characterized in terms of the transverse decay radius ($L_{xy}$) in the barrel and the longitudinal decay position ($|L_z|$) in the endcaps. The events are required to satisfy data quality requirements~\cite{DAPR-2018-01} and have a reconstructed primary vertex, as detailed in Section~\ref{sec:PVJetReco}.

The trigger demonstrates efficiency for hadronic LLP decays spanning the outer regions of the HCal to the middle stations of the MS\@. These efficiencies are derived from a subset of simulated signal events with only a single LLP decay in the MS, ensuring that trigger outcomes arise from a single burst of MS activity. The indicated uncertainties pertain solely to statistical variations. The observed discrepancies in efficiencies among benchmark samples stem from variations in LLP masses, consequently influencing their momenta and, by extension, the opening angles of the decay products. The trigger efficiency is reduced for highly boosted decays with small opening angles, where the decay products can become clustered into a single RoI, and for decays with large opening angles, where the cluster may no longer be contained within $\Delta R=0.4$ of the L1 object.

Notably, the trigger efficiency increases as LLPs decay nearer to the end of the hadronic calorimeter, situated at approximately $r \approx 4$~m for the barrel and $|z| \approx 6$~m for the endcaps. Conversely, the efficiency experiences a sharp decrease as the decay events approach the middle stations of the muon spectrometer (barrel: $r\approx 7$~m; endcaps: $|z| \approx 13$~m). Between these two regions, the more-boosted decays have larger efficiency closer to the beamline, where they have a larger distance to spread out and produce multiple RoIs, while decays with larger opening angles have their maximum efficiency at larger distances, where the RoI produced by the decay products are more likely to remain clustered together. Furthermore, material interactions with the support structures can result in non-smooth features. Through the collimation of the LLP decay products, this depends on the sample mass point. The larger flux of particles reaching the MS can increase the trigger efficiency, especially for the lower mass samples where the showers from the LLP decay are narrow and sometimes not resolved into distinct RoIs.

To assess the mis-modeling of the L1 muon trigger efficiency in MC simulation, the distributions of the number of muon-RoI clusters within a cone of $\Delta R = 0.4$ surrounding the axis of a punch-through jet in both multijet MC and data events are compared. High-energy jets in the data were identified using jet triggers with a \pT threshold of either 400~\GeV\ or 420~\GeV, depending on the specific data-taking period. A punch-through jet is required to have $p_T>30$~\GeV and at least 50 muon segments in a $\Delta R$ cone of 0.4 around the jet axis.

The MC simulation shows a muon-RoI cluster rate approximately 24\% (20\%) lower than that observed in data within the barrel (endcaps) region. Correcting this mis-modeling would have significantly increased the trigger efficiency for all samples. As this difference could be produced by mis-modeling of the punch-through jets themselves rather than the muon-RoI reconstruction alone, it is treated as a systematic uncertainty in the signal trigger efficiency instead of being applied as a correction.

\subsection{Reconstruction of MS displaced-vertices}

Using a dedicated algorithm~\cite{PERF-2013-01}, capable of reconstructing low-momentum tracks in a busy environment, MS DVs are reconstructed. This algorithm, previously employed in searches for displaced decays within the MS~\cite{HVAtlas,HVAtlas8TeV,EXOT-2017-05}, leverages the spatial separation between the multilayers within a single MDT chamber. Straight-line segments containing three or more MDT hits within a single multilayer are reconstructed using a $\chi^2$ fit. Segments from multilayer 1 are then matched with those from multilayer 2, forming ``tracklets'' with their associated track parameters.

Spatially clustered tracklets within $|\Delta\eta| < 0.7$ and $|\Delta\phi| < \pi/3$ in the detector are extrapolated backward to reconstruct the $(\eta,\phi)$ position of the MS DV using a $\chi^2$ fit. In the barrel (endcaps), at least three (four) tracklets are required for use in the fit. Detectable vertices originate from decays occurring between the outer edge of the HCal and the middle stations of the muon chambers. Given their distinct detector technology lacking spatially separated multilayers, CSC are not used in the MS DV reconstruction.

\subsection{Reconstruction of the primary vertex and prompt hadronic jets}\label{sec:PVJetReco}

Events must feature a primary vertex (PV) comprising at least two tracks, each with \pT exceeding 500~\MeV. In scenarios where multiple PV candidates are reconstructed, the candidate with the highest sum of the squares of the transverse momenta of all associated tracks is selected.

Hadronic jets are constructed by using FastJet~\cite{Fastjet} to apply the \antikt jet algorithm~\cite{Cacciari:2008gp} with a radius parameter $R=0.4$. A collection of three-dimensional topological clusters of neighboring energy deposits in the calorimeter cells containing a significant energy above a noise threshold~\cite{PERF-2014-07} provides input to the \antikt algorithm. After reconstruction, jets are calibrated using the procedure outlined in Refs.~\cite{PERF-2014-07,ATL-PHYS-PUB-2015-036}.

\subsection{Lepton reconstruction}

All candidate leptons (electrons and muons) are required to have $\pT>10~\GeV$. Electrons are required to have $\lvert\eta\rvert< 2.47 $, while muons must have $\lvert\eta\rvert<2.5$. Electrons that pass through the gap region between the barrel and endcap calorimeters at $1.37<\lvert\eta\rvert< 1.52 $ are excluded. All leptons must satisfy the requirements for \emph{Medium} identification and \emph{Loose} isolation and have a track origin consistent with the primary vertex, following the conventions defined in Refs.~\cite{EGAM-2018-01, MUON-2018-03}.

\section{Event selection}
\label{sec:eventselection}

In this search, all events undergo an event-level selection process aimed at distinguishing signal from background.

\subsection{Baseline event selection}

The events analyzed in this search are subject to common baseline selections in each channel, with the specific criteria detailed in Table~\ref{tab:baselineSelection}.

\begin{table}[!h]
\centering
\caption{Summary of the baseline selection criteria applied to both data and simulated events. The variables $n_{\textrm{MDT}}$/$n_{\textrm{RPC}}$/$n_{\textrm{TGC}}$ indicate the counts of MDT/RPC/TGC hits within the vertex cone, as explained in the text. Additionally, $\eta_{\textrm{DV}}$ represents the pseudorapidity of the DV relative to the IP.}
\begin{tabular}{p{\dimexpr0.45\textwidth-0.5\tabcolsep} p{\dimexpr0.45\textwidth-0.5\tabcolsep}}
\toprule
\toprule
\multicolumn{2}{c}{Baseline selection criteria} \\
\midrule
\multicolumn{2}{l}{Event passes data quality requirements and Muon RoI Cluster or Lepton triggers} \\
\midrule
\multicolumn{2}{l}{Event has a PV with at least two tracks with $\pT > 500$~\MeV} \\
\midrule
\multicolumn{2}{l}{Event has exactly one DV} \\
\midrule
\multicolumn{2}{l}{\emph{If muon-RoI triggered,}} \\
\multicolumn{2}{l}{DV is matched to the triggering muon-RoI cluster, $\Delta R(\text{DV}, \text{cluster})<0.4 $} \\
\midrule
\multicolumn{2}{l}{\emph{If lepton-triggered,}} \\
\multicolumn{2}{l}{Exactly one opposite-sign, same-flavor lepton pair with an invariant mass $>$ 60~\GeV and $<$ 120~\GeV} \\
\multicolumn{2}{l}{No additional leptons not used in the pair} \\
\midrule
\multicolumn{2}{l}{300 $\le n_{\textrm{MDT}} <$ 3000} \\
\hline \hline
\multicolumn{2}{c}{Additional selection criteria for barrel and endcaps} \\
\midrule
\emph{Barrel} & \emph{Endcaps} \\
\midrule
\multicolumn{2}{l}{\emph{If muon-RoI triggered,}} \\
$n_{\textrm{RPC}} \geq 800$   & $n_{\textrm{TGC}} \geq 900$   \\
\midrule
DV with $|\eta_{\textrm{DV}}| < 0.7$  & DV with $1.3 < |\eta_{\textrm{DV}}| < 2.5$ \\
\midrule
DV with $3\,\text{m} < L_{xy} < 8\,\text{m}$ & DV with $L_{xy} < 10\,\text{m}$, $5\,\text{m} < |L_z| < 15\,\text{m}$ \\
\bottomrule
\bottomrule
\end{tabular}
\label{tab:baselineSelection}
\end{table}

Events with exactly one DV are required to satisfy the data quality requirements~\cite{DAPR-2018-01} and contain a PV\@. While the PV selection minimally affects the signal efficiency, it aids in background event rejection; in simulations, the chosen PV typically corresponds to the signal interaction in approximately 95\%--99\% of the cases, depending on the sample. Despite LLPs being invisible in the ID, in the Higgs boson scalar models the scalar boson is produced with an average \pT of 20 \GeV, leaving several tracks from recoiling particles generated in the same $pp$ interaction, while in the \Zboson-associated models the PV is directly produced by the leptons originating from the prompt \Zboson decay.

Events are categorized into two non-orthogonal channels to target signatures produced with or without additional prompt objects. In the ``muon-RoI'' channel, primarily targeting the Higgs boson scalar and baryogenesis models, events must satisfy the Muon RoI Cluster trigger, where the DV is matched to the trigger-level muon-RoI cluster by requiring $\Delta R(\textrm{DV, cluster}) < 0.4$. In the ``lepton-triggered'' channel, primarily targeting the axion-like particle model with associated \Zboson bosons, events must fulfill single- or dilepton triggers, with the trigger-matched lepton having a transverse momentum greater than 27 \GeV. Additionally, the trigger-matched lepton must form exactly one pair with an opposite-sign, same-flavor lepton forming an invariant mass greater than 60 \GeV and lower than 120 \GeV to select for leptons likely to originate from the decay of a \Zboson boson. Any events with additional leptons satisfying the baseline selections that are not included in the \Zboson candidate are rejected.

A DV that originates from a displaced decay typically has many more hits than a DV from background. To take advantage of this difference, a minimum number of 300 MDT hits ($n_{\textrm{MDT}}$) and 800/900 RPC/TGC hits ($n_{\textrm{RPC}}$/$n_{\textrm{TGC}}$) is required. The $n_{\textrm{MDT}}$ hits are counted from MDT chambers whose centers lie within $\Delta \phi = 0.6$ and $\Delta \eta = 0.6$ of the DV$(\eta,\phi)$ direction. Similarly, $n_{\textrm{RPC}}$ and $n_{\textrm{TGC}}$ hits encompass the sum of hits within $\Delta R = 0.6$ of the DV\@. The requirements on $n_{\textrm{RPC}}$ and $n_{\textrm{TGC}}$ are removed for the lepton-triggered channel, where the requirement of additional prompt objects greatly reduces the overall background event rate. In both channels, a cap on the maximum number of MDT hits is enforced to eliminate background events stemming from coherent noise bursts in the MDT chambers, with negligible impact on signal events.

When a displaced decay occurs in the transition area between the MS barrel and endcaps, hits are recorded in both regions. Vertex reconstruction is carried out separately in the barrel and endcaps, with only the respective hits utilized in each region's reconstruction algorithm. Consequently, MS vertices reconstructed using either algorithm have fewer hits, being derived from a subset of the complete hit set. This results in a reduction of the reconstruction efficiency, and occasionally leads to two vertices being reconstructed from a single LLP decay. Therefore, DVs with pseudorapidity $|\eta_{\textrm{DV}}|$ between 0.8 and 1.3 are excluded from the analysis. This exclusion has a minimal impact on total signal efficiency, given that the average DV reconstruction efficiency in this range is less than 2\%.

Additionally, in the transition zone between the barrel and endcap hadronic calorimeters, $0.7 < |\eta_{\textrm{DV}}| < 1.2$, the likelihood of a jet failing minimal selection criteria for isolation and penetrating into the MS is significantly higher than in other detector regions. This area overlaps with the previously excluded MS transition region, except for $0.7 < |\eta_{\textrm{DV}}| < 0.8$. Consequently, vertices reconstructed with pseudorapidity $0.7 < |\eta_{\textrm{DV}}| < 0.8$ are also disregarded.

DVs are required to be reconstructed within the MS fiducial volume, defined by $3\,\text{m} < L_{xy} < 8\,\text{m}$ in the barrel, while in the endcaps, DVs must satisfy $L_{xy} < 10\,\text{m}$ and $5\,\text{m} < |L_z| < 15\,\text{m}$.

\subsection{Signal displaced-vertex selection}

To mitigate background vertices generated by punch-through jets, a set of vertex isolation criteria was devised. These criteria are based on the angular distance, $\Delta R$, between the direction of the tracks or jets and the vertex axis, defined as the line from the IP to the DV\@. No jets or tracks should be present in a $\Delta R$ cone around the MS DV axis.

\begin{figure}[tb]
\begin{subfigure}[b]{0.49\textwidth}
\includegraphics[width=\textwidth]{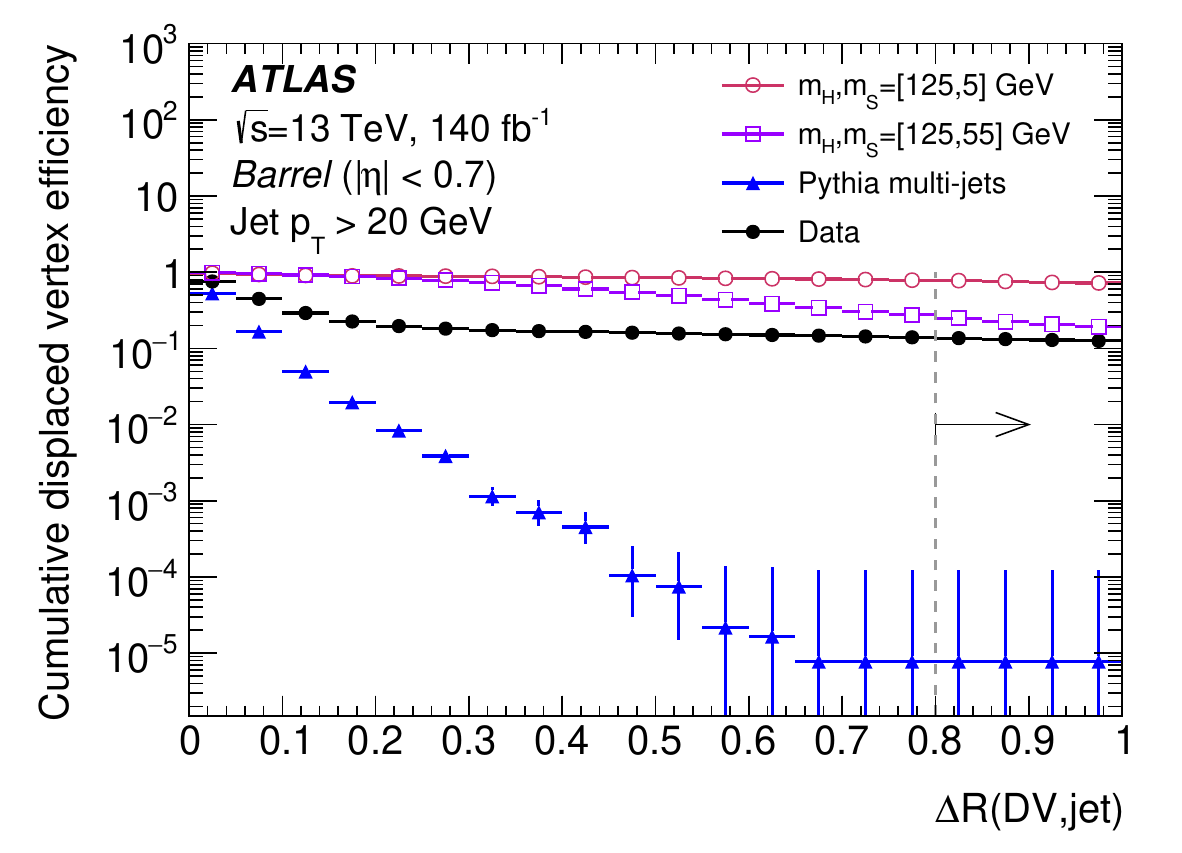}
\caption{}
\end{subfigure}
\begin{subfigure}[b]{0.49\textwidth}
\includegraphics[width=\textwidth]{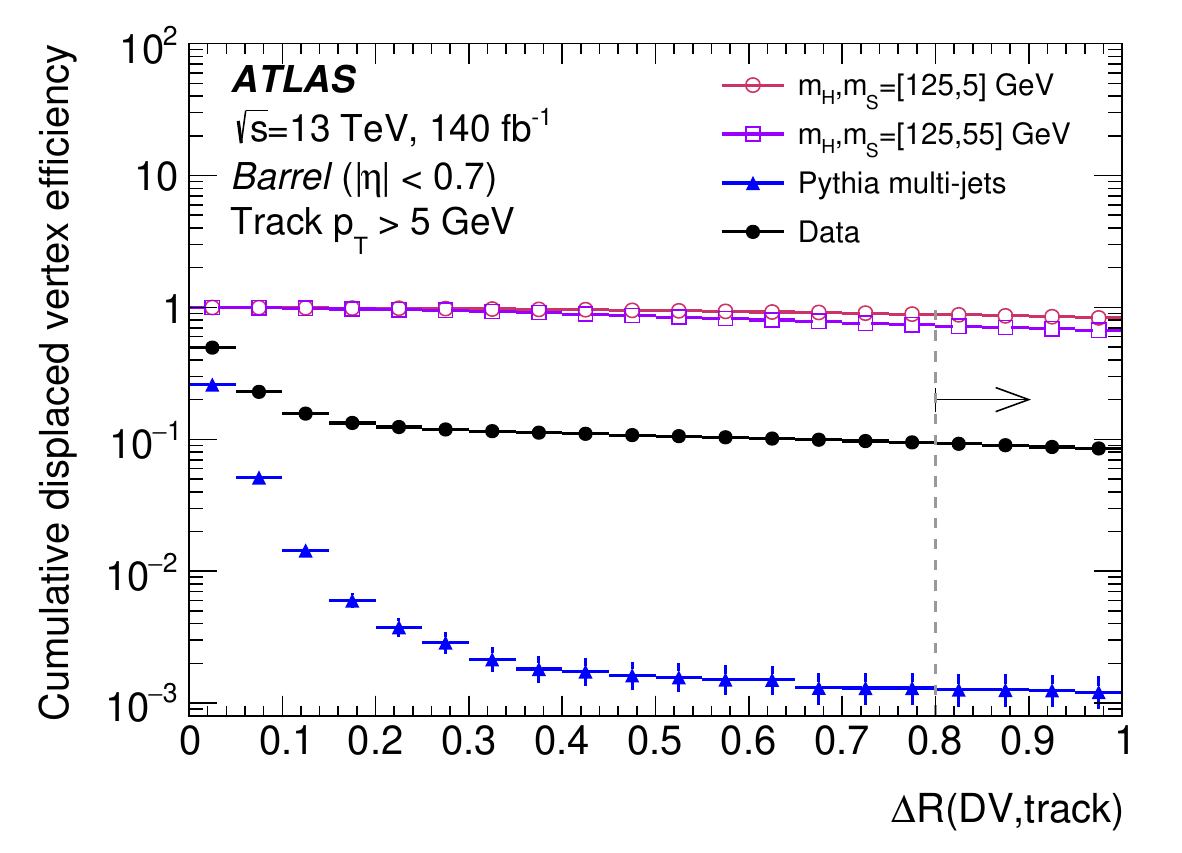}
\caption{}
\end{subfigure}
\begin{subfigure}[b]{0.49\textwidth}
\includegraphics[width=\textwidth]{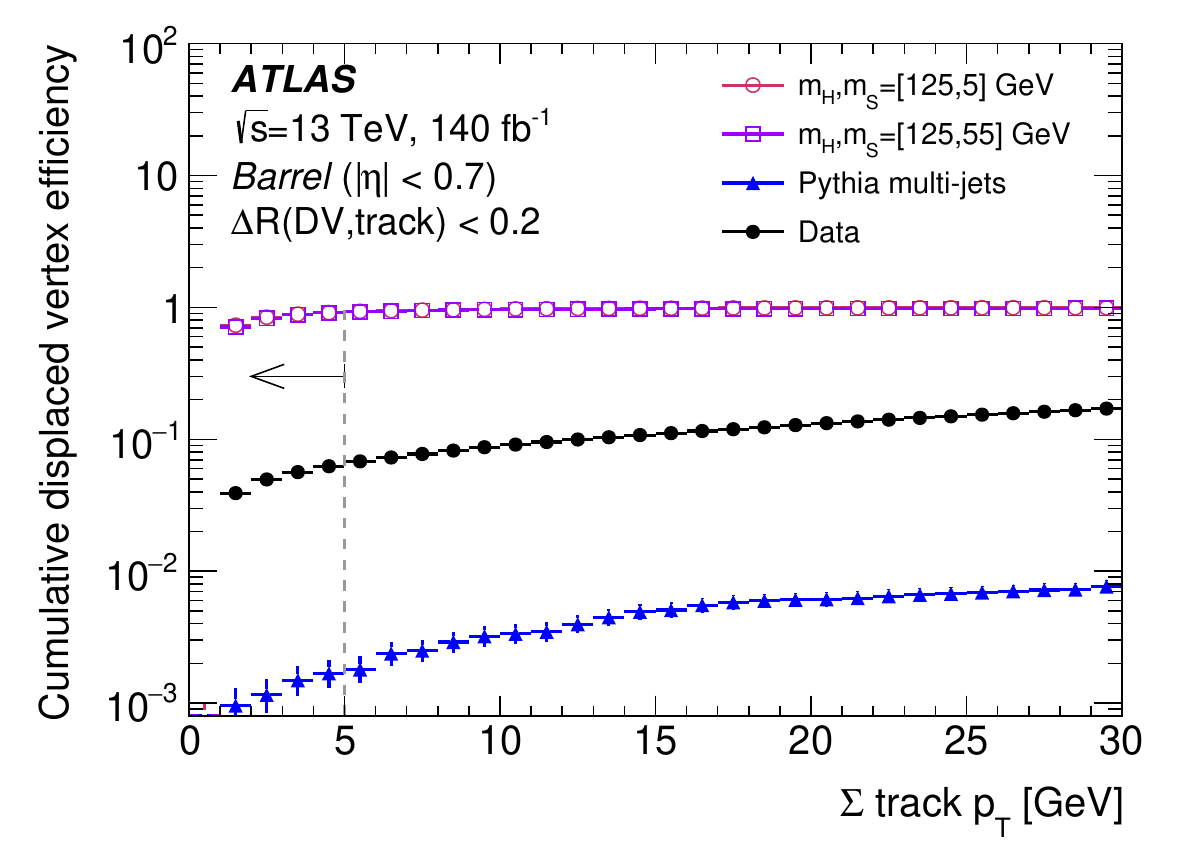}
\caption{}
\end{subfigure}
\begin{subfigure}[b]{0.49\textwidth}
\includegraphics[width=\textwidth]{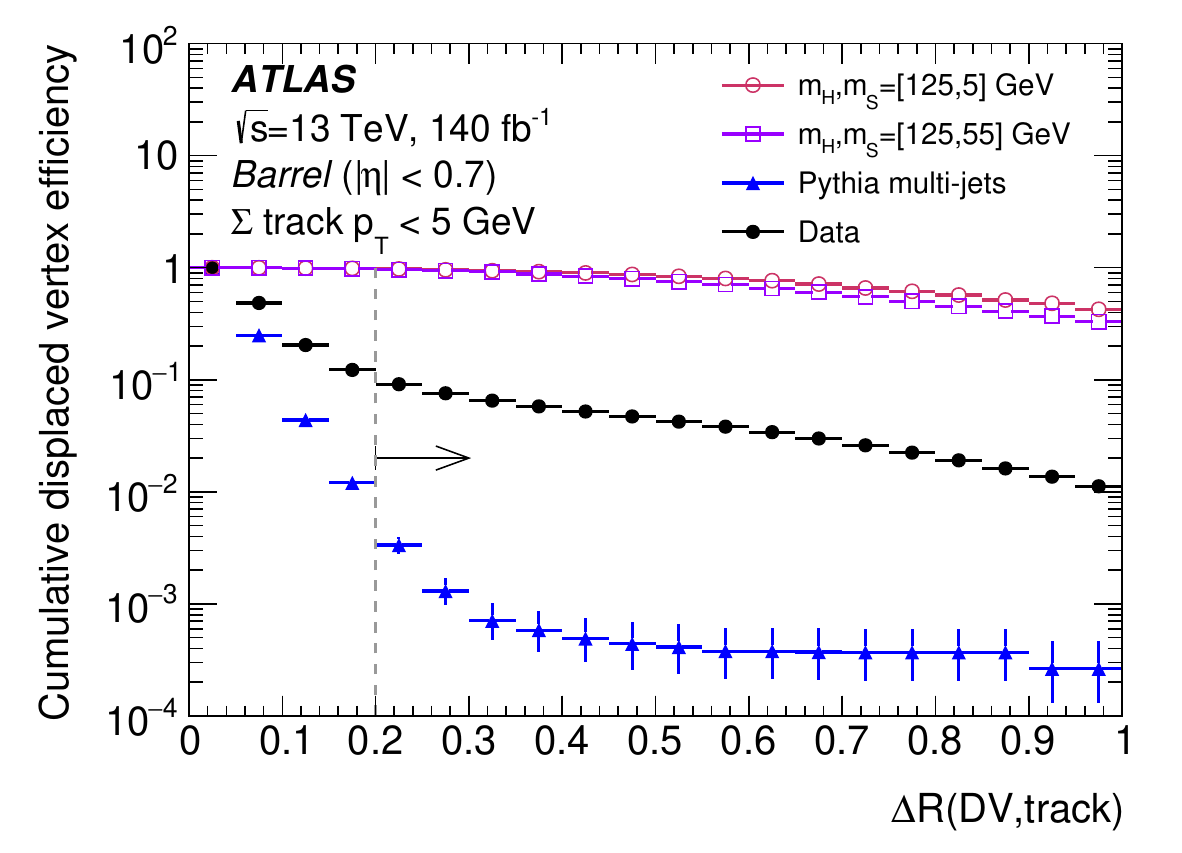}
\caption{}
\end{subfigure}
\caption{Cumulative displaced vertex efficiency in the barrel where the vertex is required to be isolated from (a) jets and (b) high-\pT tracks, depicted as a function of the selected $\Delta R$, when (c) the sum of low-\pT tracks in a $\Delta R = 0.2$ cone around the vertex direction must be less than a specified cut in \pT, and (d) when the sum of low-\pT tracks in a $\Delta R$ cone around the vertex is required to be less than $5~\GeV$, as a function of the $\Delta R$ value. The vertical lines indicate the isolation values utilized in the RoI-triggered channel. Discrepancies between \PYTHIA multijet and data distributions are attributed to non-collision background, absent in MC simulations. The $\Delta R$ selection on the jets (a) and high-\pT tracks (b) was tightened to 0.8 to suppress residual non-collision background contributions in the ABCD plane as described in Section~\ref{sec:Bkg1MSVx}.}
\label{fig:GVCPlotsBarrel}
\vspace{-0.09 cm}
\end{figure}

These isolation criteria were optimized for MC benchmark samples by comparing simulated signal events with simulated multijet events. Figure~\ref{fig:GVCPlotsBarrel} illustrates the cumulative vertex efficiency against the isolation requirements in the barrel for data events, simulated multijet events, and signal events. All jets considered for DV isolation must satisfy $\pT>20$\,\GeV.
Standard jet-quality criteria~\cite{ATL-PHYS-PUB-2015-036} are not imposed because jets failing these requirements can still generate a background MS DV, requiring their inclusion in isolation calculations. The acceptance depends on the event kinematics, with a more pronounced effect for the low-boosted $m_s={55}$~GeV sample, where the LLP decay products are more spread out, and a higher fraction of LLP pairs have a separation of $\Delta R<1$. In these cases, if the second LLP decays in the ID or calorimeters, the probability of producing prompt-like tracks and jets that contribute to the DV isolation is higher, reducing the signal acceptance.

For isolation from tracks, two criteria are employed. One targets isolation from tracks with $\pT>5$\,\GeV\, (high-\pT tracks). The other assesses the vector \pT sum of all tracks associated with the primary vertex that have $500\,\MeV<\pT<5$\,\GeV\ (low-\pT tracks) and are within a cone of $\Delta R = 0.2$ around the MS DV axis. The use of two distinct isolation criteria arises from the variance in jet composition, with some jets comprising mostly high-energy hadrons while others consist of numerous low-\pT tracks.

In models involving Higgs boson scalars and baryogenesis, single DV events typically correspond to a signal process where one LLP decays within the detector's sensitive region, while the other decays outside the detector, leading to an imbalance in the transverse energy. This imbalance is manifested as both missing transverse energy ($E_{\mathrm{T}}^{\mathrm{miss}}$)~\cite{JETM-2020-03} and missing transverse hadronic energy ($H_{\mathrm{T}}^{\mathrm{miss}}$) in signal events. While $E_{\mathrm{T}}^{\mathrm{miss}}$ includes contributions from all particles, $H_{\mathrm{T}}^{\mathrm{miss}}$ is computed as the $p_T$ sum of hadronic jets with $p_T > 20$~\GeV and $\vert \eta \vert < 3.2$. As a result, in the muon-RoI triggered channel, events are required to meet minimum selection thresholds of 20 \GeV for $E_{\mathrm{T}}^{\mathrm{miss}}$ and 40 \GeV for $H_{\mathrm{T}}^{\mathrm{miss}}$. For the lepton-triggered channel, the leptons in the primary background of $\Zboson+$ jets do not contribute to $H_{\mathrm{T}}^{\mathrm{miss}}$, so the combined $H_{\mathrm{T}}^{\mathrm{miss}}$ and $E_{\mathrm{T}}^{\mathrm{miss}}$ selections are replaced with a single requirement of $E_{\mathrm{T}}^{\mathrm{miss}} >40$ \GeV.

Compared to background vertices, particularly those from non-collision background, signal vertices generate more activity in the MS, leading to an increased presence of muon segments. Hence, for events with a vertex in the MS barrel, the number of muon segments in the outer MDT station must exceed 15. For events with a vertex in the endcap, this threshold is raised to greater than 30. In the lepton-triggered channel, the additional prompt objects reduce the non-collision background to negligible levels, and so no selection is applied to the number of outer MDT segments.

As detailed in Section~\ref{sec:Bkg1MSVx}, to estimate the expected background in the signal region, a data-driven ABCD method based on two uncorrelated variables is used. To improve the selection efficiency, a multilayer perceptron (MLP) classifier~\cite{MURTAGH1991183} is employed to define the two axes of the ABCD plane. Two MLP neural networks (NN) are trained with distinct input features to produce two NN output scores, denoted by NN1 and NN2, using the binary cross-entropy as the loss function and the Adam optimizer~\cite{kingma2017adammethodstochasticoptimization}. The NN model consists of two hidden layers. Each hidden layer has 128 neurons and uses the Rectified Linear Unit function as the activation function. The Sigmoid function is employed as the activation function in the output layer, confining NN1 and NN2 values within the range of zero to one.

The input features of NN1 are selected to have minimal correlation with those of NN2. The first set of input features primarily focuses on DV isolation and the removal of residual non-collision background contributions:

\begin{itemize}
\item Scalar sum of $\pT$ for all high-$\pT$ ($>5$~\GeV) tracks
\item Total, average, root-mean-square (RMS) and maximum energy of all calorimeter energy clusters satisfying $\Delta R(\textrm{DV,cluster}) < 0.4$
\item Calorimeter energy at each ECal and HCal sampling layer satisfying $\Delta R(\textrm{DV,cluster}) < 0.4$
\item $\Delta R$ of the high-$\pT$ ($>5$~\GeV) track or jet closest to the DV
\item Low-$\pT$ track isolation
\item Ratio of the number of muon segments in the inner and middle MDT station
\end{itemize}

The second set uses input features more directly related to the DV characteristics:
\begin{itemize}
\item Average and RMS $\Delta R (\textrm{DV,tracklets})$
\item Average and RMS $\Delta R (\textrm{DV,segments})$
\item $n_\textrm{MDT}$ + $n_\textrm{RPC}$ (in the barrel) or $n_\textrm{MDT}$ + $n_\textrm{TGC}$ (in the endcaps) associated with the DV
\item Number of tracklets satisfying $\Delta R(\textrm{DV,tracklet}) < 0.4$
\item $\Delta \phi (\textrm{DV},E_\textrm{T}^\textrm{miss})$
\item Average, RMS and maximum time of arrival of the particle depositing the energy in the calorimeter satisfying $\Delta R(\textrm{DV,cluster}) < 0.4$
\end{itemize}

Since there is no reliable simulation of all backgrounds in the MS, data events in the background-enriched region $H_\mathrm{T}^\textrm{miss} < 40$~\GeV are used as background in the training. A combination of samples with \mbox{$m_{H/\varPhi}=$ 60, 125, and 1000~\GeV} serves as the signal input to cover a relatively wide phase space. To increase the sample size, the $E_{\mathrm{T}}^{\mathrm{miss}}$ selection is omitted during training, while all other selections are applied. Training is conducted separately for events with a barrel DV and those with an endcap DV. The same neural networks are used for both the muon-RoI and lepton triggered channels.

The muon-RoI triggered signal selection criteria are summarized in Table~\ref{tab:SignalSelection}, and the lepton-triggered selections are summarized in Table~\ref{tab:ALPSignalSelection}. An MS DV that satisfies either of these selections is considered as a signal DV. After the final signal selection, background entering in the signal region may come from punch-through jets or residual non-collision backgrounds incompletely reconstructed by the ATLAS detector.

\begin{table}[tb]
\centering
\caption{Summary of signal selection for the RoI triggered channel. The NN1 and NN2 selections are described in Section~\ref{sec:Bkg1MSVx}.}
\begin{tabular}{lcc}
\toprule
Selection & Barrel & Endcaps \\
\midrule
H$^{\textrm{miss}}_T$ & \multicolumn{2}{c}{$>40$~\GeV} \\
\midrule
E$^{\textrm{miss}}_T$ & \multicolumn{2}{c}{$>20$~\GeV} \\
\midrule
Number of muon segments in the outer MDT station & $>15$ & $>30$ \\
\midrule
Isolation from high-\pT tracks ($\pT > 5$~\GeV) & $\Delta R > 0.8$ & $\Delta R >$ 0.8 \\
Isolation from low-\pT tracks $(\Sigma \pT(\Delta R<0.2))$ & $\Sigma \pT < 5$~\GeV\ & $\Sigma \pT < 5$~\GeV\ \\
Isolation from jets & $\Delta R > 0.8$ & $\Delta R > 0.8$  \\
\midrule
NN1, NN2 & $>0.5$, 0.5 & $>0.8$, 0.8 \\
\bottomrule
\end{tabular}
\label{tab:SignalSelection}
\end{table}

\begin{table}[tb]
\centering
\caption{Summary of signal selection for the lepton-triggered channel. The NN1 and NN2 selections are described in Section~\ref{sec:Bkg1MSVx}.}
\begin{tabular}{lcc}
\toprule
Selection & Barrel & Endcaps \\
\midrule
E$^{\textrm{miss}}_T$ & \multicolumn{2}{c}{$>40$~\GeV} \\
\midrule
Isolation from high-\pT tracks ($\pT > 5$~\GeV) & $\Delta R > 0.1$ & $\Delta R >$ 0.1 \\
Isolation from low-\pT tracks $(\Sigma \pT(\Delta R<0.2))$ & $\Sigma \pT < 5$~\GeV\ & $\Sigma \pT < 5$~\GeV\ \\
Isolation from jets & $\Delta R > 0.1$ & $\Delta R > 0.1$  \\
\midrule
NN1, NN2 & $>0.2$, 0.1 & $>0.2$, 0.1 \\
\bottomrule
\end{tabular}
\label{tab:ALPSignalSelection}
\end{table}

\subsection{MS displaced-vertex reconstruction efficiency}\label{sec:1MSVxEff}

\begin{figure}[h]
\begin{subfigure}[b]{0.49\textwidth}
\includegraphics[width=\textwidth]{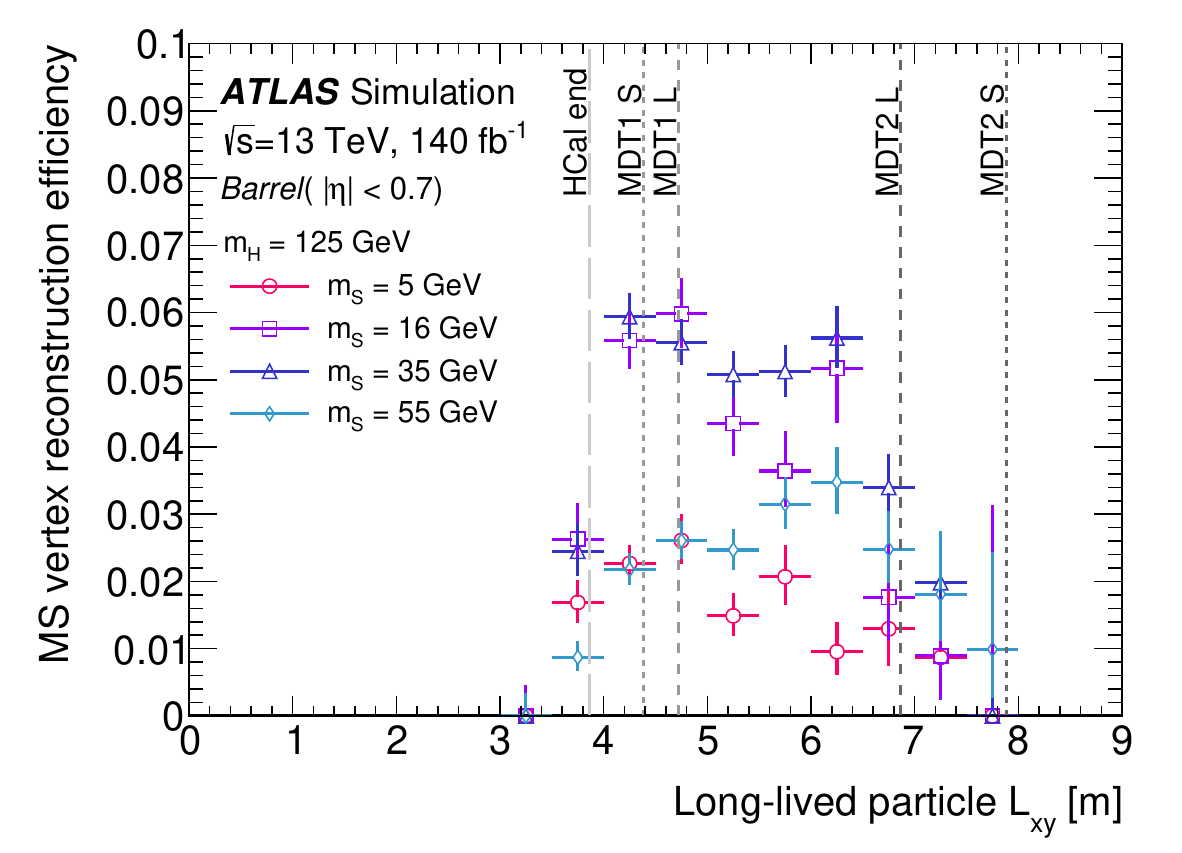}
\caption{}
\label{fig:MSTriggerVertexEffBarrel}
\end{subfigure}
\begin{subfigure}[b]{0.49\textwidth}
\includegraphics[width=\textwidth]{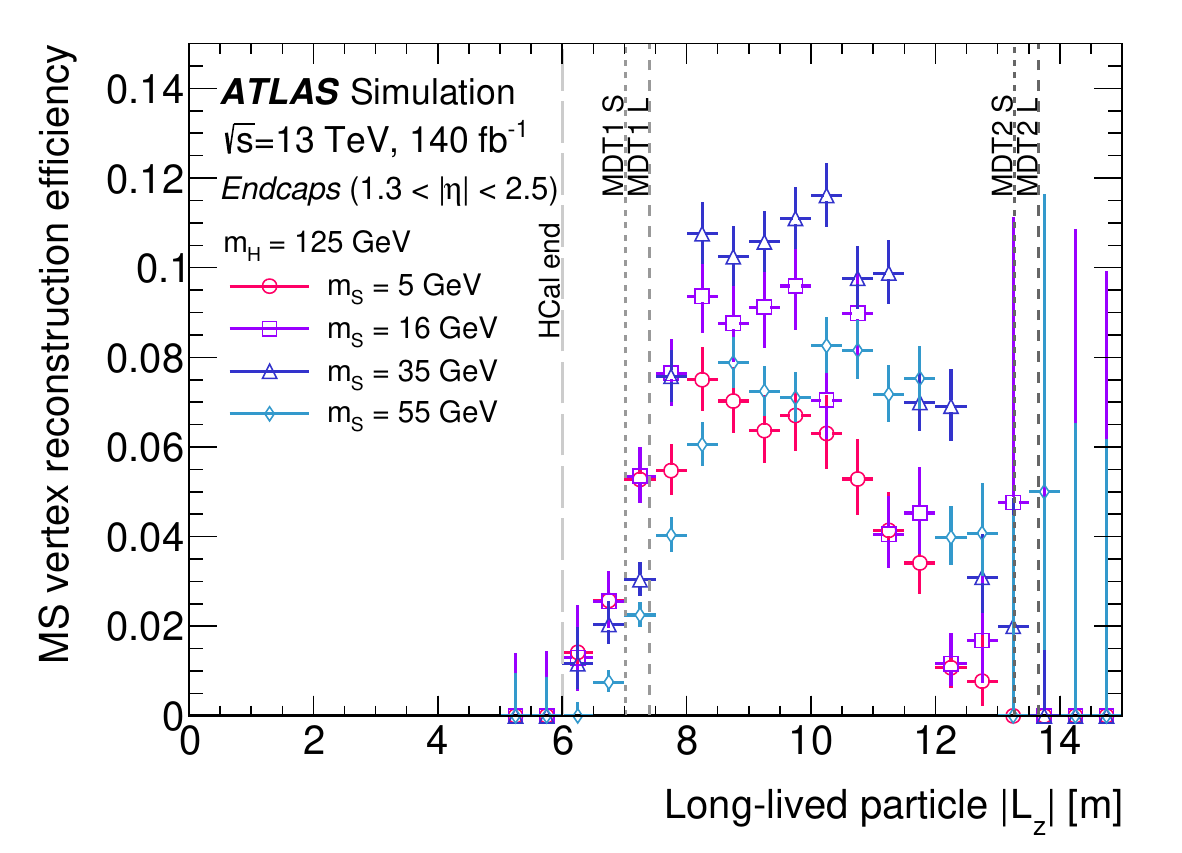}
\caption{}
\label{fig:MSTriggerVertexEffEndcap}
\end{subfigure}
\caption{Efficiency to reconstruct an MS DV for scalar portal samples with $m_H=125$~\GeV\ for vertices that satisfy the baseline and signal event selection summarized in Tables~\ref{tab:baselineSelection} and \ref{tab:SignalSelection}, respectively. (a) Efficiency to reconstruct a barrel MS DV as a function of the transverse decay position of the LLP. (b) Efficiency to reconstruct an endcap MS DV as a function of the longitudinal decay position of the LLP relative to the center of the detector. The efficiency distributions are corrected for the mismodeling detailed in Section~\ref{sec:1MSVxEff}. Vertical lines delineate the pertinent detector boundaries: ``HCal end'' marks the outer extent of the hadronic calorimeter, ``MDT 1/2'' denotes the first/second stations of MDT chambers, and ``L/S'' indicate whether they are in Large or Small sectors.}
\end{figure}

\begin{figure}[h]
\begin{subfigure}[b]{0.49\textwidth}
\includegraphics[width=\textwidth]{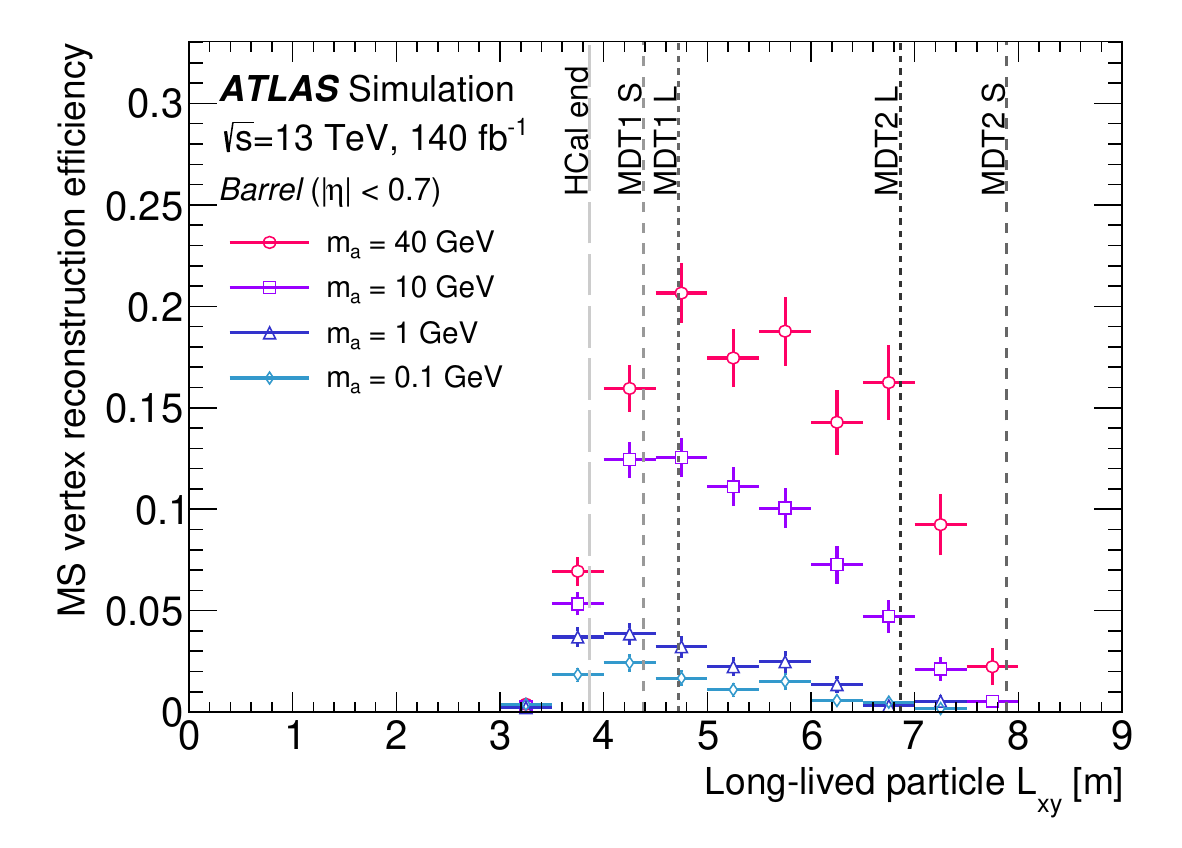}
\caption{}
\label{fig:ZALPVertexEffBarrel}
\end{subfigure}
\begin{subfigure}[b]{0.49\textwidth}
\includegraphics[width=\textwidth]{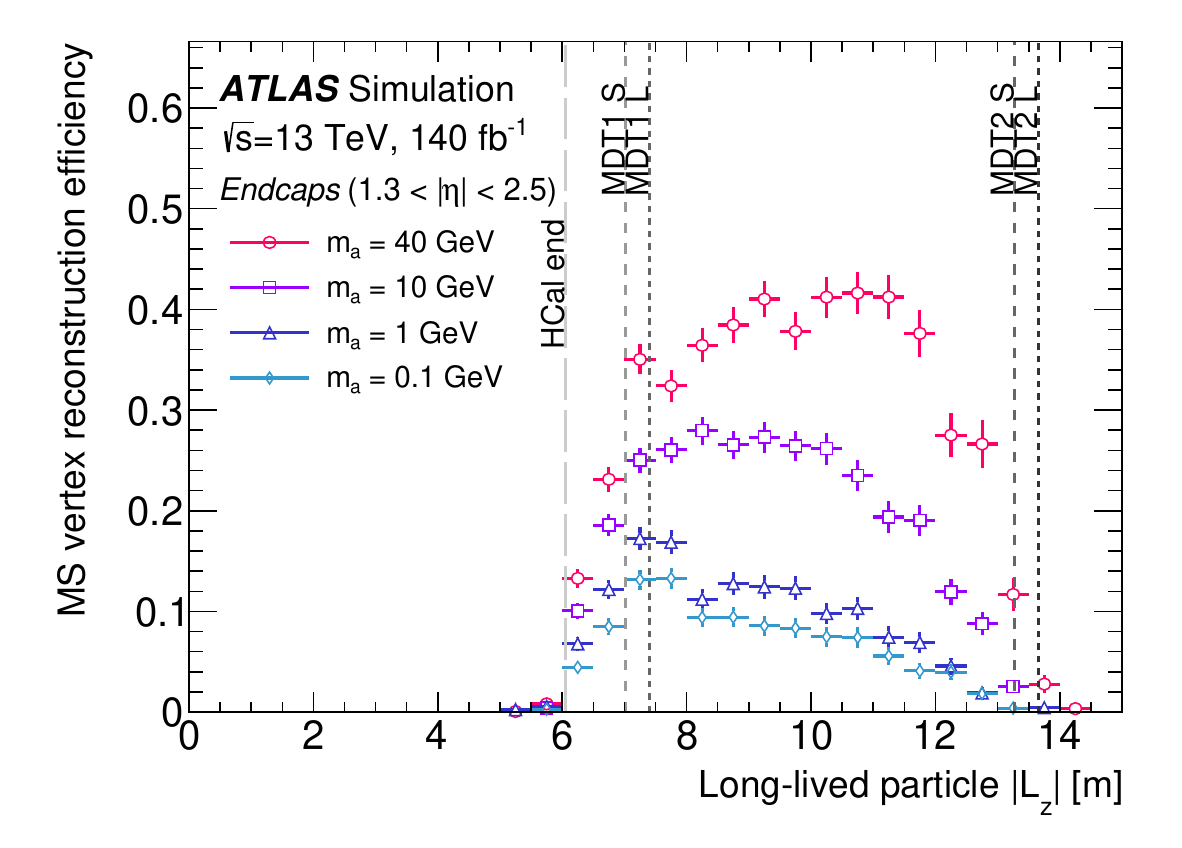}
\caption{}
\label{fig:ZALPVertexEffEndcap}
\end{subfigure}
\caption{Efficiency to reconstruct an MS DV for ALP samples with vertices that satisfy the baseline and signal event selection summarized in Tables~\ref{tab:baselineSelection} and \ref{tab:ALPSignalSelection}, respectively. (a) Efficiency to reconstruct a barrel MS DV as a function of the transverse decay position of the LLP. (b) Efficiency to reconstruct an endcap MS DV as a function of the longitudinal decay position of the LLP relative to the center of the detector. Vertical lines delineate the pertinent detector boundaries: ``HCal end'' marks the outer extent of the hadronic calorimeter, ``MDT 1/2'' denotes the first/second stations of MDT chambers, and ``L/S'' indicate whether they are in Large or Small sectors.}
\end{figure}

The efficiency for vertex reconstruction~\cite{PERF-2013-01} is defined as the fraction of simulated LLP decays in the MS fiducial volume corresponding to a reconstructed vertex that passes the baseline event selection reported in Table~\ref{tab:baselineSelection} and satisfies the signal selection criteria summarized in Table~\ref{tab:SignalSelection} for muon-RoI events and Table~\ref{tab:ALPSignalSelection} for lepton-triggered events. A reconstructed vertex is considered matched to a displaced decay if it lies within $\Delta R = 0.4$ of the simulated decay position. The MS DV efficiency is parameterized as a function of the $L_{xy}$ and $|L_z|$ LLP decay positions in the barrel and endcaps, respectively. The vertex reconstruction efficiency for the Higgs boson portal model at various LLP mass points is displayed in Figure~\ref{fig:MSTriggerVertexEffBarrel} and Figure~\ref{fig:MSTriggerVertexEffEndcap} for the MS barrel and endcaps, respectively. Similarly, the efficiencies for ALP samples are shown in Figure~\ref{fig:ZALPVertexEffBarrel} and Figure~\ref{fig:ZALPVertexEffEndcap}. The efficiency for reconstructing vertices in the lepton-triggered channel tends to be significantly higher due to the increased trigger efficiency and the relaxed selection requirements. In both cases, the effect of material interactions is visible. The additional soft tracks make the vertex reconstruction more challenging, resulting in a small efficiency reduction at particular decay positions.

For the MC samples analyzed, the MS barrel vertex reconstruction efficiency ranges from approximately 2\% to 15\% near the outer edge of the hadronic calorimeter ($r\approx4$\,m) and diminishes significantly as the decay position approaches the middle stations ($r\approx7$\,m). This decline is attributed to the lack of spatial separation between charged hadrons from the LLP decay, leading to their overlap when traversing the middle stations. This results in a reduction of the efficiencies for tracklet reconstruction and, consequently, vertex reconstruction. Additionally, the efficiencies are influenced by the mass and boost of the LLP.

The efficiency for reconstructing vertices is notably higher in the MS endcaps, reaching up to about 30\% for higher-mass benchmark models. This increased efficiency stems from a more effective selection and vertex reconstruction process. In the barrel, tracklets are influenced by a strong magnetic field, which imposes curvature constraints through charge and momentum measurements. In contrast, the region in which endcap tracklets are reconstructed features a very weak magnetic field. Consequently, the vertex reconstruction algorithm in the endcaps employs straight-line fits, ensuring that low-momentum tracks are not discarded. Meanwhile, the barrel leverages the curvature constraints and combinatorial techniques to more effectively reject misreconstructed tracks. Thus, while vertex reconstruction in the endcaps is more efficient for signal, it is also less robust in rejecting background events. Further details are available in Ref.~\cite{PERF-2013-01}.

Potential inaccuracies in MC simulation for DV reconstruction are estimated by comparing the distribution of the number of tracklets in a punch-through jet $\Delta R$ cone of $0.4$ in data and MC events, following a strategy akin to the one employed for the trigger (described in Section~\ref{sec:RoICluTrig}). MC simulation indicates a tracklet rate approximately 20\% (15\%) higher than in data in the barrel (endcaps). This mismodeling is independent of the $\eta$, $\phi$, or \pT of the jet.

The impact of this mismodeling on DV reconstruction efficiency is estimated by randomly discarding tracklets used in vertex reconstruction based on the measured mismodeling factor in the barrel and endcaps. The number of reconstructed vertices is then counted. The efficiency variation between the nominal reconstruction and that accounting for the mismodeling factors, averaged over MC benchmark samples, amounts to 27\% in the barrel and 9\% in the endcaps. Corrected efficiencies are utilized to compute the expected number of signal events. The systematic uncertainty associated with this correction is detailed in Section~\ref{sec:syst}.

Across the extrapolated lifetime range, the maximum acceptance times efficiency for vertices meeting the signal selection in the muon-RoI channel spans from 0.02\% (for the MC sample with $m_\varPhi = 60$~\GeV, $m_s = 5$~\GeV at a proper decay length of 0.5 m) to 1.4\% (for the MC sample with $m_\varPhi = 600$~\GeV, $m_s = 50$~\GeV at a proper decay length of 1.1 m). Due to the relaxed selections in the lepton-triggered channel the ALP samples have generally higher efficiencies, with maximum acceptance times efficiency for vertices meeting the signal selection ranging from 0.5\% ($m_a = 0.1$~\GeV, $C_{\tilde{G}} =10^{-2}$, proper decay length 0.0031 m) to 1.9\% ($m_a = 40$~\GeV, $C_{\tilde{G}} =10^{-7}$, proper decay length 0.48 m).

\section{Background estimation}
\label{sec:Bkg1MSVx}

\begin{figure}[tb]
\centering
\begin{subfigure}[b]{0.48\textwidth}
\includegraphics[width=\textwidth]{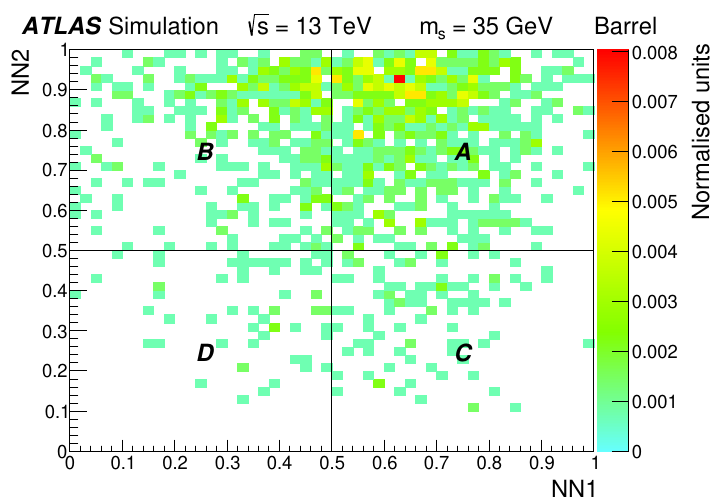}
\caption{}
\end{subfigure}
\begin{subfigure}[b]{0.48\textwidth}
\includegraphics[width=\textwidth]{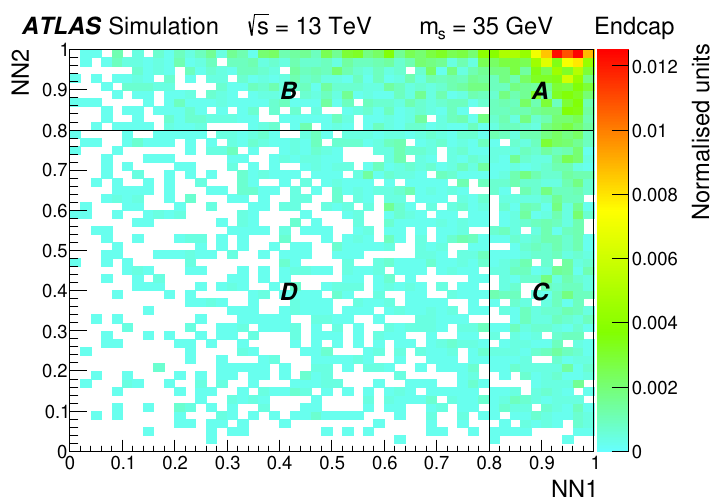}
\caption{}
\end{subfigure}
\caption{ABCD planes for the Higgs boson portal model with $m_s = 35~\text{\GeV}$ and $c\tau = 1.310$ m in the (a) barrel and (b) endcap regions of the RoI-triggered channel.}
\label{fig:abcd_mphi125ms35_SR}
\end{figure}

\begin{figure}[tb]
\centering
\begin{subfigure}[b]{0.48\textwidth}
\includegraphics[width=\textwidth]{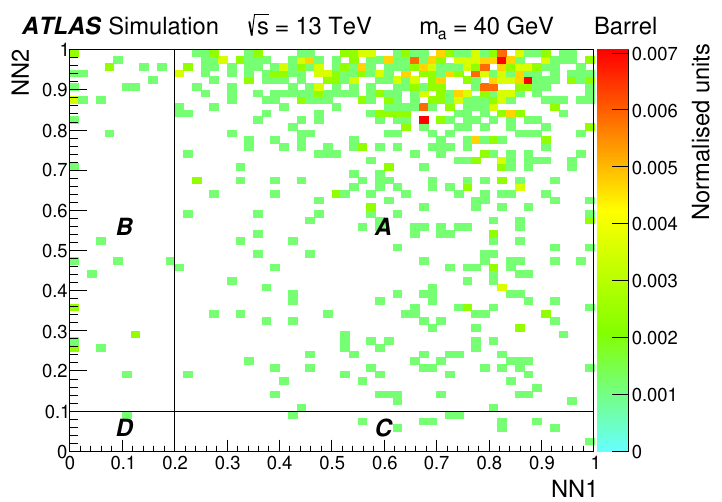}
\caption{}
\end{subfigure}
\begin{subfigure}[b]{0.48\textwidth}
\includegraphics[width=\textwidth]{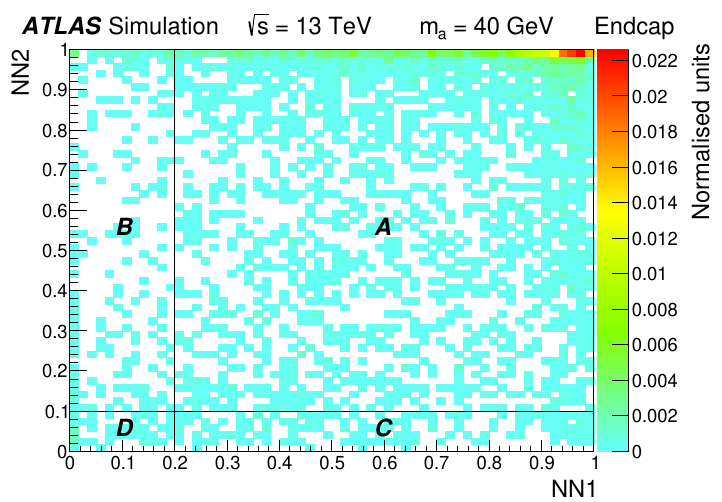}
\caption{}
\end{subfigure}
\caption{ABCD planes for the ALP model with $m_a = 40~\text{\GeV}$ and $c\tau = 0.481$ m in the (a) barrel and (b) endcap regions of the lepton-triggered channel.}
\label{fig:abcd_ma40_SR}
\end{figure}

\begin{figure}[tb]
\centering
\begin{subfigure}[b]{0.48\textwidth}
\includegraphics[width=\textwidth]{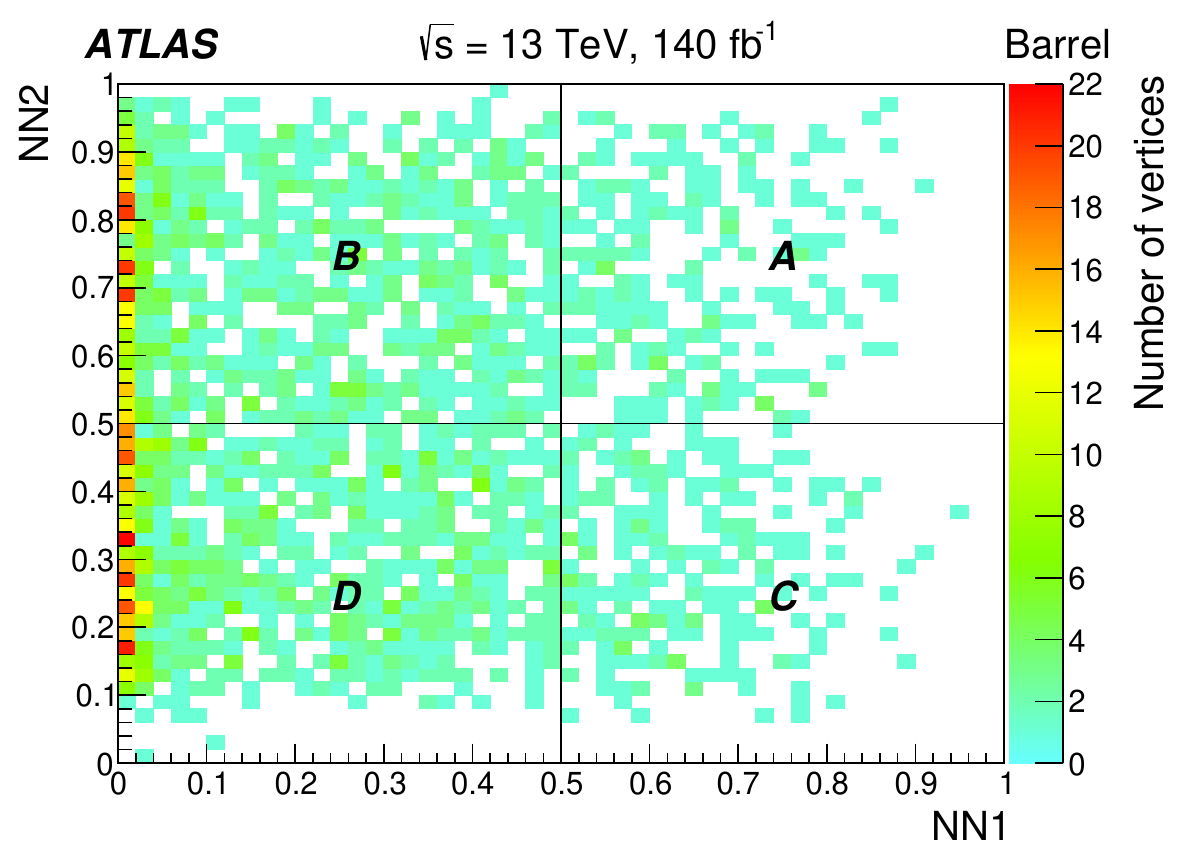}
\caption{}
\end{subfigure}
\begin{subfigure}[b]{0.48\textwidth}
\includegraphics[width=\textwidth]{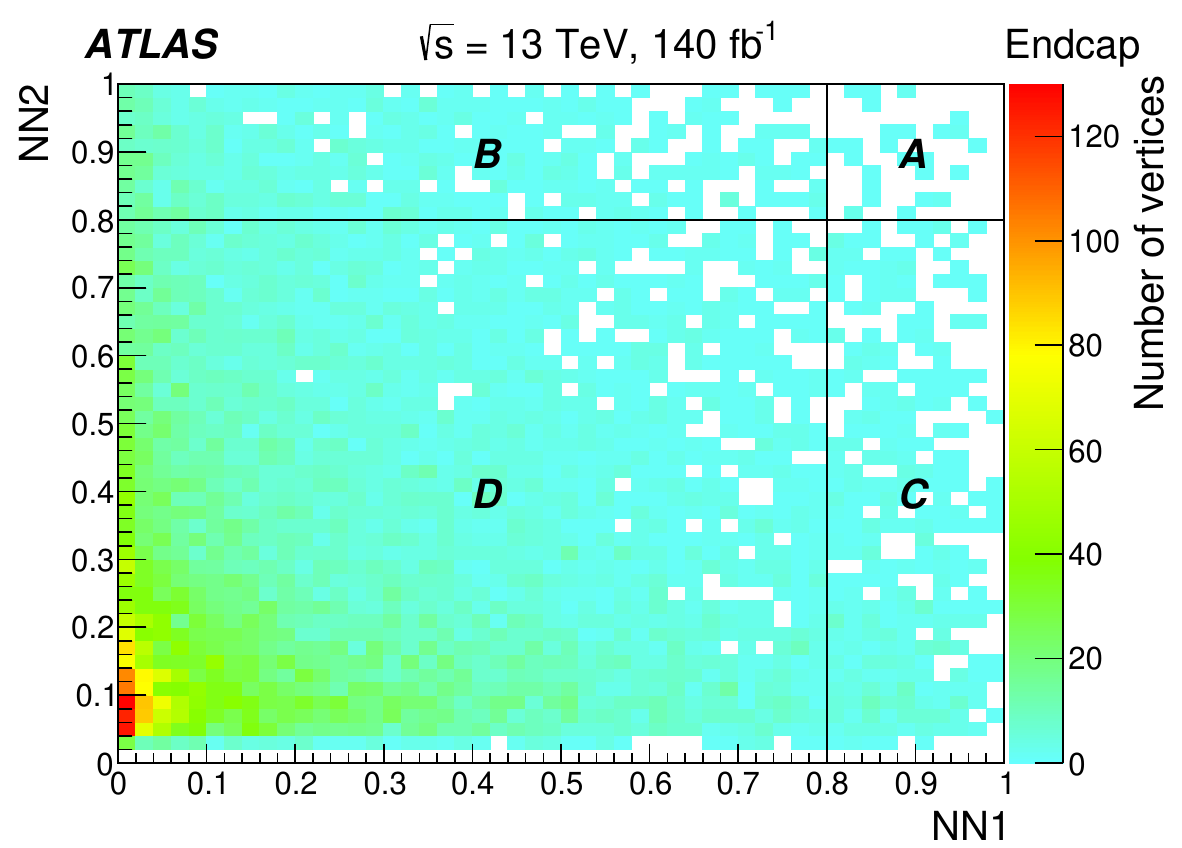}
\caption{}
\end{subfigure}
\caption{ABCD planes for Run-2 data in the (a) barrel and (b) endcap regions of the muon-RoI triggered channel. The correlation between NN1 and NN2 is found to be negligible.}
\label{fig:abcd_data_SR}
\end{figure}

\begin{figure}[tb]
\centering
\begin{subfigure}[b]{0.48\textwidth}
\includegraphics[width=\textwidth]{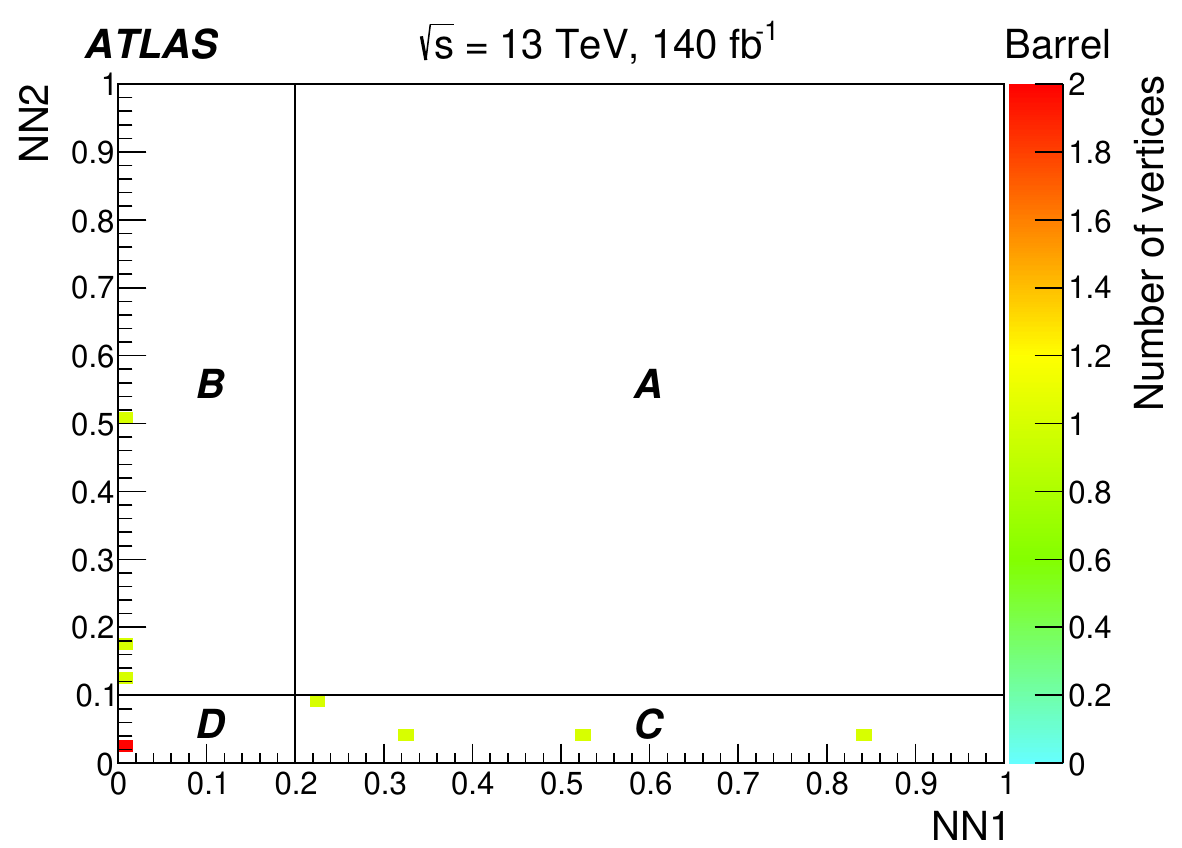}
\caption{}
\end{subfigure}
\begin{subfigure}[b]{0.48\textwidth}
\includegraphics[width=\textwidth]{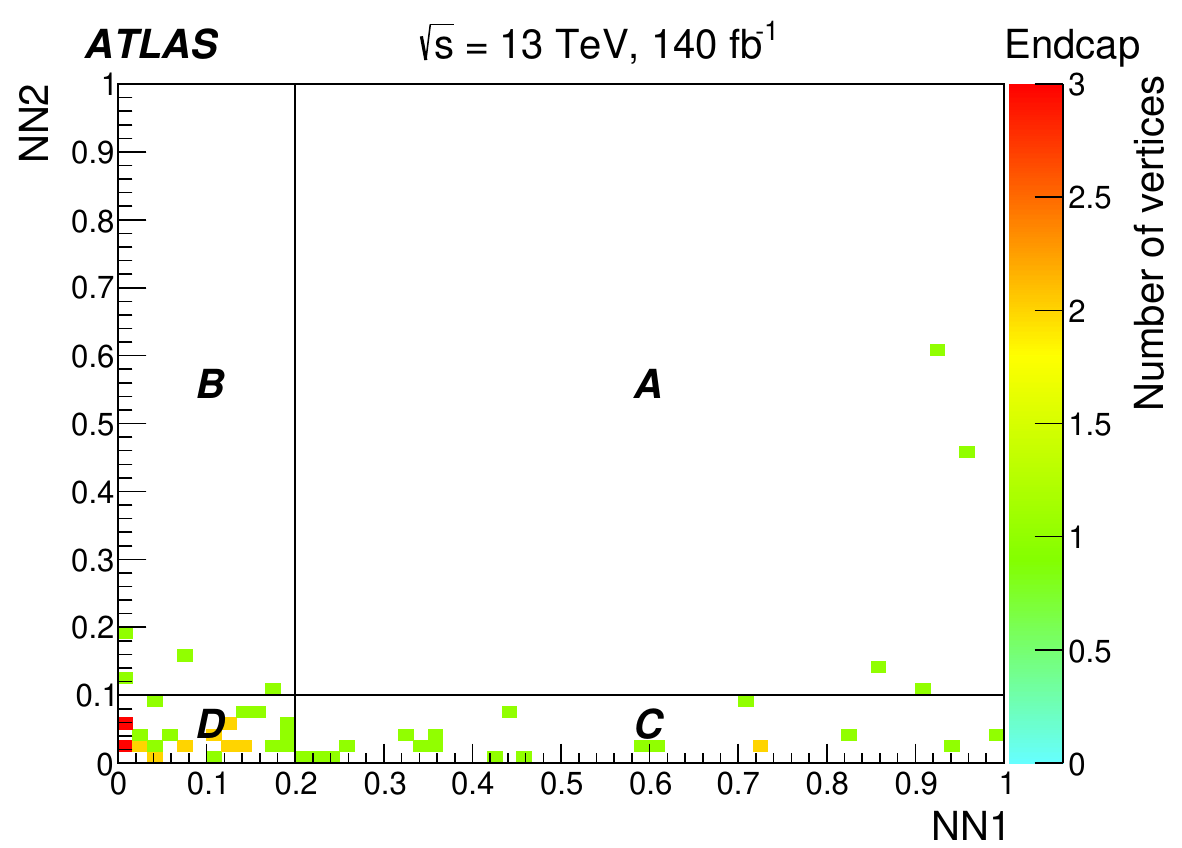}
\caption{}
\end{subfigure}
\caption{ABCD planes for Run-2 data in the (a) barrel and (b) endcap regions of the lepton-triggered channel. The criteria selecting leptons consistent with \Zboson decays greatly reduces the background event count, despite the relaxed background rejection selections. }
\label{fig:ZALPabcd_data_SR}
\end{figure}

The number of background events is estimated through the data-driven ABCD method. The ABCD method uses two variables that are chosen to be good at separating the signal from background, thereby splitting the spanned two-dimensional plane into four sub-regions, which contain an A region where most of the signal events are located and three other control regions (B, C and D) that are dominated by the background events. Provided that the axis variables are uncorrelated for the background, the expected number of background events in region A can be predicted by the numbers in the other three regions: $N_{A} = N_{B} \times N_{C}/N_{D}$. This method is only valid when there is one source of background or multiple sources of background with the same distribution in the plane. The NN1 and NN2 variables are used to define the two ABCD planes for the barrel and endcaps. Representative planes for the simulated Higgs boson portal model in the muon-RoI triggered channel and the ALP model in the lepton-triggered channel are shown in Figure~\ref{fig:abcd_mphi125ms35_SR} and Figure~\ref{fig:abcd_ma40_SR}, respectively. The Run-2 data planes for the muon-RoI triggered channel are presented in Figure~\ref{fig:abcd_data_SR} and for the lepton-triggered channel in Figure~\ref{fig:ZALPabcd_data_SR}.

In the muon-RoI triggered channel, the ABCD method is validated in a background dominated validation region (VR) with $H_{\textrm{T}}^{\textrm{miss}} < 40$ \GeV and all other requirements matching the signal region. The signal contamination in this VR is found to be negligible. The correlation between NN1 and NN2 is calculated to be less than 4\% in both of the barrel and endcap VR planes. As summarized in Table \ref{tab:ABCDVR}, the number of observed events in the validation region A is $139$ and $18$ in the barrel and endcaps, respectively, in agreement with the expected numbers $130 \pm 12 \textrm{ (stat.)}$ and $14  \pm 1 \textrm{ (stat.)}$ in the barrel and endcaps from the ABCD prediction.

\begin{table}[tb]
\centering
\caption{Event counts in each of the four regions of the muon-RoI triggered channel $H_{\textrm{T}}^{\textrm{miss}}$ VR ABCD plane. The background predictions in the barrel and endcap VRs are within one standard deviation of the observations.}
\begin{tabular}{lccccc}
\toprule
VR & A & B$\times$C/D & B & C & D  \\
\midrule
\multirow{1}*{Barrel} & $139$                 & $130 \pm \substack{12\\11} \textrm{ (stat.)}$ & 485   & 246 & 915 \\
\midrule
\multirow{1}*{Endcaps}  & $18$                  & $14  \pm \substack{1\\1} \textrm{ (stat.)}$   & 306   & 391 & 8850 \\
\bottomrule
\end{tabular}
\label{tab:ABCDVR}
\end{table}

In the lepton-triggered channel, the ABCD method is validated in two separate background-dominated VRs, one formed by inverting the $E_{\textrm{T}}^{\textrm{miss}}$ selection to require $10 < E_{\textrm{T}}^{\textrm{miss}} < 40$ \GeV (``$E_{\textrm{T}}^{\textrm{miss}}$ VR''), and the other by replacing the lepton trigger and \Zboson selection criteria with a photon trigger and selection (``$\gamma$+jets VR'') and relaxing the $E_{\textrm{T}}^{\textrm{miss}}$ requirement to $E_{\textrm{T}}^{\textrm{miss}}>10~\GeV$. As the $\Zboson+$ALP model is chosen to be photo-phobic due to strong constraints from other experiments, photons are not directly produced in association with ALPs in this model, and replacing the \Zboson selection with a photon results in negligible signal contamination. For the \Zboson-associated scalar production models, the contribution of photons from initial-state radiation was observed to be negligible in simulated signal events. Region A is defined as NN1 $>$ 0.2, NN2 $>$ 0.1 for both VRs, matching the criteria applied in the signal region. The number of observed and expected events in the $E_{\textrm{T}}^{\textrm{miss}}$ and $\gamma$+jets VRs are presented in Table~\ref{tab:ZALPABCDVR}.

\begin{table}[tb]
\centering
\caption{Event counts in each of the four regions of the lepton-triggered channel ABCD plane in the $E_{\textrm{T}}^{\textrm{miss}}$ and photon VRs. The background prediction in all validation regions is within one standard deviation of the observation.}
\begin{tabular}{lccccc}
\toprule
VR & A & B$\times$C/D & B & C & D  \\
\midrule
\multirow{1}*{$E_{\textrm{T}}^{\textrm{miss}}$ barrel} & $4$                 & $16 \pm \substack{43\\18} \textrm{ (stat.)}$ & 2   & 8 & 1 \\
\midrule
\multirow{1}*{$E_{\textrm{T}}^{\textrm{miss}}$ endcaps}  & $8$                  & $6 \pm \substack{4\\3} \textrm{ (stat.)}$   & 5   & 43 & 38 \\
\midrule
\multirow{1}*{$\gamma$+jets barrel} & $1$                 & $3 \pm \substack{7\\3} \textrm{ (stat.)}$ & 18   & 1 & 6 \\
\midrule
\multirow{1}*{$\gamma$+jets endcaps}  & $57$                  & $44 \pm \substack{13\\11} \textrm{ (stat.)}$   & 30   & 82 & 56 \\
\bottomrule
\end{tabular}
\label{tab:ZALPABCDVR}
\end{table}

The ABCD method is further validated in the muon-RoI channel by analyzing sections of the ABCD plane within the VR\@. Instead of using the entire BD region, it is vertically divided into five equal slices (VR1--VR5). The number of events in region A can then be predicted using region C and any of these slices. Similarly, another set of five VRs is defined by horizontally dividing the CD plane into five equal slices. Predictions for the number of events in region A are then made using region B and any slice from the CD plane. These sub-region predictions align well with the observed number of events. The difference between the average predictions from these sub-regions and the observed values is used to estimate the systematic uncertainty in the ABCD background estimation, calculated to be 2.5\% in the barrel and 19.1\% in the endcaps.

In the lepton-triggered VRs, the reduced event rate due to the selection on the prompt object renders the many VR slice method used for the muon-RoI triggered channel non-predictive for many of VR slices. Instead, the $E_{\textrm{T}}^{\textrm{miss}}$ and $\gamma$+jets VRs are divided into exclusive bins of $\Delta \textrm{R}_{\textrm{min}}$, the angular distance from the DV to the nearest high-\pT track or jet. This approach includes additional events that were absent in the main VR at $\Delta \textrm{R}_{\textrm{min}}$ values below the 0.1 threshold. Since the NNs are trained on samples with stricter $\Delta \textrm{R}_{\textrm{min}}$ selections, the expected and observed occupancy of the A region as a function of $\Delta \textrm{R}_{\textrm{min}}$ is used to both confirm that the lower thresholds do not introduce significant correlation between NN1 and NN2, and to provide additional regions to check for systematic excesses or deficits in the observed events. No systematic differences are observed in the $\Delta \textrm{R}_{\textrm{min}}$ scans. As the relative size of the statistical uncertainties in this channel are significantly larger than those in the muon-RoI triggered channel and no systematic differences are observed in the $\Delta \textrm{R}_{\textrm{min}}$ scan, no additional systematic uncertainties are applied to the observed background for lepton-triggered channel events.

The observed and expected number of events in region A of the muon-RoI triggered signal region ($H_\textrm{T}^\textrm{miss} > 40$~\GeV) are presented in Table \ref{tab:ABCDSR}. The predicted number of full Run~2 muon-RoI triggered background events in region A is $241 \pm 18 \textrm{ (stat.)} \pm 6 \textrm{ (syst.)}$ in the barrel and $46 \pm 2 \textrm{ (stat.)} \pm 9 \textrm{ (syst.)}$ in the endcaps. The observed and expected number of events in region A of the lepton-triggered signal region are presented in Table~\ref{tab:ZALPABCDSR}. The predicted number of full Run-2 lepton-triggered channel background events in region A is $6 \pm \substack{11\\6}\textrm{~(stat.)}$ in the barrel and $2  \pm \substack{2\\1}\textrm{~(stat.)}$ in the endcaps. No excess events above the background prediction are observed.

\begin{table}[tb]
\centering
\caption{Event counts in each of the four regions of the ABCD plane and the expected numbers in region A for the muon-RoI triggered signal region. The event counts are also reported for 2015+2016 data, 2017 data and 2018 data, separately.}
\begin{tabular}{lcccccc}
\toprule
SR & Period & A & B$\times$C/D & B & C & D  \\
\midrule
\multirow{4}*{Barrel}  & 2015 + 2016 & $65$     & $63 \pm 9 \textrm{~(stat.)} \pm 2 \textrm{~(syst.)}$ & 333   & 61 & 321 \\
& 2017         & $95$    & $80 \pm 10 \textrm{~(stat.)} \pm 2 \textrm{~(syst.)}$ & 370   & 84 & 389 \\
& 2018         & $85$     & $97 \pm 11 \textrm{~(stat.)} \pm 2 \textrm{~(syst.)}$ & 433   & 104 & 466 \\
& Run-2        & $245$   & $241 \pm 18 \textrm{~(stat.)} \pm 6 \textrm{~(syst.)}$ & 1136   & 249 & 1176 \\
\midrule
\multirow{4}*{Endcaps} & 2015 + 2016 & $12$     & $17  \pm 2 \textrm{~(stat.)} \pm 3 \textrm{~(syst.)}$   & 259   & 232 & 3510 \\
& 2017         & $11$     & $15  \pm 1 \textrm{~(stat.)} \pm 3 \textrm{~(syst.)}$   & 346   & 178 & 4133 \\
& 2018         & $8$     & $13  \pm 1 \textrm{~(stat.)} \pm 2 \textrm{~(syst.)}$   & 467   & 160 & 5547 \\
& Run-2        & $31$     & $46  \pm 2 \textrm{~(stat.)} \pm 9 \textrm{~(syst.)}$   & 1072   & 570 & 13190\\
\bottomrule
\end{tabular}
\label{tab:ABCDSR}
\end{table}

\begin{table}[tb]
\centering
\caption{Event counts in each of the four regions of the ABCD plane and the expected number in region A for the lepton-triggered signal region. The event counts are reported for the entire Run-2 data.}
\begin{tabular}{lccccc}
\toprule
SR & A & B$\times$C/D & B & C & D  \\
\midrule
\multirow{1}*{Barrel}
& 0  & $6 \pm \substack{11\\6}\textrm{~(stat.)}$ & 3   & 4 & 2 \\
\midrule
\multirow{1}*{Endcaps} & 4  & $2  \pm \substack{2\\1}\textrm{~(stat.)}$   & 4   & 19 & 31 \\
\bottomrule
\end{tabular}
\label{tab:ZALPABCDSR}
\end{table}

\section{Systematic uncertainties}
\label{sec:syst}

The systematic uncertainties in the signal efficiency are dominated by the modeling of the signal physics processes, pileup and detector response and the extrapolation of the expected number of signal events as a function of the LLP proper decay length.

An uncertainty in the NLO-reweighting of the signal samples is obtained by comparing the 125-\GeV mediator NLO \MADGRAPH predictions to next-to-next-to-leading-order (NNLO) accuracy in QCD using \POWHEGBOX v2~\cite{Hamilton:2013fea,Hamilton:2015nsa,Alioli:2010xd,Nason:2004rx,Frixione:2007vw}. This results in an additional uncertainty in the signal efficiency ranging from 0.1\% to 4\%.

One source of systematic uncertainty associated with the Muon RoI Cluster trigger is the modeling of minimum-bias interactions used to simulate pileup. Another source is the systematic uncertainty related to the PDF used to generate signal MC events. These uncertainties were estimated by varying the pileup and PDF weights according to their respective $\pm 1\sigma$ systematic uncertainties and assessing the resulting change in trigger efficiency. For the PDF-related uncertainty, the nominal PDF set was evaluated using 100 replica variations. In both cases, the systematic uncertainty was found to be negligible. The mis-modeling of the muon-RoI cluster trigger efficiency in MC simulation, as described in Section~\ref{sec:trigvtxreco}, contributes a systematic uncertainty of 20\% in the barrel and 24\% in the endcaps for the muon-RoI triggered channel. In the lepton-triggered channel, additional uncertainties arise from mismodeling in the lepton triggering efficiency, identification, and isolation. The combined lepton systematic uncertainty is found to be negligible in both the barrel and the endcap regions.

In both channels, the systematic uncertainties from the pileup and PDF contributions to the MS DV reconstruction efficiency for the signal were evaluated using a procedure similar to that used for the Muon RoI Cluster trigger. Once again, these systematic uncertainties were found to be negligible. Another source of systematic uncertainty for both channels arises from the corrected efficiencies related to MC mismodeling in vertex reconstruction, as discussed in Section~\ref{sec:1MSVxEff}. To estimate the corrected vertex reconstruction efficiencies, a second method was used. In this method, events are weighted so that the MC distribution of the number of reconstructed tracklets matches the data distribution. The efficiency was re-evaluated after reweighting each vertex according to the number of associated tracklets. The average efficiency variation was computed, and its difference from the variation calculated using the nominal method described in Section~\ref{sec:1MSVxEff} was taken as the systematic uncertainty, which is 11\% in the barrel and 13\% in the endcaps.

A systematic uncertainty associated with the efficiency extrapolation method was estimated by comparing the signal efficiency computed using fully simulated MC samples with the efficiency extrapolated using pseudo-data MC samples at the same proper decay length. The difference between these two efficiencies is taken as the systematic uncertainty, which varies from 1.9\% to 30\% depending on the mass and boost of the LLP. For several signal samples, multiple lifetime points were fully simulated: a nominal sample and secondary samples with alternate proper decay lengths, as described in Section~\ref{sec:datamc}. The secondary samples were used to cross-check the lifetime-extrapolation procedure, and good agreement was found.

The systematic uncertainty associated with the ABCD background estimation for the muon-RoI triggered channel is 2.5\% in the barrel and 19.1\% in the endcaps, whereas for the lepton-triggered channel, the systematic uncertainties are negligible.

The uncertainty in the combined 2015--2018 integrated luminosity is 0.83\%~\cite{DAPR-2021-01}, obtained using the LUCID-2 detector~\cite{LUCID2} for the primary luminosity measurements.

\section{Results}
\label{sec:result}

Upper limits on the production cross-section times branching fraction are derived using the CL$_{\textrm{s}}$ prescription~\cite{CLs}, implemented with the \texttt{pyhf}~\cite{pyhf} package using a profile likelihood function~\cite{Cowan:2010js}, with the background prediction constrained to follow the ABCD relation presented in Section~\ref{sec:Bkg1MSVx} and the signal distributed across the ABCD plane according to the signal MC\@. For the lepton-triggered channel, limits are calculated using pseudo-experiments due to the low number of events satisfying the signal selections. The reported limits exclude the $\Zboson \rightarrow \ell \bar{\ell}$ branching fraction~\cite{PDG2024}.

\begin{figure}[tb]
\centering
\begin{subfigure}[b]{0.49\textwidth}
\includegraphics[width=\textwidth]{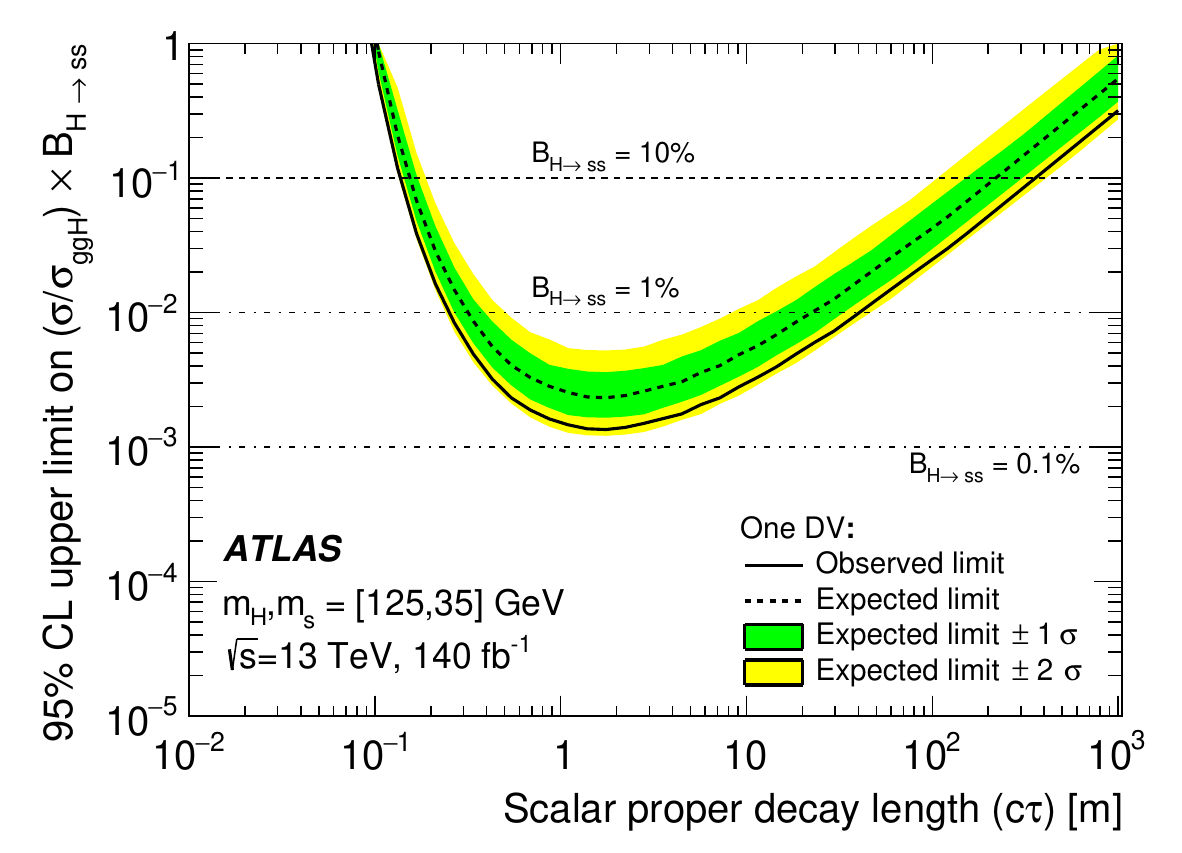}
\caption{}
\end{subfigure}
\begin{subfigure}[b]{0.49\textwidth}
\includegraphics[width=\textwidth]{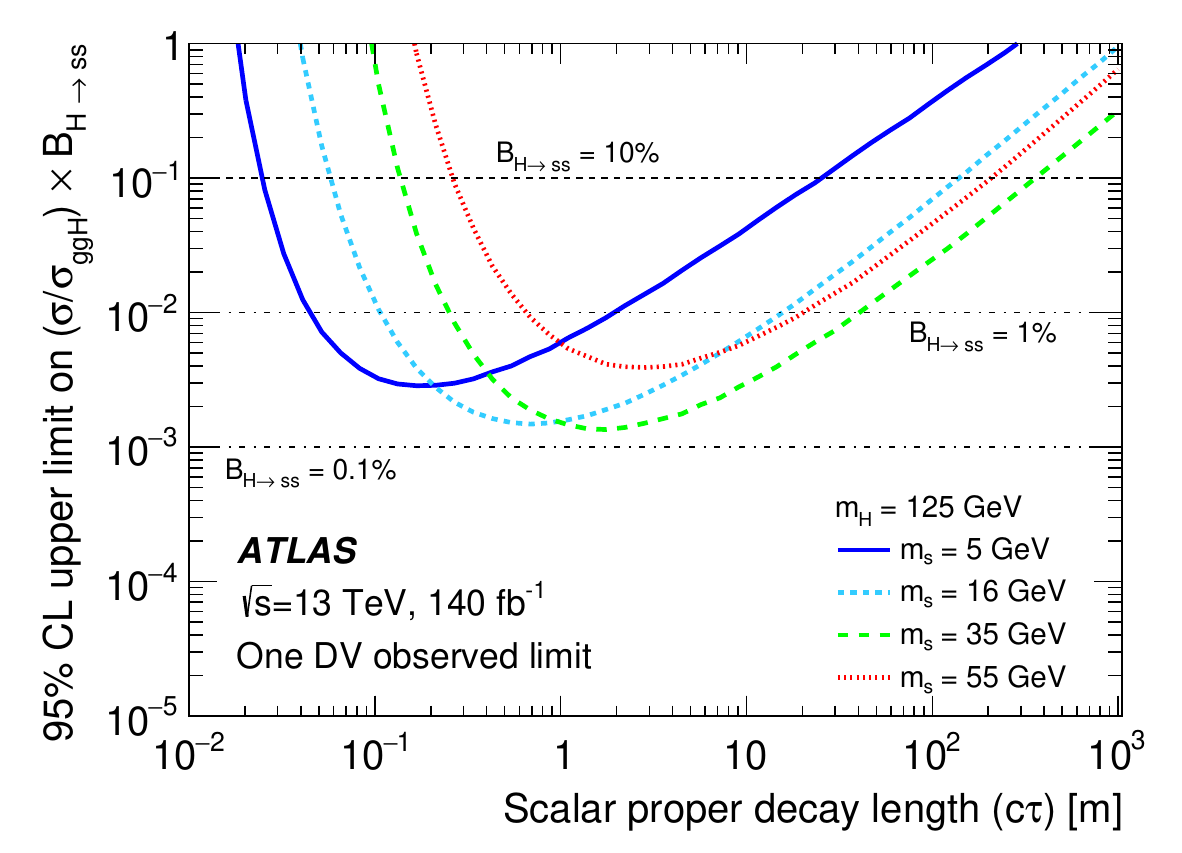}
\caption{}
\end{subfigure} \\
\vspace{0.2 cm}
\begin{subfigure}[b]{0.49\textwidth}
\includegraphics[width=\textwidth]{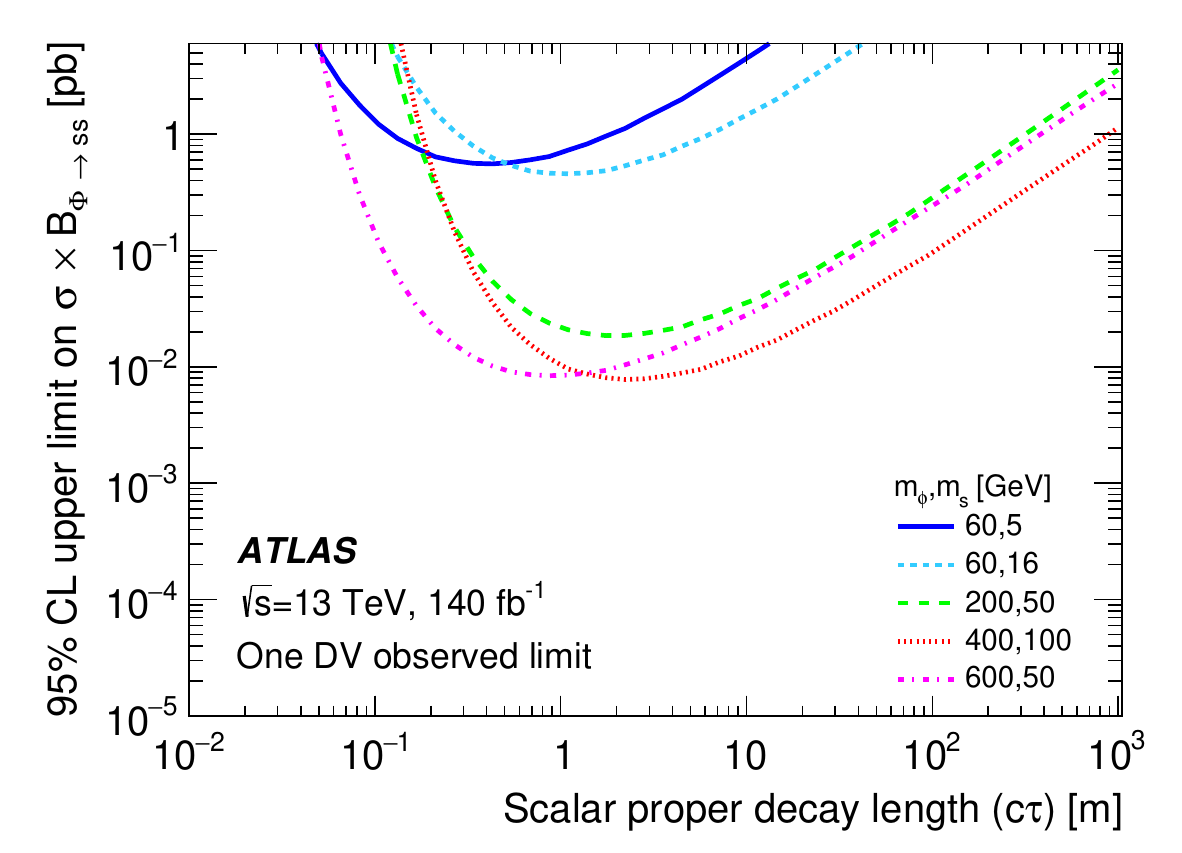}
\caption{}
\end{subfigure}
\begin{subfigure}[b]{0.49\textwidth}
\includegraphics[width=\textwidth]{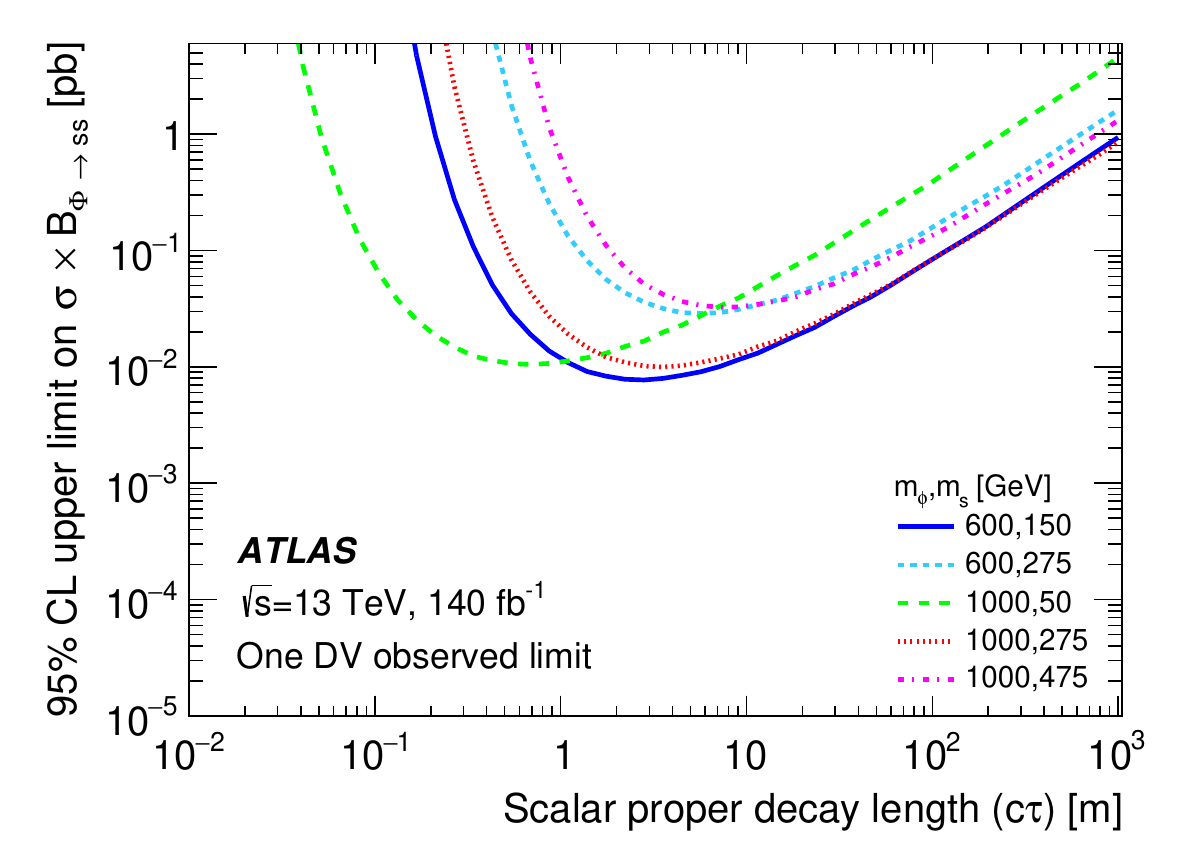}
\caption{}
\end{subfigure}
\caption{Summary of the one-DV limits for the $H/\varPhi \rightarrow ss$ model. (a) Comparison between observed and expected 95\% CL limits on $\left(\sigma/\sigma_{\textrm{ggH}}\right) \times B$ for an SM-like Higgs boson portal mediator and $m_s=35$~\GeV. (b) Observed 95\% CL limits on $\left(\sigma/\sigma_{\textrm{ggH}}\right) \times B$ for all Higgs boson portal mediator samples where the cross-section is normalized to the SM Higgs boson gluon--gluon fusion production cross-section, $\sigma_{\textrm{ggH}} = 48.61$~pb~\cite{deFlorian:2016spz}. (c)--(d) Observed 95\% CL limits on $\sigma \times B$ for $m_{\varPhi}\neq125$~\GeV. The observed limits are consistent with the expected ones within the uncertainties.}
\label{fig:ObsLimits_hss}
\end{figure}

\begin{figure}[tb]
\centering
\begin{subfigure}[b]{0.49\textwidth}
\includegraphics[width=\textwidth]{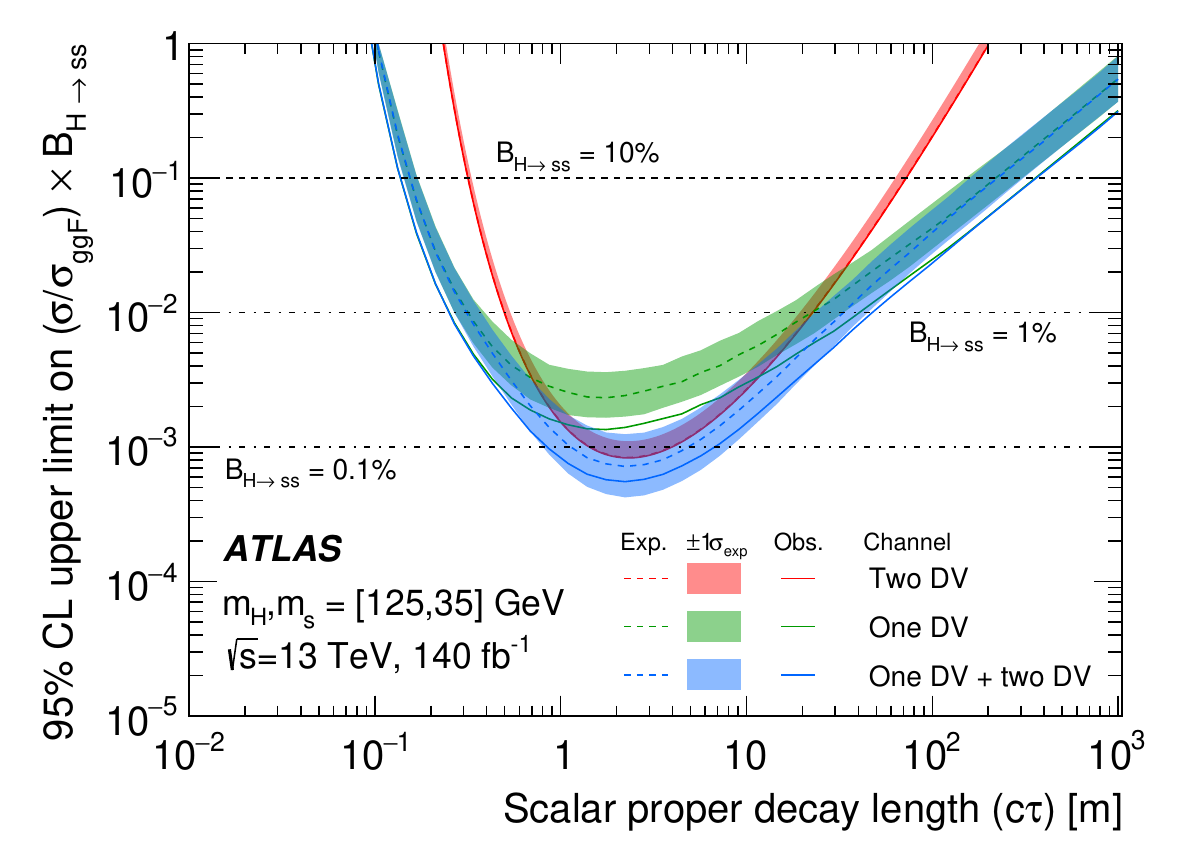}
\caption{}
\end{subfigure}
~
\begin{subfigure}[b]{0.49\textwidth}
\includegraphics[width=\textwidth]{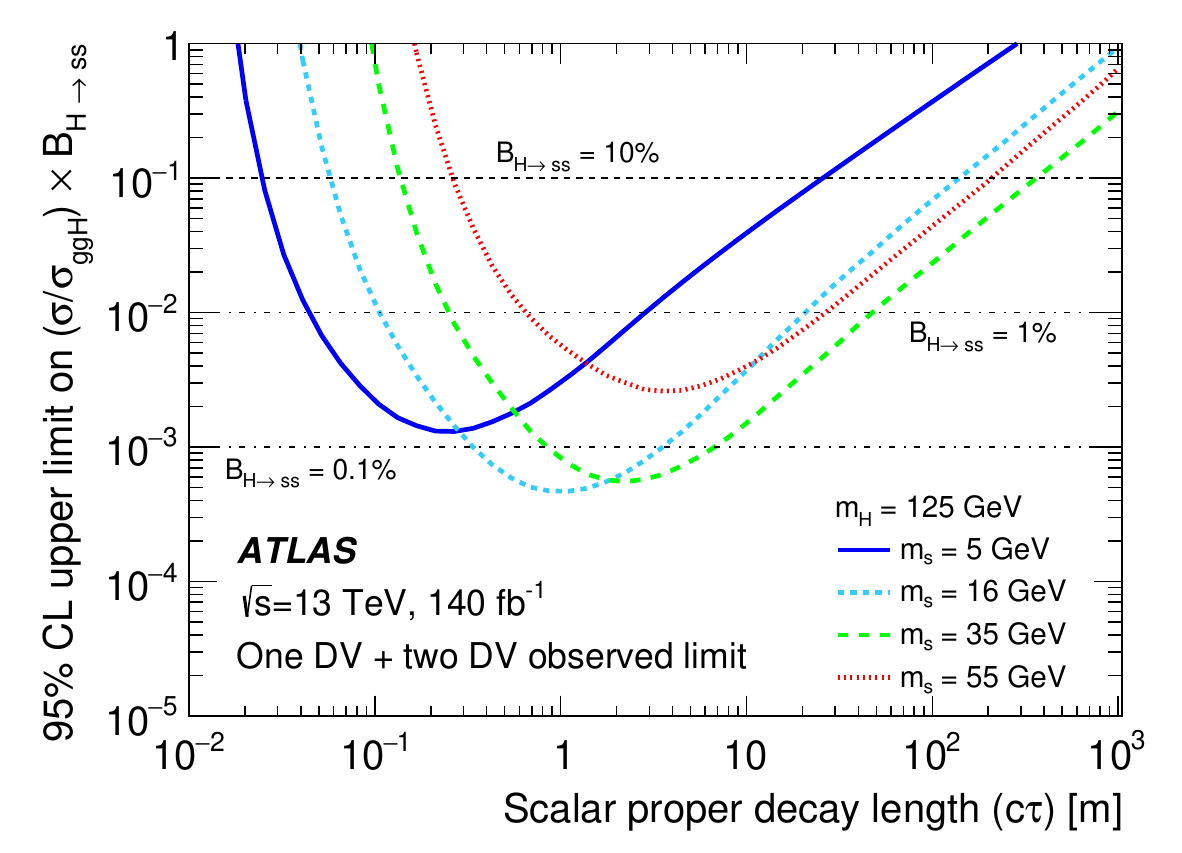}
\caption{}
\end{subfigure} \\
\vspace{0.2 cm}
\begin{subfigure}[b]{0.49\textwidth}
\includegraphics[width=\textwidth]{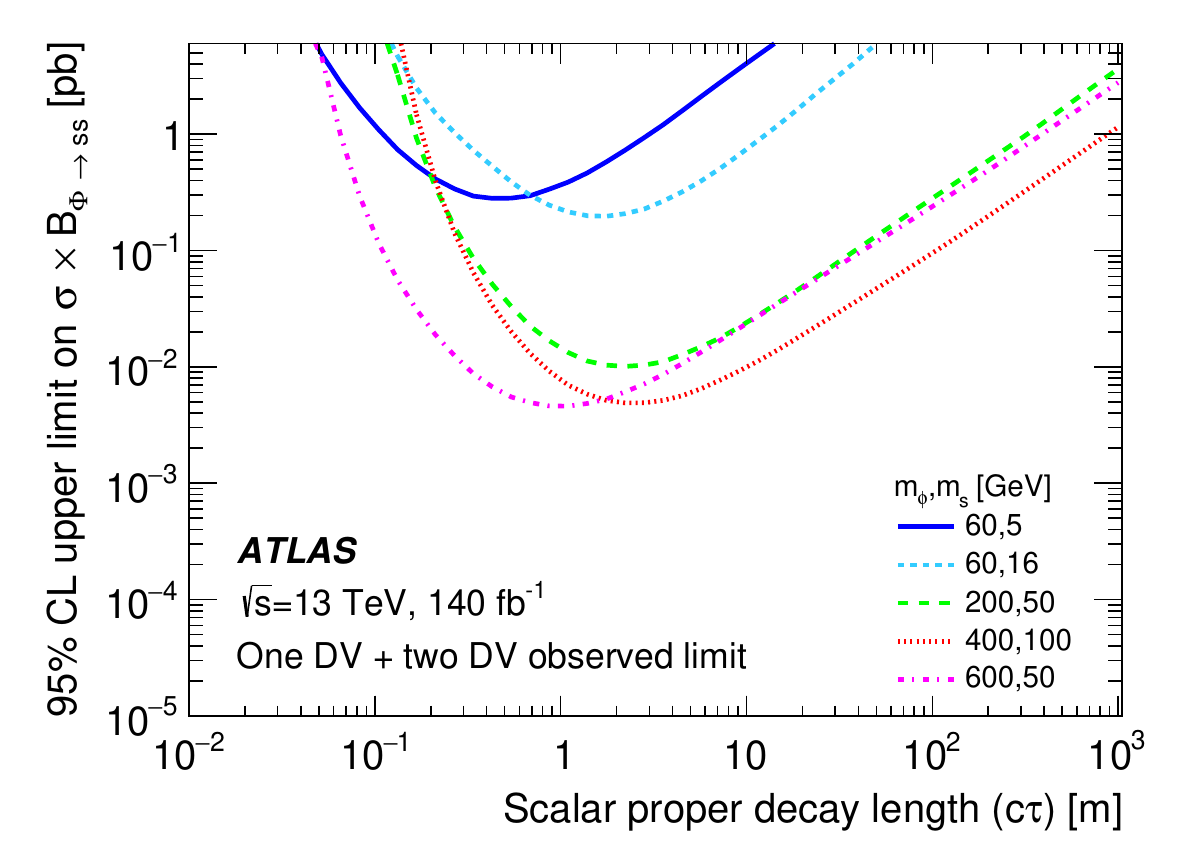}
\caption{}
\end{subfigure}
\begin{subfigure}[b]{0.49\textwidth}
\includegraphics[width=\textwidth]{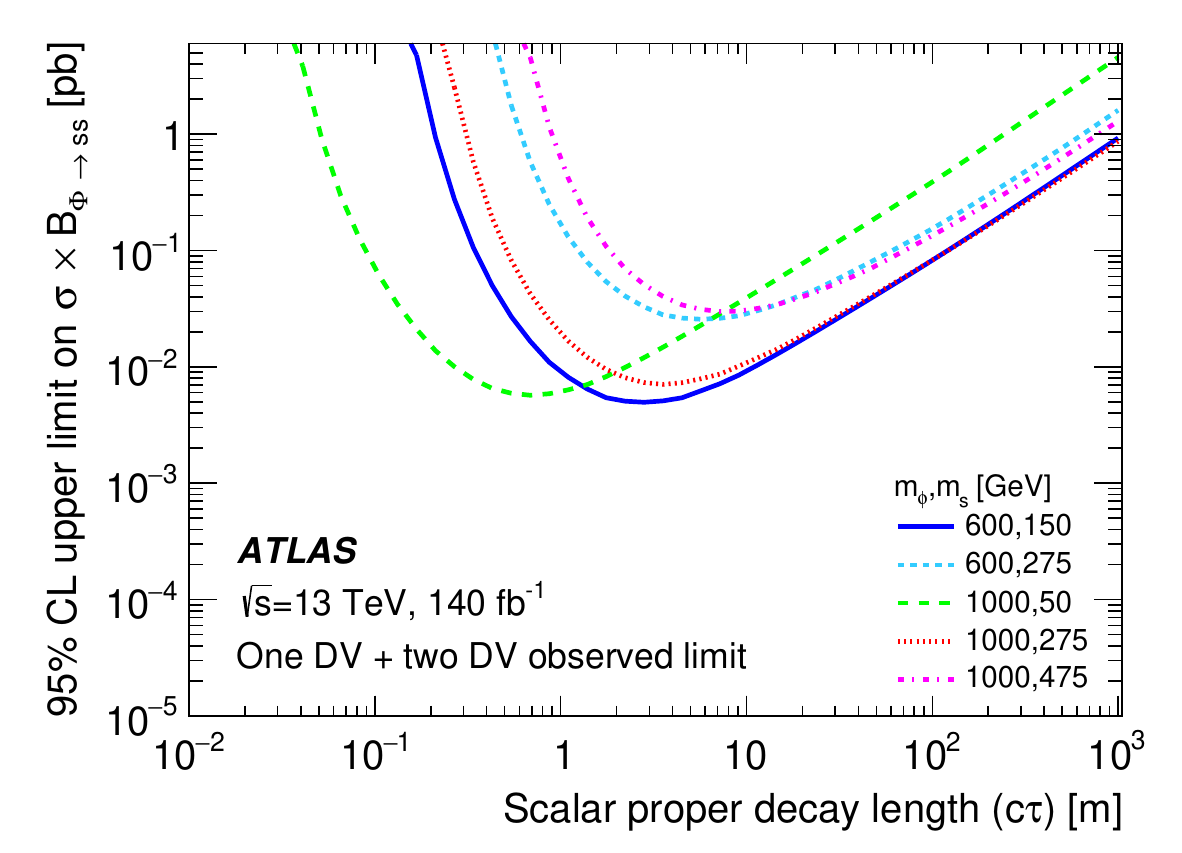}
\caption{}
\end{subfigure}
\caption{Summary of the one-DV and two-DV~\cite{EXOT-2019-24} limit combinations for the $H/\varPhi \rightarrow ss$ model. (a) Comparison between the one- and two-DV~\cite{EXOT-2019-24} results, and their combined 95\% CL limits on $\left(\sigma/\sigma_{\textrm{ggH}}\right) \times B$ for the Higgs boson portal mediator and $m_s=35$~\GeV. (b) Observed 95\% CL limits on $\left(\sigma/\sigma_{\textrm{ggH}}\right) \times B$ for all Higgs boson portal mediator samples where the cross-section is normalized to the SM Higgs boson gluon--gluon fusion production cross-section, $\sigma_{\textrm{ggH}} = 48.61$~pb~\cite{deFlorian:2016spz}. (c)--(d) Observed 95\% CL limits on $\sigma \times B$ for $m_{\varPhi}\neq125$~\GeV benchmark samples. The observed limits are consistent with the expected ones within the uncertainties.}
\label{fig:ObsLimits_Comb}
\end{figure}

For scalar boson benchmark samples with $m_{\varPhi}\neq125$~\GeV, upper limits are set on $\sigma \times B_{\varPhi \rightarrow ss}$, where $B_{\varPhi \rightarrow ss}$ is the branching fraction for $\varPhi \rightarrow ss$ assuming 100\% branching fraction of $s$ into the heaviest kinematically accessible fermion pairs. As discussed in Section~\ref{sec:models}, the long-lived scalar predominantly decays into $b\overline{b}$, except when $m_s>2m_t$ (e.g., for the sample with $m_{\varPhi}=1000$~\GeV\ and $m_s=475$~\GeV) where the dominant decay is $t\overline{t}$. For the Higgs boson portal mediator, upper limits are set on $\left(\sigma/\sigma_{\textrm{ggH}}\right) \times B_{H \rightarrow ss}$, where $\sigma_{\textrm{ggH}} = 48.61$~pb~\cite{deFlorian:2016spz} is the SM Higgs boson gluon--gluon fusion production cross-section. Figure~\ref{fig:ObsLimits_hss} compares the expected and observed 95\% CL upper limits for a representative sample and shows the observed 95\% CL limits for all scalar boson benchmarks. The observed limits are consistent with the expected ones within two standard deviations for all samples. For the Higgs boson portal mediator samples limits are stronger at intermediate LLP masses but weaken at very low and very high masses. Moreover, the mean proper decay length $c\tau$ where the limits are strongest increases with the long-lived scalar mass. These trends are correlated with variations in trigger and reconstruction efficiencies mainly driven by the LLP decay kinematics. Figure~\ref{fig:ObsLimits_Comb} shows the expected and observed limits for the one-DV search reported in this work and the two-DV search described in Ref.~\cite{EXOT-2019-24}, along with their combination for a scalar boson benchmark sample. The two searches use mutually exclusive selections. The one-DV and two-DV systematic uncertainties are assumed to be uncorrelated in the combination. The two-DV search benefits from very low background and therefore has better sensitivity than the one-DV search for $c\tau$ of the order of tens of meters. Requiring both particles to decay within the MS reduces the sensitivity for longer $c\tau$ values, where the exclusion limits on cross-sections scale as $(c\tau)^{-2}$. The one-DV search significantly extends the sensitivity to longer- and shorter-lived LLPs, with exclusion limits scaling as $(c\tau)^{-1}$. Observed limits for the one- and two-DV combination are also provided for all scalar boson benchmarks. The lifetime ranges excluded by the combination of the one- and two-DV results for branching fractions $B_{H \rightarrow ss}$ of 10\%, 1\% and 0.1\% for the scalar boson with $m_H = 125$~\GeV are summarized in Table~\ref{tab:ExcludedRange_HSS}.

\begin{table}[tb]
\centering
\caption{Ranges of mean proper decay length excluded at 95\% CL by the one-DV results and the combination of one-DV and two-DV~\cite{EXOT-2019-24} results for scalar boson benchmark models with \mbox{$m_H=125$~\GeV}, assuming $B_{H\rightarrow ss}$ equal to 10\%, 1\% and 0.1\% of the SM Higgs boson production cross-section~\cite{deFlorian:2016spz}.}\label{tab:ExcludedRange_HSS}
\begin{tabular}{c|ccc|ccc}
\toprule
& \multicolumn{6}{c}{Excluded $c\tau$ range for $s$ [m]} \\
$H \rightarrow ss$ & \multicolumn{3}{c|}{One DV} & \multicolumn{3}{c}{One DV + two DV} \\
$m_{s}$~[\GeV]  & $B = 0.1\%$ & $B = 1\%$ & $B = 10\%$ & $B = 0.1\%$ & $B = 1\%$ & $B = 10\%$  \\
\midrule
~~5  & N/A & 0.05 -- 1.96 & 0.03 -- 25.6  & N/A         & 0.05 -- 2.9~~ &  0.03 -- 25.9~~ \\
16   & N/A & 0.11 -- 15.8 & 0.06 -- 139.4 & 0.34 -- 3.6 & 0.11 -- 20.8  &  0.06 -- 141.0 \\
35   & N/A & 0.26 -- 40.1 & 0.14 -- 358.6 & 0.85 -- 6.8 & 0.26 -- 47.7  &  0.14 -- 362.7 \\
55   & N/A & 0.67 -- 20.3 & 0.27 -- 208.6 & N/A         & 0.66 -- 26.8  &  0.27 -- 208.6  \\
\bottomrule
\end{tabular}
\end{table}

\begin{figure}[tb]
\centering
\begin{subfigure}[b]{0.48\textwidth}
\includegraphics[width=\textwidth]{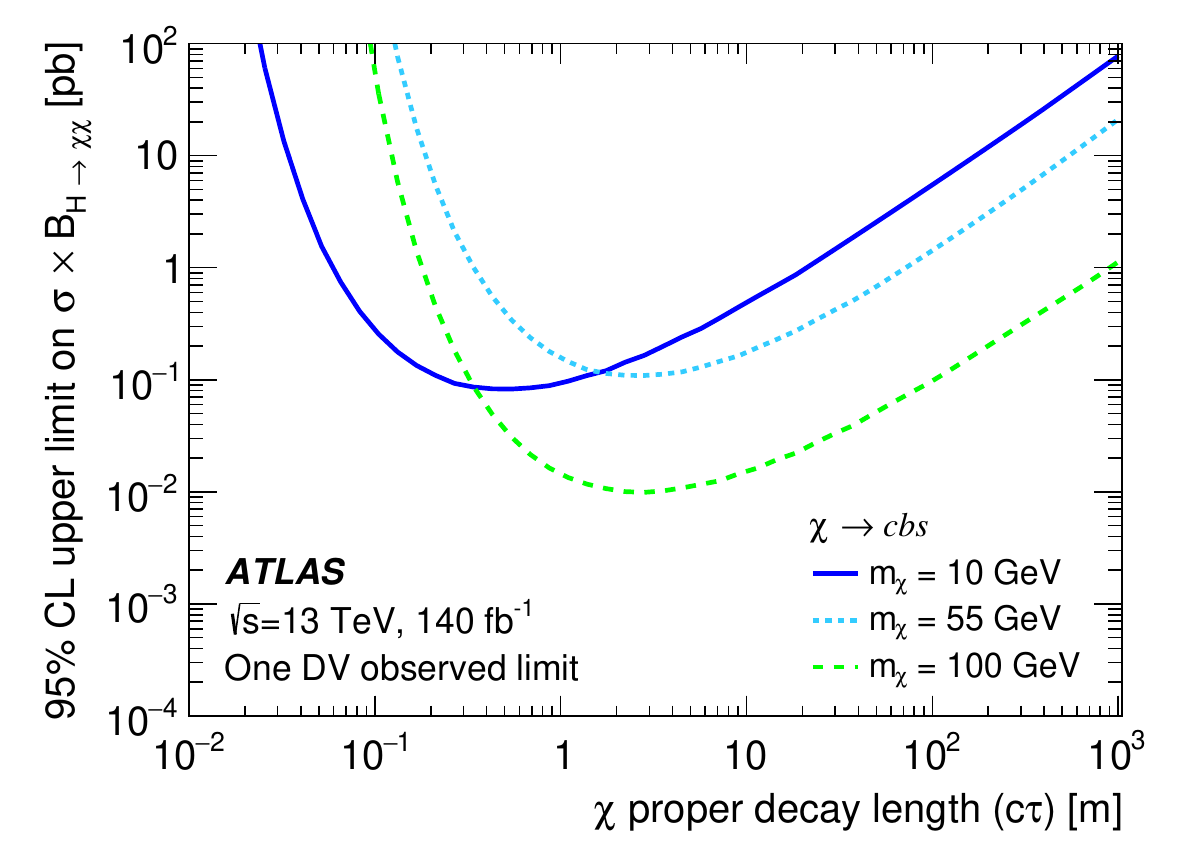}
\caption{}
\end{subfigure}
~
\begin{subfigure}[b]{0.48\textwidth}
\includegraphics[width=\textwidth]{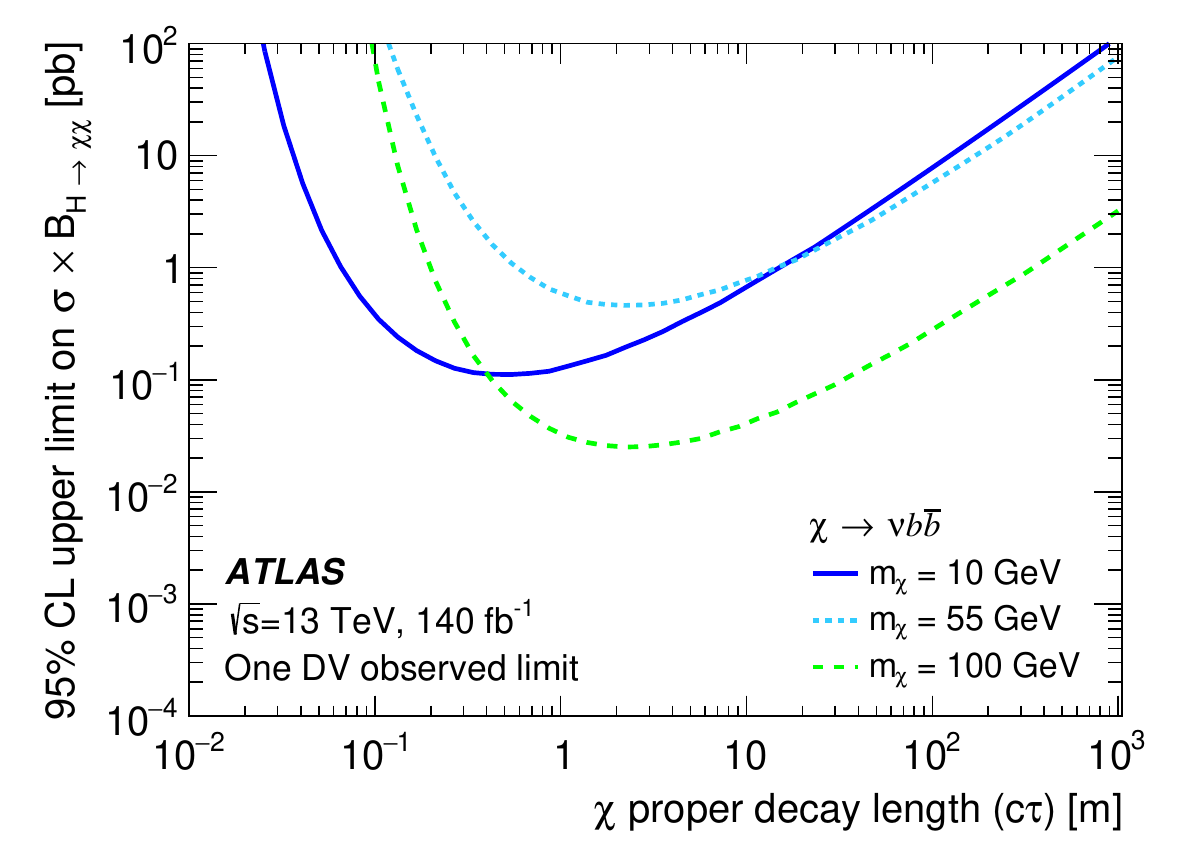}
\caption{}
\end{subfigure} \\
~
\begin{subfigure}[b]{0.48\textwidth}
\includegraphics[width=\textwidth]{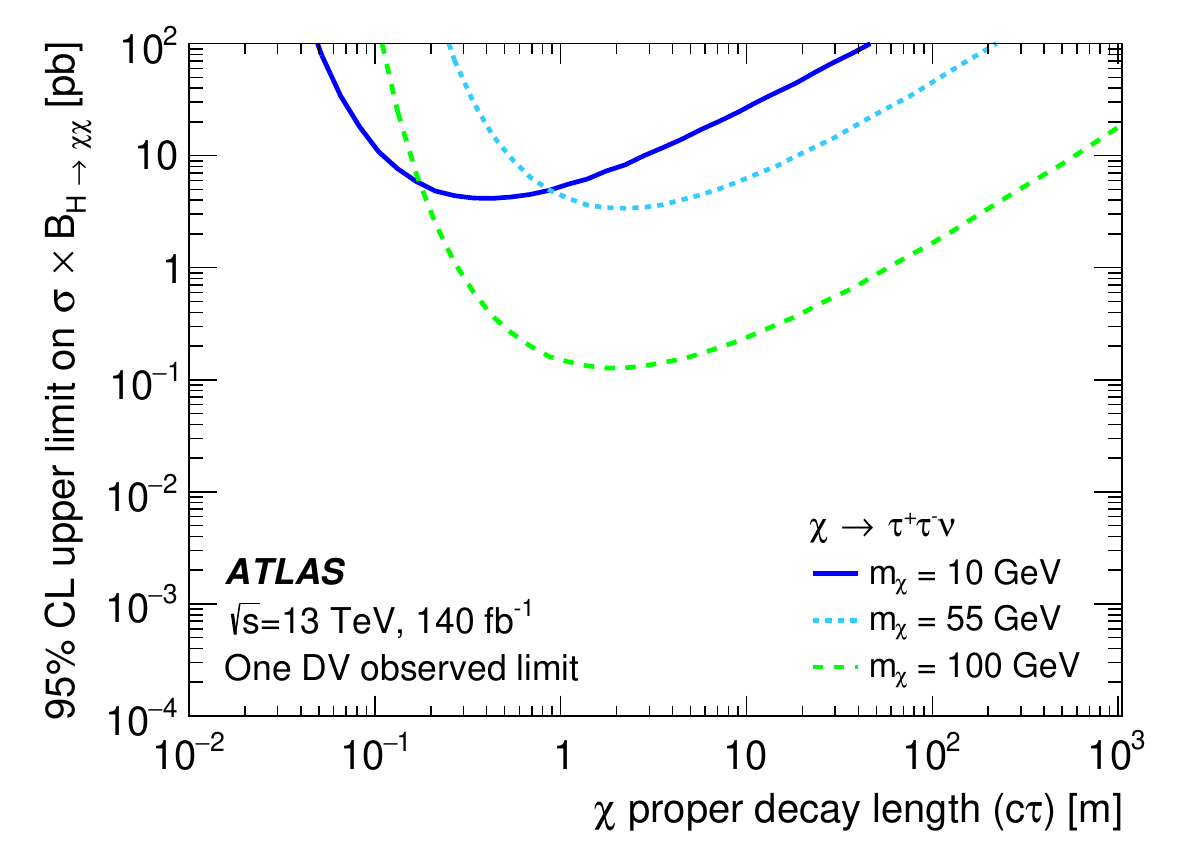}
\caption{}
\end{subfigure}

\caption{Observed 95\% CL limits on $\sigma \times B$ for baryogenesis samples for the one-DV analysis. The observed limits are consistent with the expected ones within the uncertainties.}
\label{fig:ObsLimits_bg}
\end{figure}

Figure~\ref{fig:ObsLimits_bg} shows the observed 95\% CL limits for all the baryogenesis benchmark samples. The observed limits are consistent with the expected ones within two standard deviations for all samples.

\begin{figure}[tb]
\centering
\begin{subfigure}[b]{0.48\textwidth}
\includegraphics[width=\textwidth]{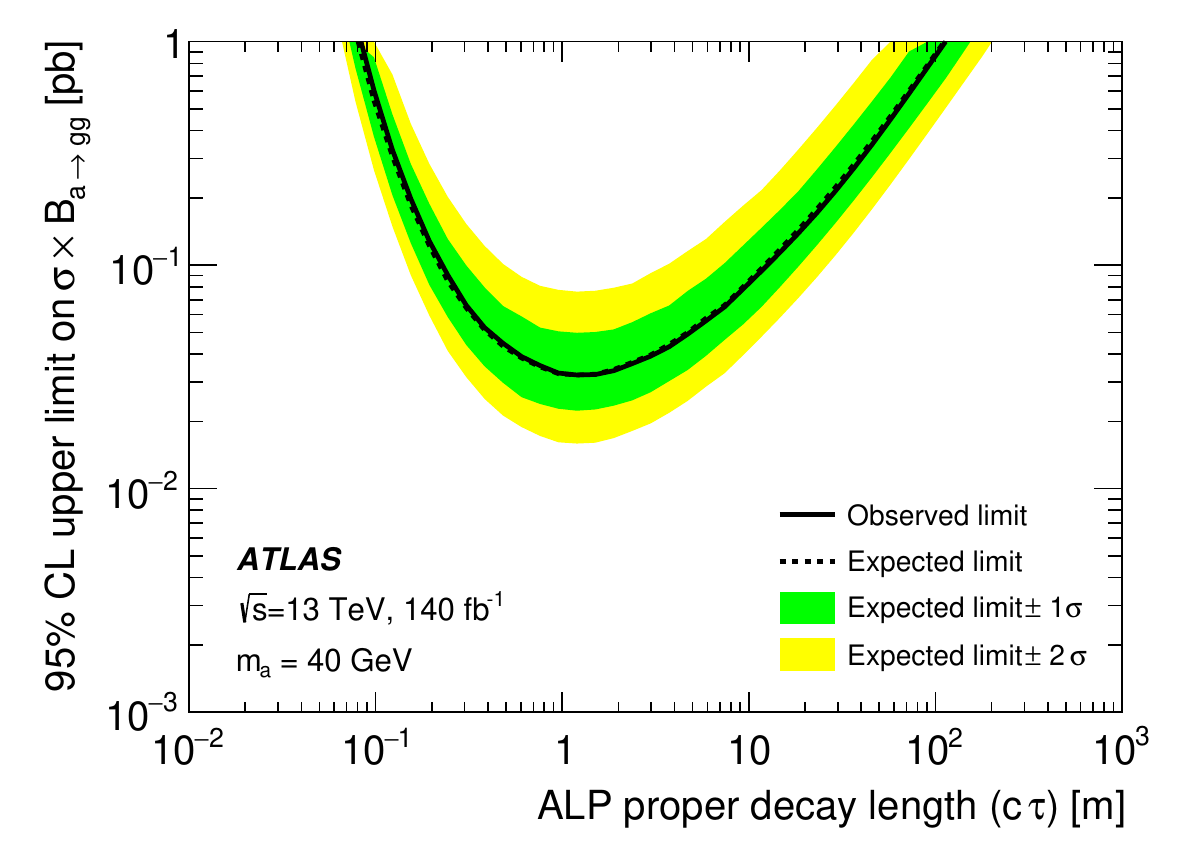}
\caption{}
\end{subfigure}
\begin{subfigure}[b]{0.48\textwidth}
\includegraphics[width=\textwidth]{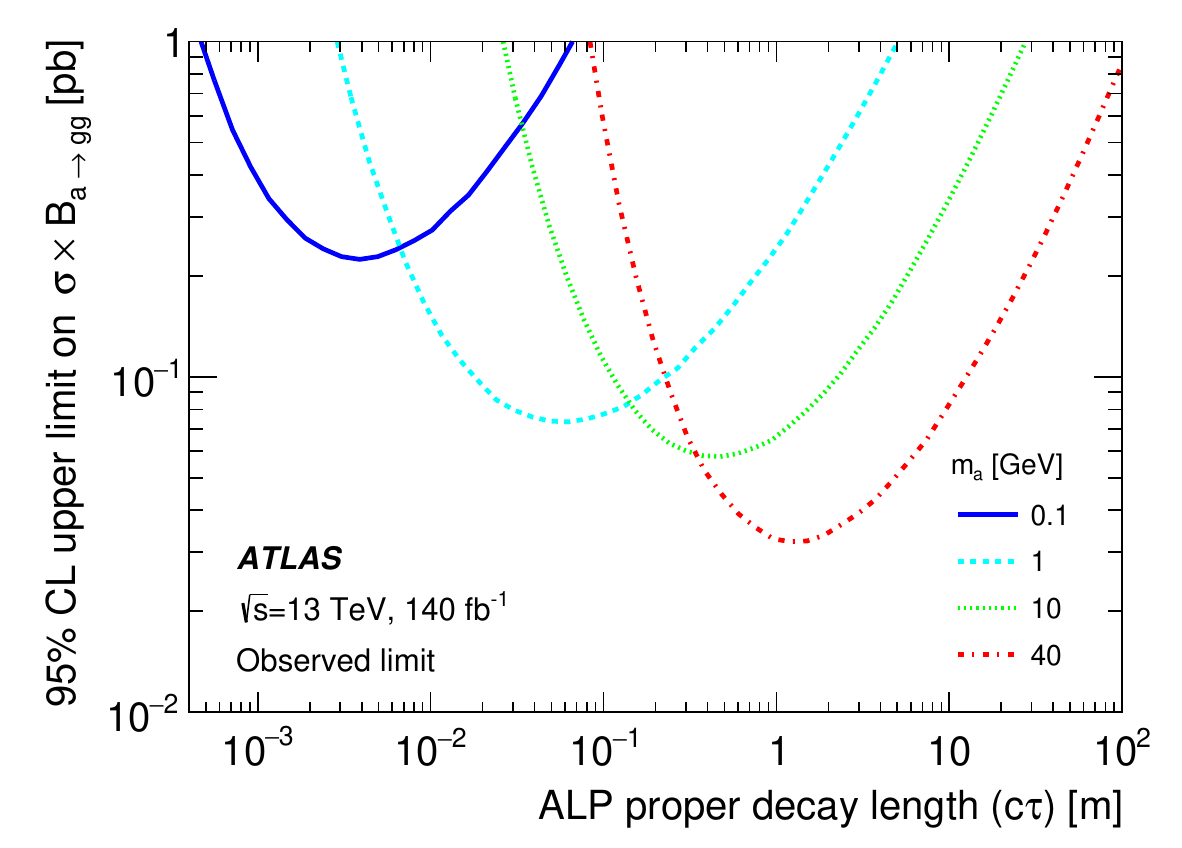}
\caption{}
\end{subfigure}
\caption{Summary of the limits for the $\Zboson+$ALP model. (a) Comparison between observed and expected 95\% CL upper limits on the $\Zboson+$ALP production cross-section $\sigma \times B_{a \rightarrow gg}$ for $m_a=$ 40~\GeV. (b) Observed 95\% CL upper limits on $\sigma \times B_{a \rightarrow gg}$ for all considered ALP mass points.}
\label{fig:ZALPLimits}
\end{figure}

For the ALP models, upper limits are set on the production $\Zboson+$ALP cross-section $\sigma \times B_{a \rightarrow gg}$, where $B_{a \rightarrow gg}$ is the branching fraction for the ALP to decay into gluons. Figure~\ref{fig:ZALPLimits} shows the observed 95\% CL upper limits on $\sigma \times B_{a \rightarrow gg}$ for all considered ALP masses. More stringent limits are observed for large ALP masses, where the reduced boost and wider decay angles increase the vertex reconstruction efficiency. The observed and expected limits are compared for a representative sample, showing consistency within two standard deviations. This agreement holds for all samples considered.

\begin{figure}[tb]
\centering
\begin{subfigure}[b]{0.48\textwidth}
\includegraphics[width=\textwidth]{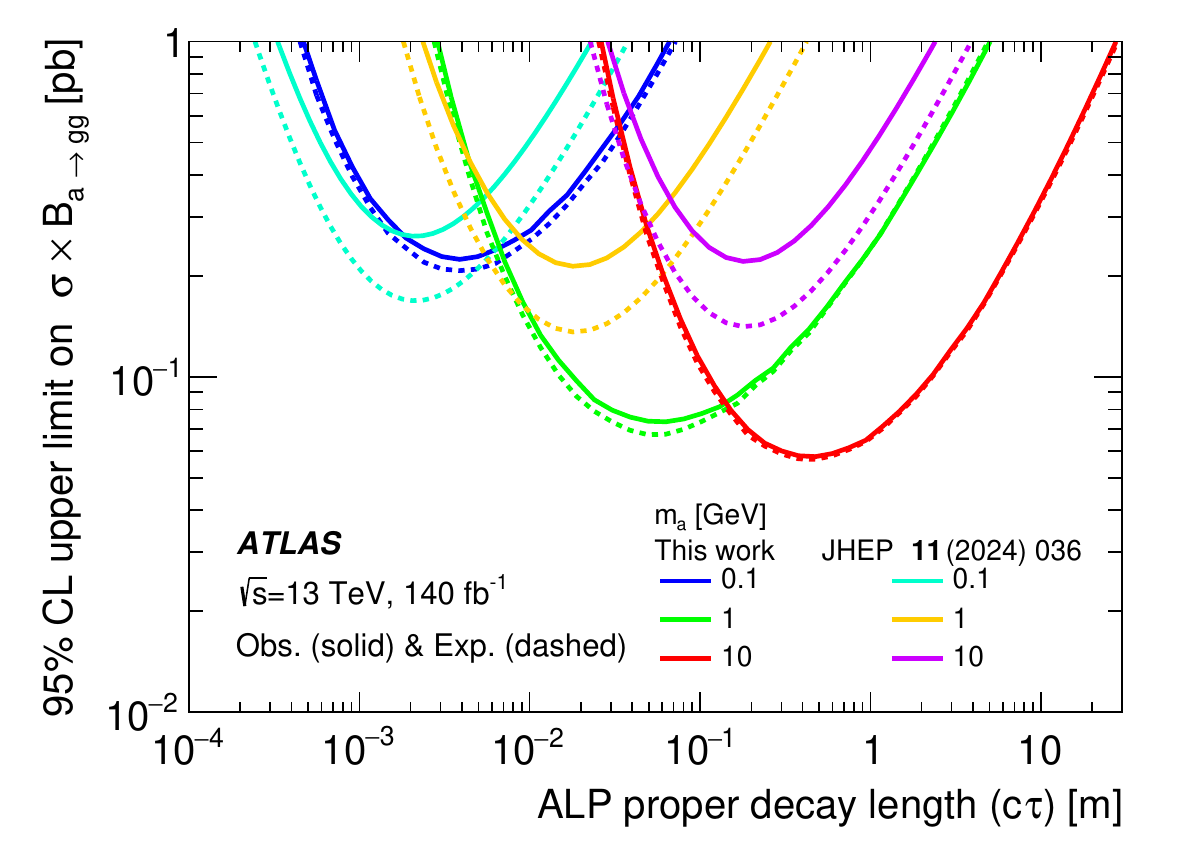}
\caption{}
\end{subfigure}
\begin{subfigure}[b]{0.48\textwidth}
\includegraphics[width=\textwidth]{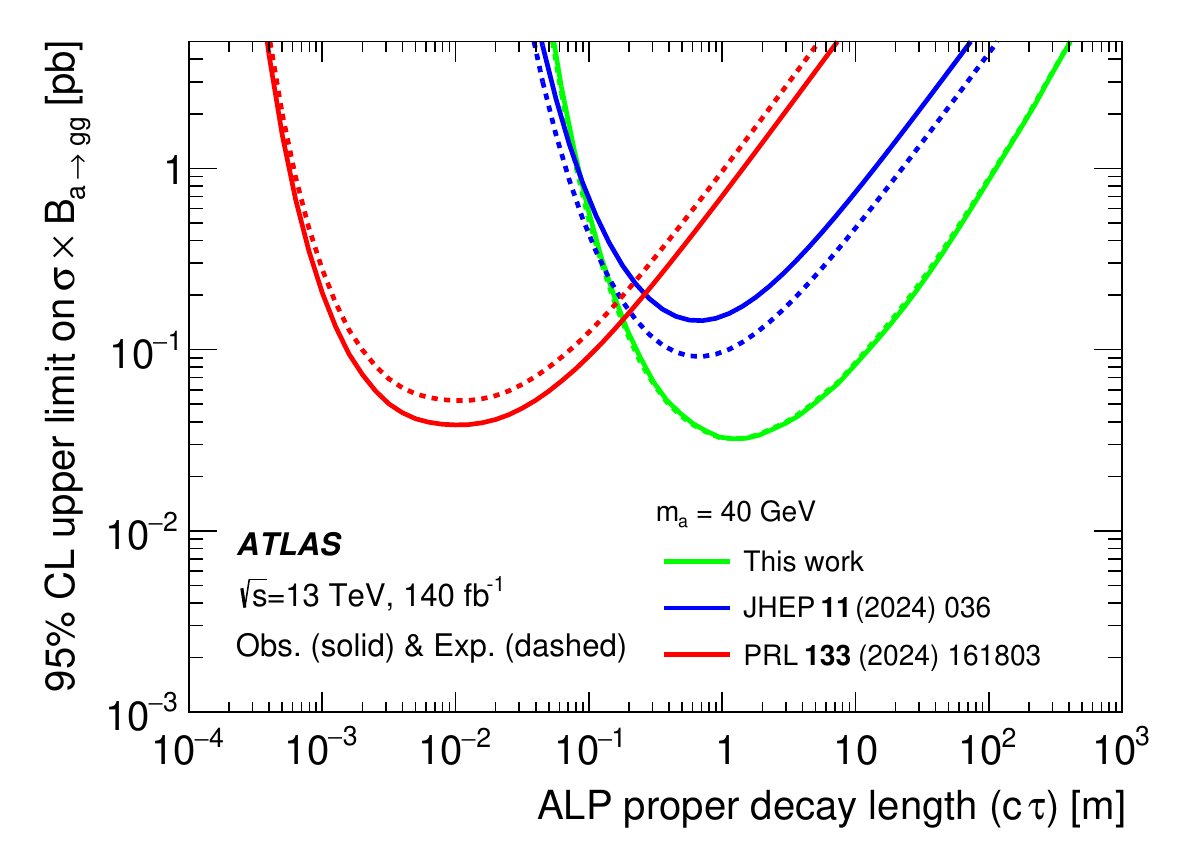}
\caption{}
\end{subfigure}
\caption{Comparison of the expected and observed 95\% CL upper limits on the $\Zboson+$ALP production cross-section for (a) \mbox{$m_a=$ 0.1, 1, and 10~\GeV} and the one of Ref.~\cite{EXOT-2022-04} and for (b) $m_a=$ 40~\GeV which is also covered in Ref.~\cite{EXOT-2021-32}. The observed limits are consistent with the expected ones within the uncertainties.}
\label{fig:ALPLimitComps}
\end{figure}

Figure~\ref{fig:ALPLimitComps} provides a comparison between the expected and observed limits for the ALP model using the result reported in this work, a previous displaced jet search in the HCal~\cite{EXOT-2022-04}, and a displaced vertex search in the ID~\cite{EXOT-2021-32}. The observed limits are consistent with the expected ones within two standard deviations for all samples.

\begin{figure}[tb]
\centering
\begin{subfigure}[b]{0.49\textwidth}
\includegraphics[width=\textwidth]{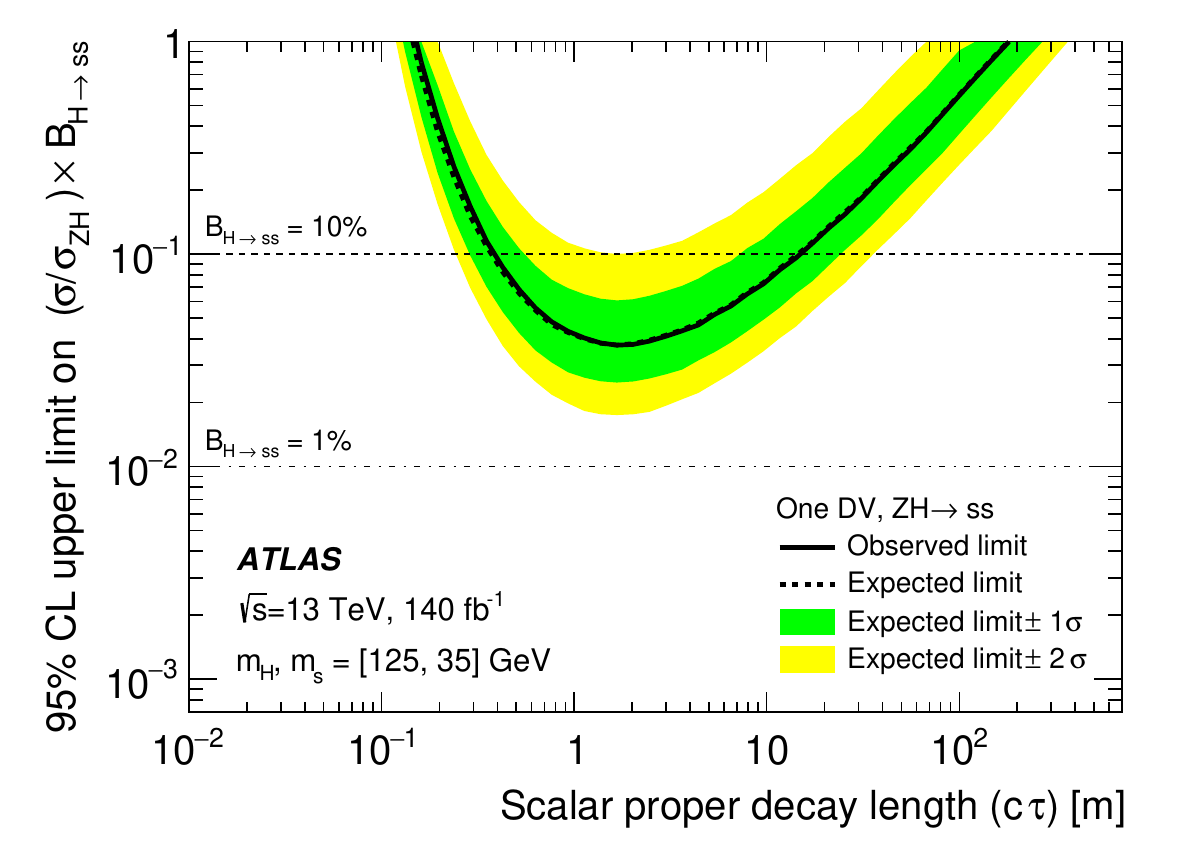}
\caption{}
\end{subfigure}
~
\begin{subfigure}[b]{0.49\textwidth}
\includegraphics[width=\textwidth]{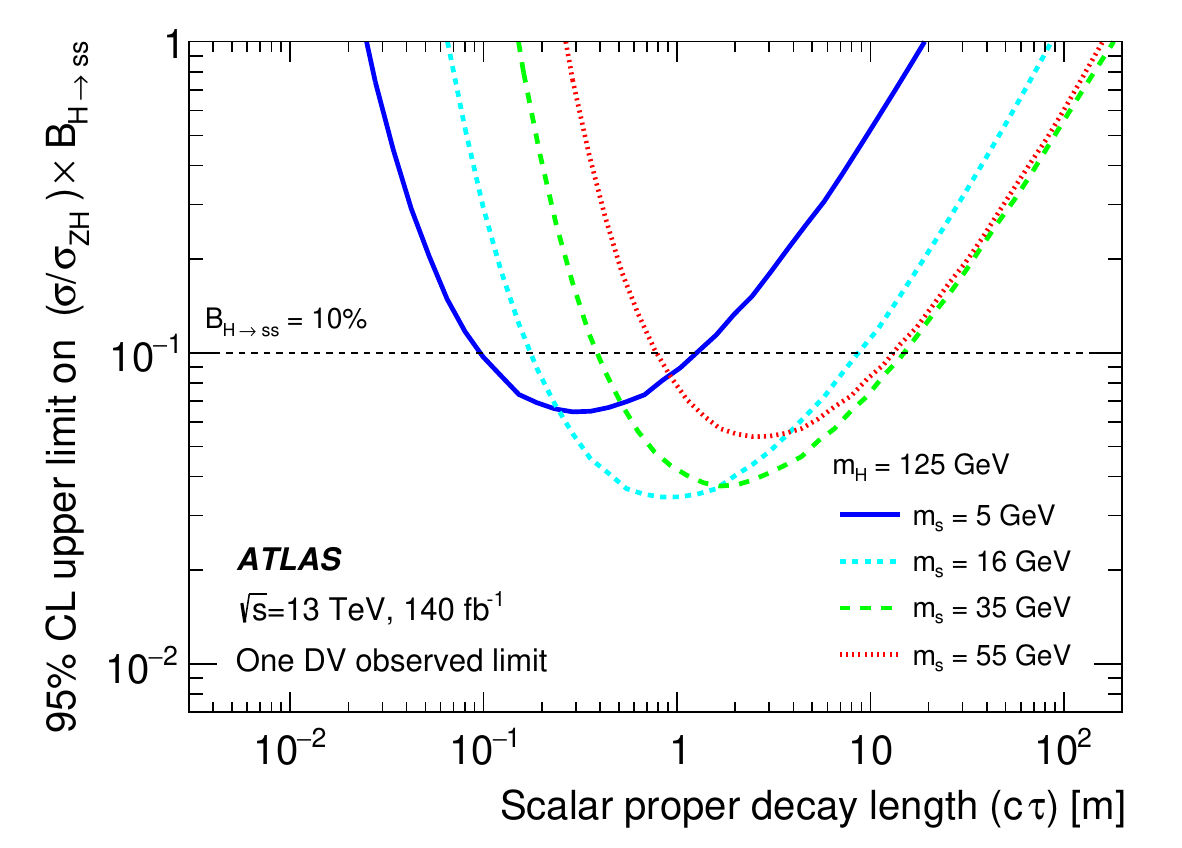}
\caption{}
\end{subfigure} \\
\vspace{0.2 cm}
\begin{subfigure}[b]{0.49\textwidth}
\includegraphics[width=\textwidth]{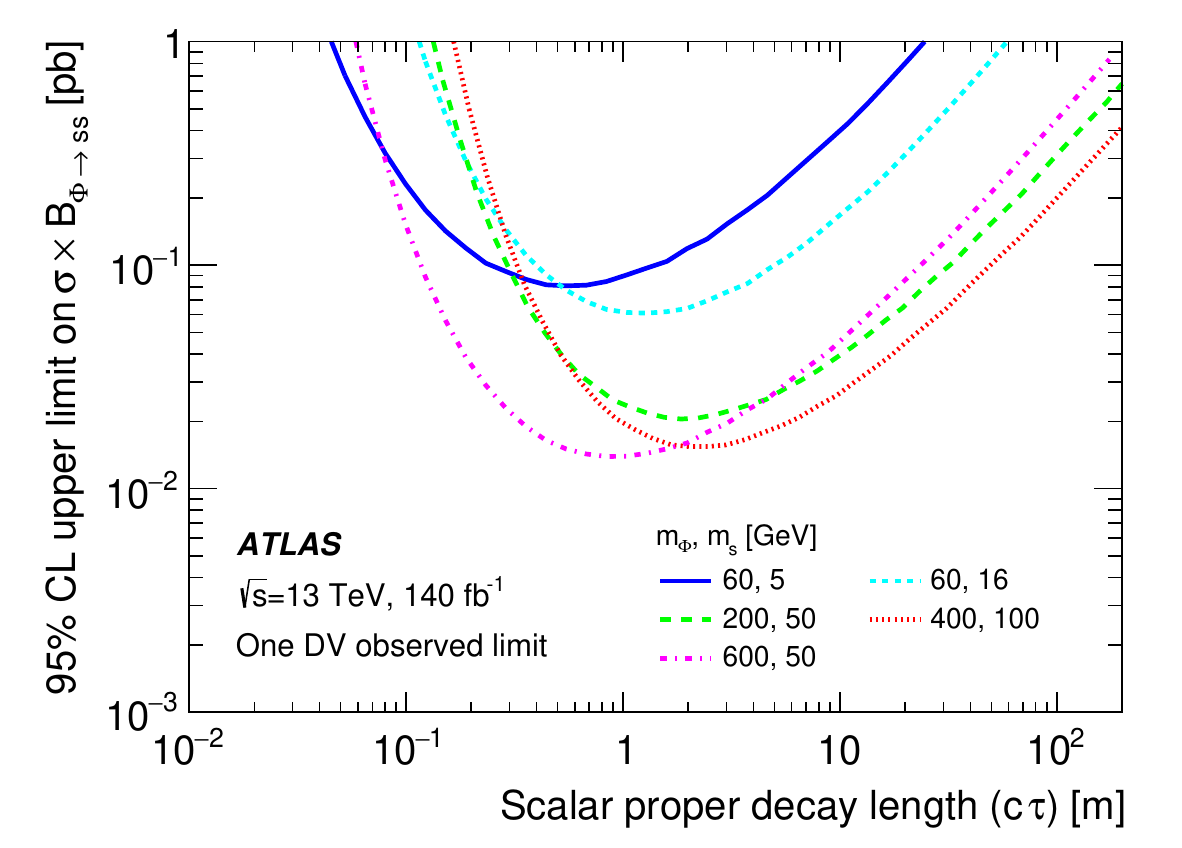}
\caption{}
\end{subfigure}
~
\begin{subfigure}[b]{0.49\textwidth}
\includegraphics[width=\textwidth]{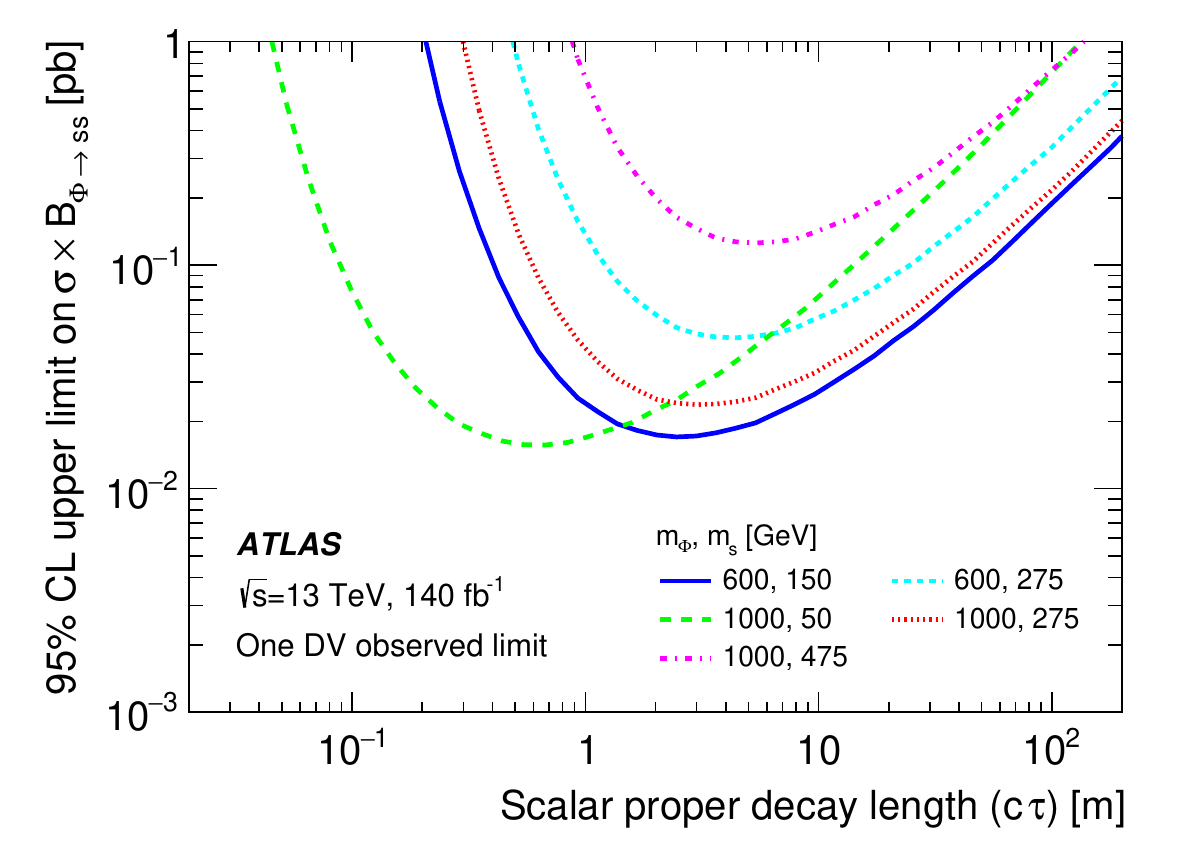}
\caption{}
\end{subfigure}
\caption{Comparison between the observed and expected 95\% CL limits (a) on $\left(\sigma/\sigma_{\textrm{ZH}}\right) \times B_{H \to ss}$ for Higgs boson portal mediator and $m_s=35$~\GeV for \Zboson-associated $H$ production with one DV\@. (b) Observed 95\% CL limits on $\left(\sigma/\sigma_{\textrm{ZH}}\right) \times B_{H \to ss}$ for all Higgs boson portal mediator samples where the cross-section is normalized to the \mbox{\Zboson($\rightarrow \ell \ell$)-associated} Higgs boson production cross-section, $\sigma_{\textrm{ZH}} = 0.089$~pb~\cite{deFlorian:2016spz}. (c)-(d) Observed 95\% CL limits on $\sigma \times B_{\varPhi \to ss}$ for $m_{\varPhi}\neq125$~\GeV. The observed limits are consistent with the expected ones within the uncertainties.}
\label{fig:ObsLimits_ZHss}
\end{figure}

The scalar boson benchmark samples with associated \Zboson production have limits set using the same method as the direct scalar production mode, where samples with $m_{\varPhi}\neq125$~\GeV have upper limits set on $\sigma \times B_{\varPhi \to ss}$, where $B_{\varPhi \to ss}$ is the branching fraction for $\varPhi \rightarrow ss$ assuming 100\% branching fraction of $s$ into the heaviest kinematically accessible fermion pairs.
For Higgs boson portal mediator samples, upper limits are set on $\left(\sigma/\sigma_{\textrm{ZH}}\right) \times B_{H \to ss}$, where $\sigma_{\textrm{ZH}} = 0.089$~pb~\cite{deFlorian:2016spz} is the SM \Zboson($\rightarrow \ell \ell$)-associated Higgs boson production cross-section calculated at NLO precision. Figure~\ref{fig:ObsLimits_ZHss} shows the expected and observed limits for a representative benchmark mass, and the observed limits for all benchmark samples. The observed limits are consistent with the expected ones within two standard deviations for all samples.

\begin{figure}[tb]
\centering
\begin{subfigure}[b]{0.48\textwidth}
\includegraphics[width=\textwidth]{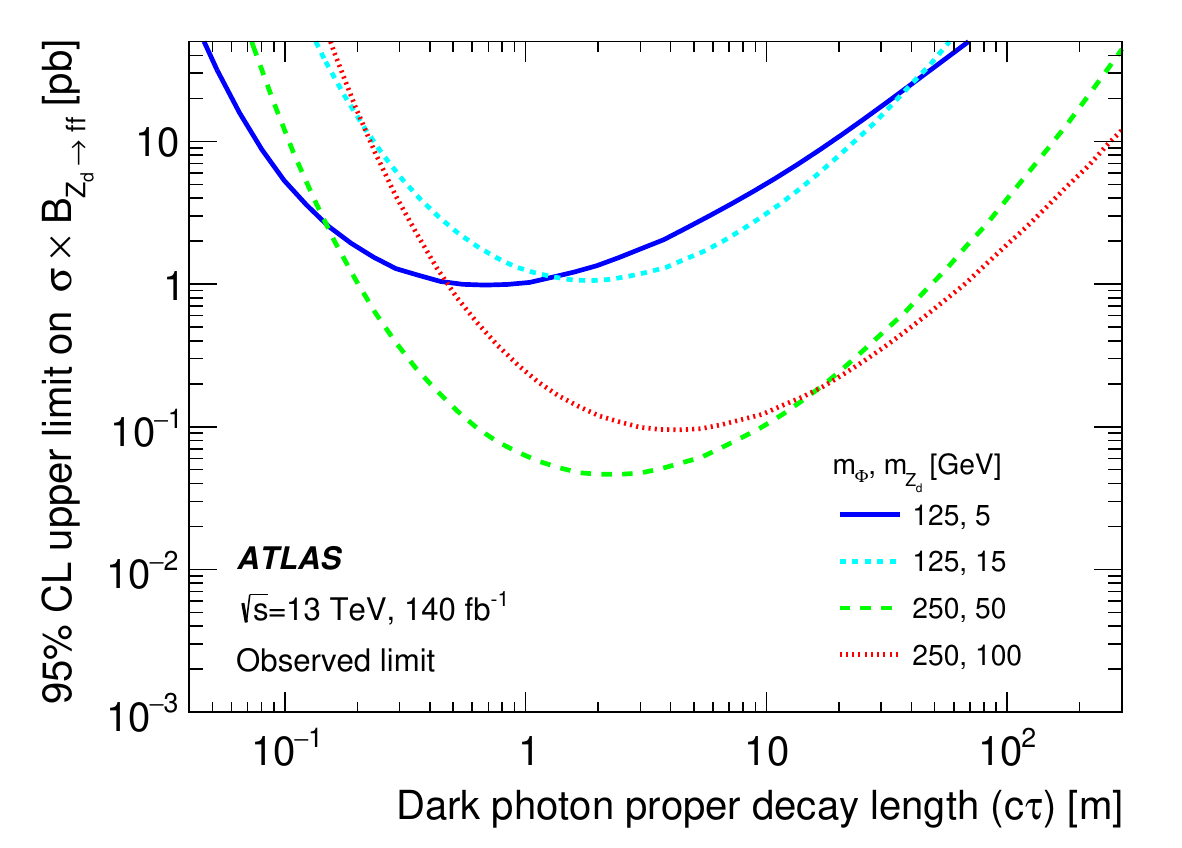}
\caption{}
\end{subfigure}
\begin{subfigure}[b]{0.48\textwidth}
\includegraphics[width=\textwidth]{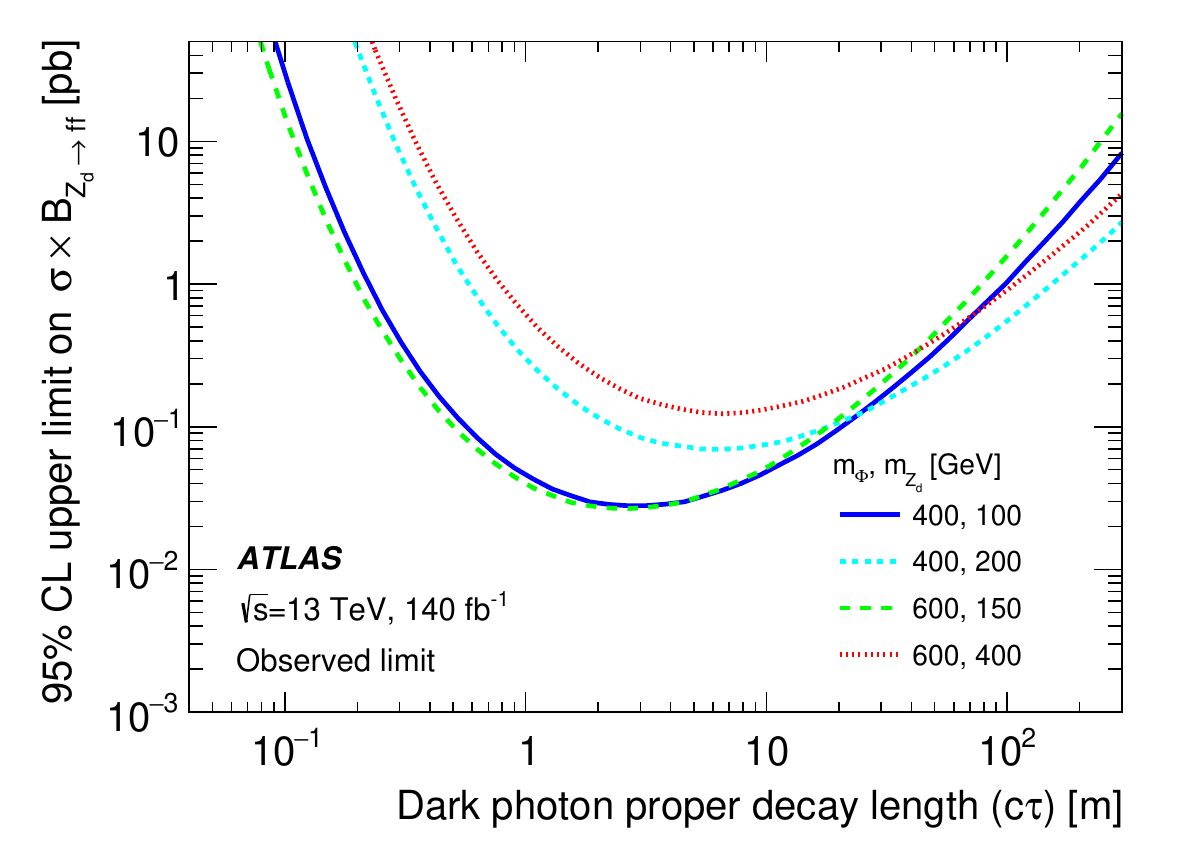}
\caption{}
\end{subfigure}
\caption{Observed 95\% CL upper limits on the $\sigma \times B_{\Zboson_{d} \rightarrow ff}$ production cross-section for dark photon $\Zboson_{d}$ benchmark samples. The observed limits are consistent with the expected ones within the uncertainties.}
\label{fig:ObsLimits_HZZd}
\end{figure}

Figure~\ref{fig:ObsLimits_HZZd} shows the observed 95\% CL limits for all $\varPhi\rightarrow \Zboson \Zboson_{d}$ samples, with limits set on $\sigma \times B_{\Zboson_{d} \rightarrow ff}$, where $B_{\Zboson_{d} \rightarrow ff}$ is the branching fraction for the dark photon to decay into a pair of fermions. The observed limits are consistent with the expected ones within two standard deviations for all samples.

\clearpage
\section{Conclusion}

A search for events featuring a single displaced vertex from pair-produced long-lived particles decaying into hadronic jets is presented. The analysis uses 140~fb$^{-1}$ of \mbox{proton--proton} collisions at $\sqrt{s}=13$~\TeV recorded by the ATLAS detector at the LHC during the 2015--2018 data-taking period. Neural networks are employed to enhance signal selection, and a data-driven method is used to estimate the expected background in the signal region. A dedicated analysis channel is included to target LLPs produced in association with prompt leptons from a \Zboson boson decay. No excess events above the background are observed in the signal regions, and exclusion limits on the LLP production cross-section as a function of its proper decay length are computed for scalar portal, baryogenesis, photo-phobic ALP, and dark photon models. For the SM Higgs boson, branching fractions above 1\% are excluded at 95\% confidence level for LLP proper decay lengths ranging from 5~cm to 40~m. For the $\Zboson+$ALP model, the most stringent ATLAS limits to date are set for proper decay lengths greater than $\mathcal{O}(10^{-1})$ m.

\section*{Acknowledgments}
%

%

%
%

%
%

We thank CERN for the very successful operation of the LHC and its injectors, as well as the support staff at
CERN and at our institutions worldwide without whom ATLAS could not be operated efficiently.

The crucial computing support from all WLCG partners is acknowledged gratefully, in particular from CERN, the ATLAS Tier-1 facilities at TRIUMF/SFU (Canada), NDGF (Denmark, Norway, Sweden), CC-IN2P3 (France), KIT/GridKA (Germany), INFN-CNAF (Italy), NL-T1 (Netherlands), PIC (Spain), RAL (UK) and BNL (USA), the Tier-2 facilities worldwide and large non-WLCG resource providers. Major contributors of computing resources are listed in Ref.~\cite{ATL-SOFT-PUB-2025-001}.

We gratefully acknowledge the support of ANPCyT, Argentina; YerPhI, Armenia; ARC, Australia; BMWFW and FWF, Austria; ANAS, Azerbaijan; CNPq and FAPESP, Brazil; NSERC, NRC and CFI, Canada; CERN; ANID, Chile; CAS, MOST and NSFC, China; Minciencias, Colombia; MEYS CR, Czech Republic; DNRF and DNSRC, Denmark; IN2P3-CNRS and CEA-DRF/IRFU, France; SRNSFG, Georgia; BMFTR, HGF and MPG, Germany; GSRI, Greece; RGC and Hong Kong SAR, China; ICHEP and Academy of Sciences and Humanities, Israel; INFN, Italy; MEXT and JSPS, Japan; CNRST, Morocco; NWO, Netherlands; RCN, Norway; MNiSW, Poland; FCT, Portugal; MNE/IFA, Romania; MSTDI, Serbia; MSSR, Slovakia; ARIS and MVZI, Slovenia; DSI/NRF, South Africa; MICIU/AEI, Spain; SRC and Wallenberg Foundation, Sweden; SERI, SNSF and Cantons of Bern and Geneva, Switzerland; NSTC, Taipei; TENMAK, T\"urkiye; STFC/UKRI, United Kingdom; DOE and NSF, United States of America.

Individual groups and members have received support from BCKDF, CANARIE, CRC and DRAC, Canada; CERN-CZ, FORTE and PRIMUS, Czech Republic; COST, ERC, ERDF, Horizon 2020, ICSC-NextGenerationEU and Marie Sk{\l}odowska-Curie Actions, European Union; Investissements d'Avenir Labex, Investissements d'Avenir Idex and ANR, France; DFG and AvH Foundation, Germany; Herakleitos, Thales and Aristeia programmes co-financed by EU-ESF and the Greek NSRF, Greece; BSF-NSF and MINERVA, Israel; NCN and NAWA, Poland; La Caixa Banking Foundation, CERCA Programme Generalitat de Catalunya and PROMETEO and GenT Programmes Generalitat Valenciana, Spain; G\"{o}ran Gustafssons Stiftelse, Sweden; The Royal Society and Leverhulme Trust, United Kingdom.

In addition, individual members wish to acknowledge support from Armenia: Yerevan Physics Institute (FAPERJ); CERN: European Organization for Nuclear Research (CERN DOCT); Chile: Agencia Nacional de Investigaci\'on y Desarrollo (FONDECYT 1230812, FONDECYT 1240864); China: Chinese Ministry of Science and Technology (MOST-2023YFA1605700, MOST-2023YFA1609300), National Natural Science Foundation of China (NSFC - 12175119, NSFC 12275265); Czech Republic: Czech Science Foundation (GACR - 24-11373S), Ministry of Education Youth and Sports (ERC-CZ-LL2327, FORTE CZ.02.01.01/00/22\_008/0004632), PRIMUS Research Programme (PRIMUS/21/SCI/017); EU: H2020 European Research Council (ERC - 101002463); European Union: European Research Council (BARD No. 101116429, ERC - 948254, ERC 101089007), European Regional Development Fund (SMASH COFUND 101081355, SLO ERDF), Horizon 2020 Framework Programme (MUCCA - CHIST-ERA-19-XAI-00), European Union, Future Artificial Intelligence Research (FAIR-NextGenerationEU PE00000013), Horizon 2020 (EuroHPC - EHPC-DEV-2024D11-051), Italian Center for High Performance Computing, Big Data and Quantum Computing (ICSC, NextGenerationEU); France: Agence Nationale de la Recherche (ANR-21-CE31-0013, ANR-21-CE31-0022, ANR-22-EDIR-0002); Germany: Baden-Württemberg Stiftung (BW Stiftung-Postdoc Eliteprogramme), Deutsche Forschungsgemeinschaft (DFG - 469666862, DFG - CR 312/5-2); China: Research Grants Council (GRF); Italy: Istituto Nazionale di Fisica Nucleare (ICSC, NextGenerationEU), Ministero dell'Università e della Ricerca (NextGeneration PRIN 153D23001490006, NextGeneration PRIN I53D23001490006, NextGenEU I53D23000820006 M4C2.1.1); Japan: Japan Society for the Promotion of Science (JSPS KAKENHI JP22H01227, JSPS KAKENHI JP22H04944, JSPS KAKENHI JP22KK0227, JSPS KAKENHI JP23KK0245, JSPS KAKENHI JP24K23939); Norway: Research Council of Norway (RCN-314472); Poland: Ministry of Science and Higher Education (IDUB AGH, POB8, D4 no 9722), Polish National Science Centre (NCN 2021/42/E/ST2/00350, NCN OPUS 2023/51/B/ST2/02507, NCN OPUS nr 2022/47/B/ST2/03059, NCN UMO-2019/34/E/ST2/00393, UMO-2020/37/B/ST2/01043, UMO-2022/47/O/ST2/00148, UMO-2023/49/B/ST2/04085, UMO-2023/51/B/ST2/00920, UMO-2024/53/N/ST2/00869); Portugal: Foundation for Science and Technology (FCT); Spain: Generalitat Valenciana (Artemisa, FEDER, IDIFEDER/2018/048), Ministry of Science and Innovation (MCIN \& NextGenEU PCI2022-135018-2, MICIN \& FEDER PID2021-125273NB, RYC2019-028510-I, RYC2020-030254-I, RYC2021-031273-I, RYC2022-038164-I); Sweden: Carl Trygger Foundation (Carl Trygger Foundation CTS 22:2312), Swedish Research Council (Swedish Research Council 2023-04654, VR 2021-03651, VR 2022-03845, VR 2022-04683, VR 2023-03403, VR 2024-05451), Knut and Alice Wallenberg Foundation (KAW 2018.0458, KAW 2022.0358, KAW 2023.0366); Switzerland: Swiss National Science Foundation (SNSF - PCEFP2\_194658); United Kingdom: Leverhulme Trust (Leverhulme Trust RPG-2020-004), Royal Society (NIF-R1-231091); United States of America: U.S. Department of Energy (ECA DE-AC02-76SF00515), Neubauer Family Foundation.

%
%


%
\clearpage
\printbibliography
\clearpage
 
\begin{flushleft}
\hypersetup{urlcolor=black}
{\Large The ATLAS Collaboration}

\bigskip

\AtlasOrcid[0000-0002-6665-4934]{G.~Aad}$^\textrm{\scriptsize 104}$,
\AtlasOrcid[0000-0001-7616-1554]{E.~Aakvaag}$^\textrm{\scriptsize 17}$,
\AtlasOrcid[0000-0002-5888-2734]{B.~Abbott}$^\textrm{\scriptsize 123}$,
\AtlasOrcid[0000-0002-0287-5869]{S.~Abdelhameed}$^\textrm{\scriptsize 119a}$,
\AtlasOrcid[0000-0002-1002-1652]{K.~Abeling}$^\textrm{\scriptsize 55}$,
\AtlasOrcid[0000-0001-5763-2760]{N.J.~Abicht}$^\textrm{\scriptsize 49}$,
\AtlasOrcid[0000-0002-8496-9294]{S.H.~Abidi}$^\textrm{\scriptsize 30}$,
\AtlasOrcid[0009-0003-6578-220X]{M.~Aboelela}$^\textrm{\scriptsize 45}$,
\AtlasOrcid[0000-0002-9987-2292]{A.~Aboulhorma}$^\textrm{\scriptsize 36e}$,
\AtlasOrcid[0000-0001-5329-6640]{H.~Abramowicz}$^\textrm{\scriptsize 157}$
% ...
\bigskip
\\

% ...
$^{17}$Department for Physics and Technology, University of Bergen, Bergen; Norway.\\
% ...
$^{30}$Physics Department, Brookhaven National Laboratory, Upton NY; United States of America.\\
% ...
$^{36}$$^{(a)}$Facult\'e des Sciences Ain Chock, Universit\'e Hassan II de Casablanca;$^{(b)}$Facult\'{e} des Sciences, Universit\'{e} Ibn-Tofail, K\'{e}nitra;$^{(c)}$Facult\'e des Sciences Semlalia, Universit\'e Cadi Ayyad, LPHEA-Marrakech;$^{(d)}$LPMR, Facult\'e des Sciences, Universit\'e Mohamed Premier, Oujda;$^{(e)}$Facult\'e des sciences, Universit\'e Mohammed V, Rabat;$^{(f)}$Institute of Applied Physics, Mohammed VI Polytechnic University, Ben Guerir; Morocco.\\
% ...
$^{45}$Physics Department, Southern Methodist University, Dallas TX; United States of America.\\
% ...
$^{49}$Fakult\"{a}t Physik , Technische Universit{\"a}t Dortmund, Dortmund; Germany.\\
% ...
$^{55}$II. Physikalisches Institut, Georg-August-Universit\"{a}t G\"ottingen, G\"ottingen; Germany.\\
% ...
$^{104}$CPPM, Aix-Marseille Universit\'e, CNRS/IN2P3, Marseille; France.\\
% ...
$^{119}$$^{(a)}$New York University Abu Dhabi, Abu Dhabi;$^{(b)}$United Arab Emirates University, Al Ain; United Arab Emirates.\\
% ...
$^{123}$Homer L. Dodge Department of Physics and Astronomy, University of Oklahoma, Norman OK; United States of America.\\
% ...
$^{157}$Raymond and Beverly Sackler School of Physics and Astronomy, Tel Aviv University, Tel Aviv; Israel.\\
% ...
$^{a}$ Also at Affiliated with an institute formerly covered by a cooperation agreement with CERN.\\
$^{b}$ Also at An-Najah National University, Nablus; Palestine.\\
$^{c}$ Also at Borough of Manhattan Community College, City University of New York, New York NY; United States of America.\\
$^{d}$ Also at Center for Interdisciplinary Research and Innovation (CIRI-AUTH), Thessaloniki; Greece.\\
$^{e}$ Also at Centre of Physics of the Universities of Minho and Porto (CF-UM-UP); Portugal.\\
$^{f}$ Also at CERN, Geneva; Switzerland.\\
% ...

\end{flushleft}


%

%
%
\clearpage

\end{document}